\documentclass[prd, aps, reprint, amsfonts, amssymb, amsmath, showpacs, nofootinbib, superscriptaddress, preprintnumbers, ]{revtex4-1}
\usepackage{graphicx, multirow,soul}
\usepackage[citecolor=blue]{hyperref}
\usepackage[all]{hypcap}
\usepackage{url}
\usepackage{minibox}
\usepackage{color}
\usepackage{subfigure}
\usepackage{slashed} 
\usepackage{float}
\usepackage{amssymb}
\usepackage{lipsum}
\newcommand{\overbar}[1]{\mkern 1.1mu\overline{\mkern-1.1mu#1\mkern-1.1mu}\mkern 1.1mu}
\newcommand{\equaref}[1]{Eq.~(\ref{#1})}

\newcommand{\figref}[1]{Fig.~\ref{#1}}

\newcommand{\secref}[1]{Section~\ref{#1}}
\newcommand{\appref}[1]{Appendix~\ref{#1}}
\newcommand{\tabref}[1]{Table~\ref{#1}}

\begin{document}

\title{Three-Flavoured Non-Resonant Leptogenesis at Intermediate Scales}

\author{K. Moffat}\email{kristian.p.moffat@durham.ac.uk}
\affiliation{Institute for Particle Physics Phenomenology, Department of
Physics, Durham University, South Road, Durham DH1 3LE, United Kingdom.}
\author{S. Pascoli}\email{silvia.pascoli@durham.ac.uk}
\affiliation{Institute for Particle Physics Phenomenology, Department of
Physics, Durham University, South Road, Durham DH1 3LE, United Kingdom.}
\author{S.T. Petcov}\email{petcov@sissa.it}
\affiliation{SISSA/INFN, Via Bonomea 265, I-34136 Trieste, Italy.}
\affiliation{Kavli IPMU (WPI), University of Tokyo, 5-1-5 Kashiwanoha, 277-8583 Kashiwa, Japan.}	
\author{H. Schulz}\email{schulzhg@ucmail.uc.edu}
\affiliation{Department of Physics, University of Cincinnati, Cincinnati, OH 45219, USA.}
\author{J. Turner}\email{jturner@fnal.gov}
\affiliation{Theoretical Physics Department, Fermi National Accelerator Laboratory, P.O. Box 500, Batavia, IL 60510, USA.}

\date{\today}

\begin{abstract}
Leptogenesis can successfully explain the matter-antimatter asymmetry via out-of-equilibrium decays of heavy Majorana neutrinos in the early Universe. In this article, we focus on non-resonant thermal leptogenesis and  the possibility of lowering its scale. In order to do so,
we calculate the lepton asymmetry produced from the 
decays of one and two heavy Majorana neutrinos using 
 three-flavoured density matrix equations  in an exhaustive exploration of the model parameter space. 
We find regions of the parameter space where thermal leptogenesis is viable at intermediate scales, $T\sim 10^{6}$ GeV. However, the viability of thermal leptogenesis at such scales requires a certain degree of cancellation between the tree and one-loop level contribution to the light neutrino mass matrix and we quantify such fine-tuning.
\end{abstract}
%
\preprint{\minibox[]{IPPP/18/25\\ FERMILAB-PUB-18-100-T\\ IPMU18-0062\\ SISSA 17/2018/FISI}}
 \pacs{98.80.cq,14.60.Pq}
\maketitle

\section{Introduction}\label{sec:introduction}
There is overwhelming experimental evidence for an excess of matter over antimatter in the Universe. This asymmetry  remains a fundamental and unresolved mystery
 whose explanation demands new physics beyond the Standard Model (SM).
The baryon asymmetry may be parametrised by the baryon-to-photon ratio, $\eta_B$, which is 
defined to be
\[
\eta_B \equiv \frac{n_B-n_{\overbar{B}}}{n_\gamma},
\]
where $n_B$, $n_{\overbar{B}}$ and $n_\gamma$ are the number densities of baryons, anti-baryons and photons, respectively.
This quantity can be measured using two independent methods that
probe the Universe at different stages of its evolution. Big-Bang nucleosynthesis, BBN,  \cite{Patrignani:2016xqp} and 
Cosmic Microwave Background radiation, CMB, data  \cite{Ade:2015xua} give
\[
\begin{aligned}
{\eta_{B}}_{\text{BBN}}  & = \left(5.80-6.60\right)\times 10^{-10}, \\
{\eta_{B}} _{\text{CMB}}& = \left(6.02-6.18\right)\times 10^{-10},
\end{aligned}
\]
at 95$\%$ CL, respectively. As the  uncertainties of the CMB measurement are smaller than those from BBN, we shall apply the CMB value throughout this
work.

In order to dynamically produce the observed baryon asymmetry in the early Universe, the mechanism of interest must satisfy the Sakharov conditions \cite{Sakharov:1967dj}\footnote{This statement implicitly assumes quantum field theory is  CPT invariant.}: $B$ (or $L$) violation; C/CP violation and
a departure from thermal equilibrium. Leptogenesis \cite{Fukugita:1986hr} satisfies these conditions and produces a lepton asymmetry which is subsequently  partially converted to a baryon asymmetry via $B+L$ violating sphaleron processes \cite{Khlebnikov:1988sr}.

Leptogenesis is particularly appealing as it typically takes place in models of neutrino masses, simultaneously explaining the baryon asymmetry and the smallness of the neutrino masses. In its simplest realisation, the lepton asymmetry is generated via out-of-equilibrium decays of heavy Majorana neutrinos \cite{Covi:1996wh,Covi:1996fm,Pilaftsis:1997jf,Buchmuller:1997yu}. This process occurs approximately when the temperature, $T$, of the Universe equals the mass scale of the decaying heavy Majorana neutrino.

In general, the scale of thermal leptogenesis is not explored below the Davidson-Ibarra (DI) bound, $M_1\approx 10^{9}$~GeV \cite{Davidson:2002qv}. Davidson and Ibarra found an upper bound, proportional to $M_{1}$, on the absolute value of the CP-asymmetry of the decays of the lightest heavy Majorana neutrino. This constrains the regions of parameter space in which successful leptogenesis may occur as a function of $M_1$. This translates into the DI bound on $M_{1}$ itself as the minimum value required for successful leptogenesis.
There have been a number of in-depth numerical studies which support this bound and require $M_{1}\geq10^{9}$ GeV \cite{Buchmuller:2002rq,Ellis:2002xg} in conjunction with a bound on the lightest
neutrino mass, $m_{1}\leq 0.1$ eV \cite{Buchmuller:2003gz,Buchmuller:2002rq,Buchmuller:2004nz}. 

The original derivation of this bound makes some simplifying analytical assumptions and hence is subject to three caveats: only the lightest heavy Majorana neutrino decays; the heavy Majorana neutrino mass spectrum is hierarchical; and flavour effects, which account for the differing interaction rates of the charged-lepton decay products of the heavy Majorana neutrinos, are ignored. In this work, we shall investigate scenarios of three-flavoured
thermal leptogenesis in a more general setting than these conditions allow. We shall then consider lower heavy Majorana neutrino masses at scales $M_1 \approx 10^6 \text{ GeV}$. Given the existence of low-scale leptogenesis models at the TeV scale, we shall refer to this as ``intermediate" scale leptogenesis.

There are several reasons to explore leptogenesis at intermediate scales. Firstly, the introduction of heavy neutrinos to the SM leads to a correction to the Higgs mass which may potentially be unnaturally large. This is because the correction to the electroweak parameter $\mu^2$ (the negative of the coefficient in the quadratic term of the Higgs potential), is proportional to the light neutrino masses and to $M^3$, with $M$ the heavy Majorana neutrino mass scale \cite{Vissani:1997ys}. In order to avoid corrections to $\delta \mu^2$ larger than say $1 \text{ TeV}^2$ one requires the lightest pair of Majorana neutrino masses to have $M_{1}<4\times10^{7}$ GeV and $M_{2}<7\times10^{7}$ GeV \cite{Clarke:2015gwa}. Secondly, there is a tendency for baryogenesis models to reside at the GeV- or GUT-scales which leaves intermediate scales relatively unexplored. Finally, thermal leptogenesis at intermediate scales may resolve a problem that arises in the context of supersymmetric models which include gravitinos in their particle spectrum. Gravitinos have interaction strengths that are suppressed by the Planck scale and consequently are long-lived and persist into the nucleosynthesis era. The decay products of the gravitinos can destroy ${}^\text{4}\text{He}$ and D nuclei \cite{Khlopov:1984pf,Ellis:1984eq} and ruin the successful predictions of nucleosynthesis. Thus, in order reduce the number of gravitinos present at this stage, one requires a reheating temperature less than a few times $10^9$ GeV depending on the gravitino mass~\cite{Kawasaki:2008qe}. 

The scale of leptogenesis may be lowered through the introduction of a symmetry to the SM. In \cite{Racker:2012vw}, non-resonant thermal leptogenesis is explored at intermediate scales in the context of small $B-L$ violation. It is shown that the DI bound may be evaded because, in the context of this near-symmetry, the lepton number conserving part of the CP asymmetries can be enhanced as they are not connected to light neutrino masses. It is shown that the scale may be lowered to $10^6$ GeV. An alternative symmetry-based approach is to introduce supersymmetry in which one may also reduce the scale of leptogenesis to intermediate scales. In this context, the bound on the absolute value of the CP-asymmetry found by Davidson and Ibarra is greatly enhanced. Consequently, the DI bound is lowered thus allowing for the possibility of intermediate scale leptogenesis~ \cite{Raidal:2004vt}.

Beyond the application of supersymmetry and heavy pseudo-Dirac neutrinos, there are other means of lowering the scale of leptogenesis; if the decaying heavy Majorana neutrinos are near-degenerate in mass,
the indirect CP-violation may be resonantly enhanced \cite{Covi:1996wh,Covi:1996fm,Pilaftsis:1997jf,Buchmuller:1997yu} and  subsequently the mechanism may be lowered to the TeV scale. This has been explored in the context of type-I \cite{Mohapatra:1979ia,GellMann:1980vs,Yanagida:1979as,Minkowski:1977sc}, II \cite{Mohapatra:1980yp,Magg:1980ut,Lazarides:1980nt,Wetterich:1981bx} and III \cite{Foot:1988aq,Ma:1998dn} seesaw mechanisms. Another mechanism, proposed by \cite{Akhmedov:1998qx}, is one in which
 leptogenesis is achieved via CP-violating heavy Majorana neutrino oscillations~\cite{Drewes:2017zyw, Hernandez:2016kel, Dev:2017trv, Antusch:2017pkq}. The generation of the lepton asymmetry takes place  close to the electroweak scale and the associated GeV-scale heavy Majorana neutrinos may be searched for at a variety of experiments such as LHCb \cite{Antusch:2015mia,Milanes:2016rzr}, BELLE II \cite{Canetti:2014dka} and the proposed facility, SHiP \cite{Anelli:2015pba,Alekhin:2015byh,Graverini:2015dka}. 
 Although, leptogenesis via oscillations is a testable and plausible explanation of the baryon asymmetry, it has been shown its simplest formulations require a  certain amount ($\sim 10^5$) of fine-tuning \cite{Shuve:2014zua}.

In this article, we revisit the question: \emph{how low can the scale of thermal leptogenesis go?} 
We focus solely on the possibility that the heavy neutrinos are Majorana in nature and find thermal leptogenesis is possible at intermediate scales without resonant effects. In addition, we present an in-depth numerical study of the dependence of the baryon asymmetry produced from non-supersymmetric thermal leptogenesis on the low and high-scale model parameters.

The work presented in this paper is structured as follows: in \secref{sec:numass} we review the origins of light neutrino masses in the type-I seesaw framework, further we review the Casas-Ibarra parametrisation of the Yukawa matrix and then introduce a modification of this parametrisation in the presence of large radiative corrections. We end this section by introducing a measure of fine-tuning in the context of the neutrino masses.
In \secref{sec:TL} we discuss the motivations for and some theoretical aspects of thermal leptogenesis. 
We follow in \secref{sec:DME} with the density matrix equations we shall solve to calculate the lepton asymmetry. We demonstrate in \secref{sec:3flavours}, that the fully flavoured Boltzmann equations, which do not incorporate flavour oscillations, 
  may significantly qualitatively differ from the lepton asymmetry calculated from the density matrix equations and  justify the use of semi-classical density matrix equations rather than kinetic equations derived from first principles non-equilibrium quantum field theory (NE-QFT). Our numerical methods are described in \secref{sec:NT}. The results of our numerical study for one and two decaying heavy Majorana neutrinos
  are presented in \secref{sec:1DS} and  \secref{sec:2DS} respectively. In \secref{sec:ftdiscus} we explore the analytical consequences of the numerical results and provide an explanation for the fine-tuning. Finally, we summarise and make some  concluding remarks in \secref{sec:discus}.
\begin{figure*}\label{fig:treeandloopdiagrams}
\includegraphics[width=0.95\textwidth]{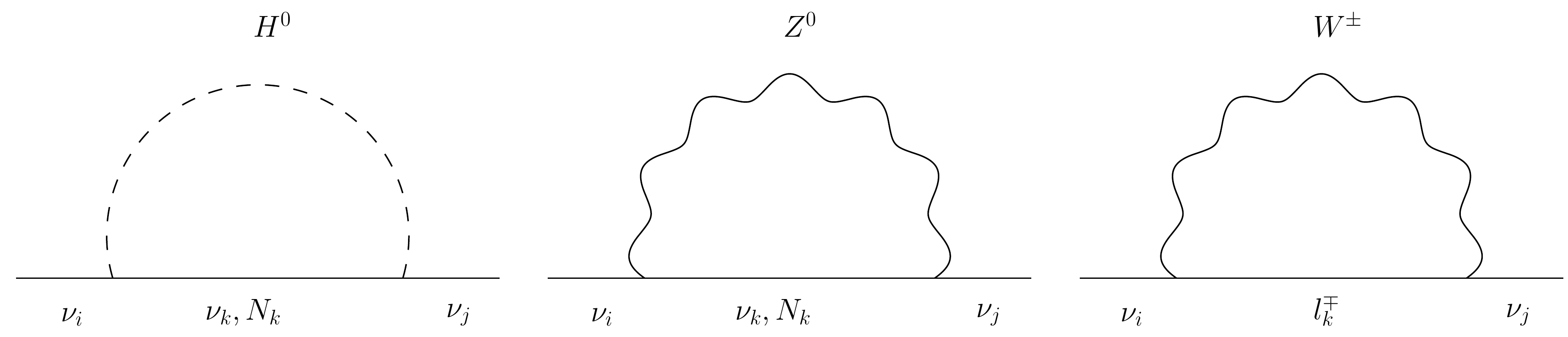}
\caption{One-loop level diagrams showing the physical particle contributions to neutrino mass at one-loop. The $W$-boson contribution proportional to the momentum is neglected. When, the external momentum is taken to be zero, the $Z$- and Higgs-boson contributions ($Z^0$, $H^0 \equiv \phi^0 - v$) together provide a correction to the tree-level mass that is finite and independent of the choice of renormalisation scale.}
\end{figure*}
%
\section{Neutrino Masses and the Type-I Seesaw Mechanism}\label{sec:numass}
\begin{table}[h!]\label{tab:gfd}
\centering
\begin{tabular}{ c | c   }
   &   best-fit  $\pm $1$\sigma$ range  \\
 \hline \hline
$\theta_{13}(^\circ)$ & $8.52^{+0.15}_{-0.15}$ \\
$\theta_{12}(^\circ)$ & $33.63^{+0.78}_{-0.75}$ \\
$\theta_{23}(^\circ)$ & $48.7^{+1.4}_{-6.9}$ \\
$\quad\delta(^\circ)$ & $228^{+51}_{-33}$ \\
$\frac{\Delta m^2_{21}}{10^{-5}\text{eV}^2}$ & $7.40^{+0.21}_{-0.20}$ \\
$\frac{\Delta m^2_{31}}{10^{-3}\text{eV}^2} $ (NO)& $2.515^{+0.035}_{-0.035}$ \\
$\frac{\Delta m^2_{32}}{10^{-3}\text{eV}^2} $  (IO)& $-2.483^{+0.034}_{-0.035}$
\end{tabular}\caption{Best fit and 1$\sigma$ ranges from global fits to neutrino data \cite{Esteban:2016qun}. }\label{tab:table}
\end{table}
 In addition to the excess of matter over antimatter, the SM cannot account for non-zero neutrino masses which 
were discovered through the observation of neutrino oscillations by Super-Kamiokande  twenty years ago~\cite{Fukuda:1998mi} and were subsequently confirmed by a large number of other oscillation experiments. These oscillations occur because the neutrino flavour and mass eigenstates do not coincide. Such a misalignment between bases may be described by the Pontecorvo-Maki Nakagawa-Sakata, PMNS, matrix $U$, which relates the flavour and mass eigenstates of the light neutrinos through
\[
\nu_{\alpha L} = \sum_{i=1,2,3} U_{\alpha i} \nu_i,
\]
and is conventionally parametrised as
 \footnote{We have adopted the PDG
parametrisation of the PMNS matrix \cite{Patrignani:2016xqp}.}
\[
\begin{aligned}
U =&\begin{pmatrix}
1 & 0 & 0 \\
0 & c_{23} & s_{23}  \\
0 &- s_{23} & c_{23} 
\end{pmatrix}
\begin{pmatrix}
c_{13} & 0 & s_{13}e^{-i\delta} \\
0 & 1 & 0 \\
-s_{13}e^{i\delta} &0 & c_{13}
\end{pmatrix}\\
\times&\begin{pmatrix}
c_{12} & s_{12} & 0\\
-s_{12} & c_{12} & 0\\
0 & 0 & 1
\end{pmatrix}
\begin{pmatrix}
1 & 0 & 0\\
0&e^{i\frac{\alpha_{21}}{2}} & 0\\
0 & 0 &  e^{i\frac{\alpha_{31}}{2}}
\end{pmatrix},
\end{aligned}
\]
 where $c_{ij} \equiv \cos\theta_{ij}$, $s_{ij} \equiv \sin\theta_{ij}$, $\delta$ is the Dirac  phase and $\alpha_{21}$, $\alpha_{31}$ are the Majorana phases \cite{Bilenky:1980cx}
 which are physical if and only if neutrinos are Majorana in nature. 
 
The current best-fit and 1$\sigma$ range of the neutrino parameters~\cite{Esteban:2016qun} are provided in \tabref{tab:gfd}.
The Dirac CP-violating phase, $\delta$, enters the neutrino oscillation probabilities sub-dominantly 
and remains mostly unconstrained by experimental data. However, as has been anticipated~\cite{Huber:2004ug}, the complementarity of long-baseline experiments such as T2K \cite{Abe:2011ks} and NO$\nu$A \cite{Ayres:2004js} with reactor experiments like Daya-Bay \cite{An:2012eh}, RENO \cite{Ahn:2012nd} and Double-Chooz \cite{Ardellier:2006mn}, have begun to show slight sensitivity to the value of $\delta$.

With information on neutrino masses given by oscillation experiments, another crucial question is whether they are Dirac or Majorana particles. The nature of neutrinos is of fundamental importance as it relates to lepton number violation. The most sensitive process to this is neutrinoless double beta decay which can also provide some information on the neutrino mass spectrum~\cite{Bilenky:2001rz,Bilenky:1996cb,Alfonso:2015wka,KlapdorKleingrothaus:2000sz,Agostini:2018tnm,Bongrand:2015yfa,Yang:2017lnf,Ouellet:2016moa,Pattavina:2016kqn,Yu:2018due,Gomez-Cadenas:2013lta,Lozza:2016rwo,Torre:2014tma} and the Majorana phases \cite{Bilenky:1980cx,Schechter:1980gr,Pascoli:2001by,Pascoli:2002xq,Pascoli:2002qm,Pascoli:2002ae,Barger:2002vy} ($\alpha_{21}, \alpha_{31} = [0,720]^{\circ}$). Constraints on the absolute mass scale can be derived from $^3H \beta$-decay experiments \cite{Weinheimer:2003fj} such as Mainz \cite{Kraus:2004zw} and Troitzk \cite{Lobashev:2001uu} which place an upper limit of the electron antineutrino mass of $2.2$ eV. The experiment KATRIN~\cite{Osipowicz:2001sq} will be able to reduce this limit by an order of magnitude to $0.2$ eV or make a measurement if the mass is larger than $0.35$ eV.

In addition cosmology provides complementary bounds on the sum of the neutrino masses thanks to the imprint that neutrinos leave on the CMB and large-scale structure (LSS) in the early Universe. These bounds are significantly more stringent than the bounds from tritium-decay experiments with the $95\%$ CL constraint $\sum m_{\nu}\leq0.2\,\text{ eV}$~\cite{Ade:2015xua}.
In order to account for different analyses and underlying cosmological models we shall impose a constraint\footnote{For our best-fit points we find none that exceed $\sum m_{\nu} = 0.63$ eV (see \appref{sec:2loops}) and thus all are in within the more relaxed cosmological bound $\sum m_{\nu} < 0.72$ eV provided by \textit{Planck} TT + lowP~\cite{Ade:2015xua}.}
\[
\sum m_{\nu} \leq 1.0\, \text{ eV},
\]
throughout this work.
%
%
As discussed previously, the SM cannot explain neutrino masses in its minimal form. Arguably the simplest extension of the SM that incorporates small neutrino masses is the type-I seesaw mechanism \cite{Minkowski:1977sc,Yanagida:1979as,GellMann:1980vs}. This mechanism introduces a set of heavy Majorana neutrino fields $N_i$ and augments the SM Lagrangian to include the following terms
\begin{equation}
\mathcal{L} = i\overline{N_{i}}\slashed{\partial}N_{i}  -Y_{\alpha i}\overline{L_{\alpha}}\tilde{\Phi}N_{i}-\frac{1}{2}M_{i}\overline{N^c_{i}}N_{i} + \text{h.c.},
\end{equation}
in which $Y$ is the Yukawa coupling and $\Phi$ the Higgs $SU(2)$ doublet, $\Phi^T = \left( \phi^+, \phi^0 \right)$ and $\tilde{\Phi} = i \sigma_2 \Phi^*$, and $L^T = \left( \nu^T_L, l^T_L  \right)$ is the leptonic $SU(2)$ doublet. For convenience we have chosen, without loss of generality, the basis in which the Majorana mass term is diagonal. In our work, we shall assume that there are three heavy Majorana neutrinos $N_i$, with a mildly hierarchical mass spectrum in which $M_1<M_2<M_3$.

After electroweak symmetry breaking, at the tree-level, the light neutrino mass matrix (at first order in the seesaw expansion) is~\footnote{Here we have chosen to work with a convention in which this term lacks the usual minus sign. We choose the sign of the one-loop contribution to be consistent with this.}
\begin{equation}\label{eq:treemass}
m^{\text{tree}} \approx m^{}_D M^{-1} m_D^T,
\end{equation}
in which $m_D = v Y$ is the Dirac mass matrix that develops once the Higgs acquires the vacuum expectation value $v$. 

The tree-level mass matrix is not generically an accurate approximation of the light neutrino mass matrix over all regions of the parameter space. This is because there is no guarantee that the radiative corrections to the neutrino self-energy are negligible. Indeed there exist regions of parameter space in which radiative corrections are comparable to, or larger than, the tree-level contribution to the mass (see \tabref{tab:mtm1lm2l}). For this reason, we find it necessary to incorporate the effects of the one-loop contribution to the masses given by \cite{LopezPavon:2012zg}
\[
\begin{aligned}
& m^{\text{1-loop}} = \\
& - m^{}_D \left( \frac{M}{32 \pi^2 v^2} \left(\frac{\log\left(\frac{M^2}{m_H^2}\right)}{\frac{M^2}{m_H^2}-1} + 3 \frac{\log\left(\frac{M^2}{m_Z^2}\right)}{\frac{M^2}{m_Z^2}-1}\right) \right) m_D^T,\\
& = - \frac{1}{32 \pi^2 v^2} m_D \text{diag} \left(g\left(M_{1}\right), g\left(M_{2}\right), g\left(M_{3}\right) \right) m_D^T,
\end{aligned}
\]
with
\[
g\left(M_i\right) \equiv M_i \left(\frac{\log\left(\frac{M_i^2}{m_H^2}\right)}{\frac{M_i^2}{m_H^2}-1} + 3 \frac{\log\left(\frac{M_i^2}{m_Z^2}\right)}{\frac{M_i^2}{m_Z^2}-1}\right),
\]
giving a total light neutrino mass of
\[
m_{\nu} = m^{\text{tree}}+m^{\text{1-loop}}.
\]
The contribution from two-loop corrections is usually small as these will be suppressed by an extra factor of the Yukawa couplings squared and a further factor $\mathcal{O}(10^{-2})$ from the loop integral. This is discussed in more detail and estimated in \appref{sec:2loops}.

The matrix $m_{\nu}$ is rewritten in the Takagi factorised form through the PMNS matrix with
\[
m_{\nu}  = U \hat{m}_{\nu} U^T,
\]
where $\hat{m}_{\nu}$ is the positive diagonal matrix of light neutrino masses.

By analogy with the parametrisation of Casas and Ibarra (CI)~\cite{Casas:2001sr}, the Yukawa matrix can be written in the following way to include the loop-level corrections~\cite{Lopez-Pavon:2015cga}
\begin{equation}\label{eq:CIloop}
Y=\frac{1}{v} m_D=\frac{1}{v}U\sqrt{\hat{m}_{\nu}}R^T\sqrt{f(M)^{-1}},
\end{equation}
with $R$ a complex orthogonal matrix and
\[
\begin{aligned}
f(M) & \equiv M^{-1} - \frac{M}{32 \pi^2 v^2} \left(\frac{\log\left(\frac{M^2}{m_H^2}\right)}{\frac{M^2}{m_H^2}-1} + 3 \frac{\log\left(\frac{M^2}{m_Z^2}\right)}{\frac{M^2}{m_Z^2}-1}\right) \\
 & = 
 \text{diag} \left(\frac{1}{M_{1}},\frac{1}{M_{2}},\frac{1}{M_{3}} \right) \\
& -\frac{1}{32 \pi^2 v^2} \text{diag} \left(g\left(M_{1}\right), g\left(M_{2}\right), g\left(M_{3}\right) \right).
\end{aligned}
\]

This parametrisation expresses the Yukawas in terms of both low energy measurable parameters (in $m_{\nu}$ and $U$) and high energy, currently untestable parameters (in the complex orthogonal matrix $R$ and the Majorana mass matrix $M$). An advantage to using this parametrisation is that one can automatically achieve the correct structure of mass-squared differences.

Naturally, when the loop-corrections are negligible we may replace $f(M)^{-1}$ with $M$ and \equaref{eq:CIloop} reduces to the usual CI parametrisation. Throughout the remainder of this work, we shall apply the parametrisation of \equaref{eq:CIloop} to ensure radiative corrections are accounted for.
We parametrise the $R$-matrix in the following way
\begin{equation}
R=\begin{pmatrix}
1 & 0 & 0 \\
0 & c_{\omega_{1}} & s_{\omega_{1}} \\
0 &- s_{\omega_{1}} & c_{\omega_{1}} 
\end{pmatrix}
\begin{pmatrix}
c_{\omega_{2}} & 0 & s_{\omega_{2}} \\
0 & 1 & 0\\
-s_{\omega_{2}} & 0 & c_{\omega_{2}} 
\end{pmatrix}\\
\begin{pmatrix}
c_{\omega_{3}} & s_{\omega_{3}} & 0\\
-s_{\omega_{3}} & c_{\omega_{3}} & 0\\
0 & 0 & 1
\end{pmatrix},
\end{equation}
where $c_{\omega_{i}} \equiv \cos\omega_{i}$, $s_{\omega_{i}} \equiv \sin\omega_{i}$  and the complex angles are given by $\omega_{i} \equiv x_{i}+iy_{i}$ with $\lvert x_{i}\rvert, \lvert y_{i}\rvert\leq180^{\circ}$ for $i= 1, 2, 3$.

In general the structure of the $R$-matrix cannot be constrained; however in \cite{Chen:2016ptr}, the authors demonstrated  if the heavy Majorana neutrino mass matrix is invariant under a
residual CP-symmetry, the $R$-matrix is constrained to be real or purely imaginary \cite{Chen:2016ptr}. It was shown in \cite{Pascoli:2006ie,Pascoli:2006ci} that the PMNS phases were a sufficient source of CP-violation to generate the observed baryon asymmetry if thermal leptogenesis occurred during an era in which flavour effects were non-negligible. 

Both the light and heavy Majorana neutrino mass matrices  determine  the  structure of the Yukawa matrix. Once
 the value of the lightest neutrino mass is fixed, we shall  assume the best-fit value for the solar and atmospheric mass squared splittings
 and hence the light neutrino mass matrix is determined. In addition,
we constrain the sum of the neutrino masses such that it is within experimental bounds, specifically $\sum m_{\nu} \leq 1.0$ eV.
In order to ensure the lepton asymmetry does not become \emph{resonantly enhanced} \cite{Pilaftsis:2003gt}, we choose the right-handed neutrino mass spectrum to be mildly hierarchical: $M_{2}>3M_{1}$ and $M_{3}>3M_{2}$ \cite{Blanchet:2008pw}. In summary, the \emph{model parameter space} of leptogenesis, as given by the Casas-Ibarra parametrisation of \equaref{eq:CIloop}, is 18-dimensional and we shall henceforth 
denote this quantity as $\mathbf{p}$. 

In anticipation of our results, we shall define a parameter that quantifies the degree of fine-tuning for a given solution. First we define the fine-tuning measure 
to be
\begin{equation}\label{eq:FT}
\text{F.T.} \equiv \frac{\sum_{i=1}^{3}\text{SVD}[m^{\text{1-loop}}]_\text{i}}{\sum_{i=1}^{3}\text{SVD}[m_{\nu}]_\text{i}},
\end{equation}
where $\text{SVD}[m^{\text{1-loop}}]_\text{i}$ and $\text{SVD}[m_{\nu}]_\text{i}$ denote the $i$th singular values of the $m^{\text{1-loop}}$ and $m_{\nu}$ neutrino mass matrices respectively.
As the neutrino mass matrix is the sum of the tree- and one-loop contributions, a cancellation between the two leads to the fine-tuning measure being larger than unity. In the limit that the tree-level contribution dominates, the fine-tuning measure tends to zero.

We  declare a technical limitation that we shall accept in this work. In lowering the value of $M_1$, we find fine-tuned solutions in which the tree-level and one-loop contributions cancel to produce a neutrino mass matrix smaller than either alone. However, the higher-order radiative corrections cannot be assumed to perform a similar cancellation and thus we should take care that the two-loop contribution is not too large in comparison with the one-loop correct light neutrino mass matrix (see \appref{sec:2loops}).

%
\section{Thermal Leptogenesis}\label{sec:TL}
Minimal thermal leptogenesis \cite{Fukugita:1986hr} proceeds via the out-of-equilibrium decays of the heavy Majorana neutrinos. CP violation, arising from the interference between tree- and loop-level diagrams, causes the CP-asymmetric decays of the heavy Majorana neutrinos which induce a lepton asymmetry. The production of the asymmetry from decays competes with a washout from inverse decays of the heavy Majorana neutrinos. The final lepton asymmetry is partially reprocessed to a baryon asymmetry via electroweak sphaleron processes which occur at unsuppressed rates at temperatures above the electroweak scale \cite{Khlebnikov:1988sr}. 

The  time evolution of the lepton asymmetry may be calculated using semi-classical or NE-QFT methods.
 In both approaches, 
 these kinetic equations  account for the decay of the heavy Majorana neutrino and washout processes. In the simplest formulation, these kinetic equations are in the one-flavoured regime, in which only a single flavour of charged lepton is  accounted for. This regime is only realised at sufficiently high temperatures ($T\gg 10^{12}$ GeV) when the rates of processes mediated by the charged lepton Yukawa couplings are out of thermal equilibrium and therefore there is a single charged lepton flavour state  which is a coherent superposition of the three flavour eigenstates. However, if leptogenesis occurs at lower temperatures ($10^{9}\ll T \ll10^{12}$ GeV), 
scattering induced by the tau  Yukawa couplings can cause the single charged lepton flavour to decohere and the dynamics of leptogenesis must be described in terms of two flavour eigenstates.
  In such a regime, a density matrix formalism \cite{Barbieri:1999ma, Abada:2006fw,DeSimone:2006nrs,Blanchet:2006ch,Blanchet:2011xq} allows for a more general description than semi-classical Boltzmann equations, since it is possible to calculate the asymmetry in intermediate regimes where the one and two-flavoured treatments are inadequate.

\subsection{Density Matrix Equations}\label{sec:DME}
As previously mentioned, the most basic leptogenesis calculations were performed in the single lepton flavour regime.  
The one-flavoured regime is realised at high temperatures ($T\gg10^{12}$ GeV) where the leptons and anti-leptons that couple to the right-handed neutrinos, $N_{i}$, maintain their coherence throughout the era of lepton asymmetry production. This implies there is a single lepton (anti-lepton) flavour, $\ell_{1}$ ($\overline{\ell_{1}}$), which may be described as a coherent superposition of charged lepton flavour-states, ($e$, $\mu$, $\tau$),
\[
\begin{aligned}
 \vert \ell_{1}\rangle &  \equiv \sum_{\alpha=e, \mu, \tau}c_{1\alpha} \vert \ell_{\alpha}\rangle, \quad  c_{1\alpha} \equiv \langle \ell_\alpha \vert \ell_{1}\rangle, \\
 \vert \overline{\ell_{1}}\rangle & \equiv \sum_{\alpha=e, \mu, \tau}\overline{c_{1\alpha}} \vert \overline{\ell_{\alpha}} \rangle, \quad  \overline{c_{1\alpha}} \equiv \langle  \overline{\ell_{\alpha}} \vert \overline{\ell_{1}}\rangle,
\end{aligned}
\]
where the amplitudes $c_{i\alpha}$ are functions of the Yukawa matrix, $Y$.  In such a regime, the interaction rate mediated by the SM lepton Yukawas are out of thermal equilibrium ($\Gamma_{\alpha}<H$) and this implies there are no means of distinguishing between the three leptonic flavours. However, if leptogenesis occurs at  lower scales ($10^{9}\lesssim T \text{ (GeV)} \lesssim 10^{12}$), the interactions mediated by the tau charged lepton  Yukawa come into thermal equilibrium ($\Gamma_{\tau} > H$) and the Universe may distinguish between $\tau$ and  $\tau^{\prime}$, where  $\tau^{\prime}$ is a linear combination of the electron and muon flavoured leptons orthogonal to $\tau$. The one-  and fully two-flavoured description of leptogenesis is appropriate at $T\gg10^{12}$ GeV and $10^{9}\ll T \ll 10^{12}$ GeV, respectively. There exists the possibility that thermal leptogenesis occurs at even lower temperatures, $T<10^{9}$ GeV, during
which the  interactions mediated by the muon have equilibrated. In such a regime the kinetic equations should be given in terms of all three lepton flavours. In this work, we shall focus on this particular scenario. 
 Our discussion of the density matrix equations and notation will closely follow the prescription of \cite{Blanchet:2011xq} where the theoretical background and derivation of the density matrix equations are fully presented. We refrain from rederiving the details of this formalism and instead refer the interested reader to the aforementioned reference.  The most general form of the density matrix equations, assuming three decaying heavy Majorana neutrinos,  is given by
\begin{equation}\label{eq:full3}
\begin{aligned}
\frac{dn_{N_{1}}}{dz}=&-D_{1}(n_{N_{1}}-n^\text{eq}_{N_{1}})\\
\frac{dn_{N_{2}}}{dz}=&-D_{2}(n_{N_{2}}-n^\text{eq}_{N_{2}})\\
\frac{dn_{N_{3}}}{dz}=&-D_{3}(n_{N_{3}}-n^\text{eq}_{N_{3}})\\
 \frac{dn_{\alpha\beta}}{dz} =&\epsilon^{(1)}_{\alpha\beta}D_{1}(n_{N_{1}}-n^\text{eq}_{N_{1}})-\frac{1}{2}W_{1}\left\{P^{0(1)},n\right\}_{\alpha\beta} \\
+& \epsilon^{(2)}_{\alpha\beta}D_{2}(n_{N_{2}}-n^\text{eq}_{N_{2}})-\frac{1}{2}W_{2}\left\{P^{0(2)},n\right\}_{\alpha\beta} \\
+&\epsilon^{(3)}_{\alpha\beta}D_{3}(n_{N_{3}} -n^\text{eq}_{N_{3}})-\frac{1}{2}W_{3}\left\{P^{0(3)},n\right\}_{\alpha\beta} \\
-&\frac{\Im(\Lambda_{\tau})}{Hz}\left[\begin{pmatrix}1&0&0\\ 0&0&0 \\
0&0&0 \end{pmatrix},\left[\begin{pmatrix}1&0&0\\ 0&0&0 \\
0&0&0 \end{pmatrix},n\right]\right]_{\alpha\beta}\\
-&\frac{\Im(\Lambda_{\mu})}{Hz}\left[\begin{pmatrix}0&0&0\\ 0&1&0 \\
0&0&0 \end{pmatrix},\left[\begin{pmatrix}0&0&0\\ 0&1&0 \\
0&0&0 \end{pmatrix},n\right]\right]_{\alpha\beta},
\end{aligned}
\end{equation}
where  Greek letters denote flavour indices, $n_{N_{i}}$ ($i=1, 2, 3$) is the abundance of the $i$th heavy Majorana neutrino\footnote{This quantity is normalised to a co-moving volume containing one right-handed neutrino which is ultra-relativistic and in thermal equilibrium.}, $n^\text{eq}_{N_{i}}$ the equilibrium distribution of the $i$th heavy Majorana neutrino, $D_i$ ($W_i$) denotes the decay (washout) corresponding to the $i$th heavy Majorana neutrino, and are given by \cite{Buchmuller:2004nz}
\begin{equation}
D_{i}(z) = K_{i} x_{i} z \frac{\mathcal{K}_1 (z_i)}{\mathcal{K}_2 (z_i)},
\end{equation}
and
\begin{equation}\label{eq:washout}
W_{i}(z) = \frac{1}{4} K_{i} \sqrt{x_i} \mathcal{K}_1 (z_i) z^3_i,
\end{equation}
with $\mathcal{K}_1$ and $\mathcal{K}_2$ the modified Bessel functions of the second kind with
\[
x_{i} \equiv M^2_{i}/M^2_{1}, \quad z_i \equiv \sqrt{x_i} z,
\]
and
\begin{equation}
K_{i}\equiv\frac{\tilde{\Gamma}_{i}}{H (T=M_i)}, \quad \tilde{\Gamma}_i = \frac{M_i \left(Y^{\dagger} Y\right)_{ii}}{8 \pi}.
\end{equation}
$H$ is the Hubble expansion rate and $\Lambda_{\alpha}$ is the self-energy of $\alpha$-flavoured leptons. The thermal widths $\Lambda_{\tau}$, $\Lambda_{\mu}$ of the charged leptons is given by the imaginary part of the self-energy correction to the lepton propagator in the plasma (see \appref{sec:thermalwidth}).  Finally, the $P^{0(i)}_{\alpha \beta} \equiv c_{i \alpha} c_{i \beta}^*, $ denotes the  projection matrices which describe how a given flavour of lepton is washed out and the CP-asymmetry matrix describing the decay asymmetry generated by  $N_{i}$ is denoted by  $\epsilon^{(i)}_{\alpha\beta}$. These CP-asymmetry parameters may be written as \cite{Covi:1996wh,Blanchet:2011xq,Blanchet:2011xq,Abada:2006ea,DeSimone:2006nrs} 
\begin{equation}
\begin{aligned}
\epsilon^{(i)}_{\alpha\beta}&=\frac{3}{32\pi\left(Y^{\dagger} Y\right)_{ii}}\\
&\sum_{j\neq i}\Bigg\{ i[Y_{\alpha i}Y^{*}_{\beta j}(Y^{\dagger}Y)_{ji}-Y_{\beta i}Y^{*}_{\alpha j}(Y^{\dagger}Y)_{ij}] f_1\left(\frac{x_{j}}{x_{i}}\right) \\
&+i[Y_{\alpha i}Y^{*}_{\beta j}(Y^{\dagger}Y)_{ij}-Y_{\beta i}Y^{*}_{\alpha j}(Y^{\dagger}Y)_{ji}] f_2\left(\frac{x_{j}}{x_{i}}\right) \Bigg\},
\label{eq:CPoff}
 \end{aligned}
\end{equation}
where
\begin{equation}\label{eq:CPa2}
\begin{aligned}
f_1\left(\frac{x_{j}}{x_{i}}\right)&=\frac{\xi\left(\frac{x_{j}}{x_{i}}\right)}{\sqrt{\frac{x_{j}}{x_{i}}}}, \quad
 f_2\left(\frac{x_{j}}{x_{i}}\right)=\frac{2}{3\left(\frac{x_{j}}{x_{i}}-1\right)}\\
\end{aligned}
\end{equation}
and
\begin{equation*}
 \xi\left(x\right) = \frac{2}{3}x\left[ \left(1+x\right)\log\left( \frac{1+x}{x}  \right) -\frac{2-x}{1-x}   \right].
\end{equation*}
 \equaref{eq:full3} may be used to calculate the lepton asymmetry in all flavour regimes and even accurately describes the transitions between them \cite{Barbieri:1999ma, Abada:2006fw,DeSimone:2006nrs,Blanchet:2006ch,Blanchet:2011xq}.  The off-diagonal entries of $n_{\alpha \beta}$, which in general may be complex, allow for a quantitative description of these transitions. If leptogenesis proceeds at temperatures $10^{9}\lesssim T \text{ (GeV)} \lesssim 10^{12}$, (for example), the terms $\Im\left(\Lambda_{\tau}\right)/Hz$ damp the evolution of the off-diagonal elements of $n_{\alpha \beta}$. This reflects the loss of coherence of the tau states when the SM tau Yukawa couplings come in to equilibrium. The remaining equations provide a description of leptogenesis in terms of Boltzmann equations for the diagonal entries of $n_{\alpha \beta}$ and for $n_{N_i}$. Although a more accurate treatment of leptogenesis is provided by the NE-QFT approach, the density matrix equations that we choose to solve are accurate so long as we are in the strong washout regime. In \appref{sec:decayparam}, we demonstrate that this is the case and thus justify our use of the density matrix formalism.

In general, $n_{\alpha\beta}$ is a $3\times3$ matrix whose trace gives the total lepton asymmetry:
\[
n_{B-L} \equiv \sum\limits_{\alpha=e,\mu,\tau}n_{\alpha\alpha}.
\]
The latter is then multiplied by a factor $a/f\approx 0.01$, where $a=28/79$ describes the partial conversion of the $B-L$ asymmetry into a baryon asymmetry by sphaleron processes, and $f \equiv n^{\text{rec}}_{\gamma}/n^{*}_{\gamma}=2387/86$ accounts for the dilution of the asymmetry due the change of photon densities ($n_{\gamma}$) between leptogenesis ($n_{\gamma}=n^{*}_{\gamma}$) and recombination ($n_{\gamma}=n^{\text{rec}}_{\gamma}$) : $\eta_{B}\simeq10^{-2} n_{B-L}$\cite{Buchmuller:2004nz}.

\subsection{Thermal Leptogenesis with Three Flavours}\label{sec:3flavours}
In this section we demonstrate, in the case of one decaying heavy Majorana neutrino, the need to solve the density matrix equations, rather than the more approximate Boltzmann equations (in which the off-diagonal entries of the density matrix are set to zero). Although we shall give explicit expressions for the density matrix equations in this section, we emphasise that our main results are always found by solving \equaref{eq:full3} in which all three heavy Majorana neutrinos decay. The Boltzmann equations, with one decaying heavy Majorana neutrino, are written as
\begin{table}[t]\label{tab:benchmarks}
  \centering
  \begin{tabular}{c|c|c|c|c|c|c|c|c|c}
       & $\delta(^\circ)$    & $\alpha_{21}(^\circ)$ &  $\alpha_{31}(^\circ)$ &  $x_{1}(^\circ)$    & $y_{1}(^\circ)$ &  $x_{2}(^\circ)$ & $y_{2}(^\circ)$    & $x_{3}(^\circ)$ &  $y_{3}(^\circ)$   \\
          \hline\hline
  BP1 & $180$              & 	$0$ & $0$ & $100$ & $45$ & $150$ & $25$ & $45$ & $35$ \\
  BP2 & $270$ &  $90$ & $180$ & $10$ & $60$ & $55$ & $25$ & $70$ & $-15$ \\
  BP3 & $330$& $40$ & $220$ & $0$ & $100$ & $-1$ & $10$ & $1$ & $-75$ \\
  BP4 & $320$& $450$ & $450$ & $15$ & $180$ & $-90$ & $2$ & $144$ & $-175$ 
      \end{tabular}
  \caption{Benchmark points used to test the three-flavoured equations against the density matrix equations.}
\end{table}
\begin{equation}\label{eq:threefull}
\begin{aligned}
\frac{dn_{N_{1}}}{dz}=&-D_{1}(n_{N_{1}}-n^\text{eq}_{N_{1}})\\
\frac{dn_{\tau \tau}}{dz} = & \epsilon^{(1)}_{\tau\tau}D_{1}(n_{N_{1}}-n^\text{eq}_{N_{1}}) -\mathcal{W}_{1}\left(\left|Y_{\tau 1}\right|^2 n_{\tau\tau}\right) \\
\frac{dn_{\mu \mu}}{dz} = & \epsilon^{(1)}_{\mu\mu}D_{1}(n_{N_{1}}-n^\text{eq}_{N_{1}}) -\mathcal{W}_{1}\left(\left|Y_{\mu 1}\right|^2 n_{\mu\mu}\right) \\
\frac{dn_{ee}}{dz} = & \epsilon^{(1)}_{ee}D_{1}(n_{N_{1}}-n^\text{eq}_{N_{1}}) -\mathcal{W}_{1}\left(\left|Y_{e 1}\right|^2 n_{ee}\right),
\end{aligned}
\end{equation}
where we have used the abbreviation $\mathcal{W}_{1}=\frac{W_{1}}{(Y^\dagger Y )_{11}} $. This set of equations is appropriate for $M_{1} \ll 10^{9}\,\text{GeV}$, when the flavour-components of the charged leptons each experience strong and distinct interactions with the early Universe plasma.  
The density matrix equations, with specified flavour indices,  follow straightforwardly from an explicit expansion of the commutators  in \equaref{eq:full3}. The resulting equations, for a single decaying heavy Majorana neutrino are
\begin{widetext}
\begin{equation}\label{eq:longDM}
\begin{aligned}
\frac{dn_{N_{1}}}{dz}&=-D_{1}(n_{N_{1}}-n^\text{eq}_{N_{1}})\\
 \frac{dn_{\tau\tau}}{dz} &= \epsilon^{(1)}_{\tau \tau}D_{1} (n_{N_{1}}-n^\text{eq}_{N_{1}})   -\mathcal{W}_{1} \left\{
 \left| Y_{\tau 1}\right| {}^2   n_{\tau \tau } +  \Re\left[Y^*_{\tau 1} \left( Y_{e1}n_{\tau e}+Y_{\mu 1} n_{\tau \mu}\right)\right] \right\}\\
 \frac{dn_{\mu\mu}}{dz} &= \epsilon^{(1)}_{\mu \mu}D_{1} (n_{N_{1}}-n^\text{eq}_{N_{1}}) -\mathcal{W}_{1} \left\{
\left| Y_{\mu 1}\right| {}^2   n_{\mu \mu } + \Re\left[Y^*_{\mu 1}\left(  Y_{e1}n_{\mu e} + Y_{\tau 1}   n^*_{\tau \mu}  \right)\right] \right\} \\
 \frac{dn_{ee}}{dz} &= \epsilon^{(1)}_{ee}D_{1} (n_{N_{1}}-n^\text{eq}_{N_{1}}) -\mathcal{W}_{1} \left\{
 \left| Y_{e 1}\right| {}^2   n_{ee }  + \Re\right[Y^*_{e1} \left(Y_{\mu 1}n^*_{\mu e} + Y_{\tau 1} n^*_{\tau e} \right )\left] \right\} \\
 \frac{dn_{\tau \mu}}{dz}& = \epsilon^{(1)}_{\tau \mu}D_{1} (n_{N_{1}}-n^\text{eq}_{N_{1}})  -\frac{\mathcal{W}_{1}}{2} \left\{   n_{\tau \mu} \left( \left| Y_{\tau 1}\right| {}^2+ \left| Y_{\mu 1}\right| {}^2 \right)+ Y^*_{\mu 1}Y_{\tau 1} \left( n_{\tau \tau } + n_{\mu \mu}\right) + Y^*_{e1}Y_{\tau 1} n^*_{\mu e} + Y^*_{\mu 1}Y_{e1}n_{\tau e}  \right\}\\
 &	-\left(\frac{\Im \left( \Lambda_{\tau}  \right)}{Hz} + \frac{\Im \left( \Lambda_{\mu}  \right)}{Hz}\right)n_{\tau \mu}  \\
 \frac{dn_{\tau e}}{dz} &= \epsilon^{(1)}_{\tau e}D_{1} (n_{N_{1}}-n^\text{eq}_{N_{1}}) -\frac{\mathcal{W}_{1}}{2} \left\{  n_{\tau e}\left( \left| Y_{e 1}\right| {}^2+ \left| Y_{\tau 1}\right| {}^2\right) +  Y^*_{e1}Y_{\tau 1}\left( n_{ee} + n_{\tau \tau }  \right)   + Y^*_{\mu 1}Y_{\tau 1} n_{\mu e} + Y^*_{e1}Y_{\mu 1}n_{\tau \mu}
\right\} \\
&-\frac{\Im \left( \Lambda_{\tau}  \right)}{Hz}n_{\tau e}\\ 
 \frac{dn_{\mu e}}{dz} &= \epsilon^{(1)}_{\mu e}D_{1} (n_{N_{1}}-n^\text{eq}_{N_{1}}) -\frac{\mathcal{W}_{1}}{2} \left\{  n_{\mu e}\left( \left| Y_{e 1}\right| {}^2+ \left| Y_{\mu 1}\right| {}^2\right) + Y^*_{e1}Y_{\mu 1}\left( n_{ee} + n_{\mu \mu }  \right)   + Y^*_{e1}Y_{\tau 1}n^*_{\tau \mu} + Y^*_{\tau 1} Y_{\mu 1} n_{\tau e}
\right\}\\
&-\frac{\Im \left( \Lambda_{\mu}  \right)}{Hz}n_{\mu e}.   \end{aligned} 
\end{equation}
\end{widetext}
The Boltzmann equations, \equaref{eq:threefull}, are recovered in the limit $\Im \left( \Lambda_{\mu}  \right)/Hz, \Im \left( \Lambda_{\tau}  \right)/Hz \rightarrow \infty$ as the off-diagonal density matrix elements become fully damped. This limit is only valid for $M_{1}\ll10^{9}$ GeV.  However, for a given point in the model parameter space, $\mathbf{p}$, it is not \textit{a priori} obvious if these Boltzmann equations well approximate the density matrix equations.

We illustrate the quantitative difference between the density matrix equations (\equaref{eq:longDM}) and the Boltzmann equations (\equaref{eq:threefull}) by solving both for four benchmark points: BP1, BP2, BP3 and BP4 with a vanishing initial abundance of $N_{1}$, (see  \tabref{tab:benchmarks}). In these scenarios, the light mass spectrum is chosen to be normally ordered, $m_1=10^{-2}\,\text{eV}$, $M_{1}$ is allowed to vary with  $M_{2}=3.5\,M_{1}$ $M_{3}=3.5\,M_{2}$ which satisfies a mildly hierarchical mass spectrum.

\begin{figure*}[t]\label{fig:DMBE}
\includegraphics[width=0.48\textwidth]{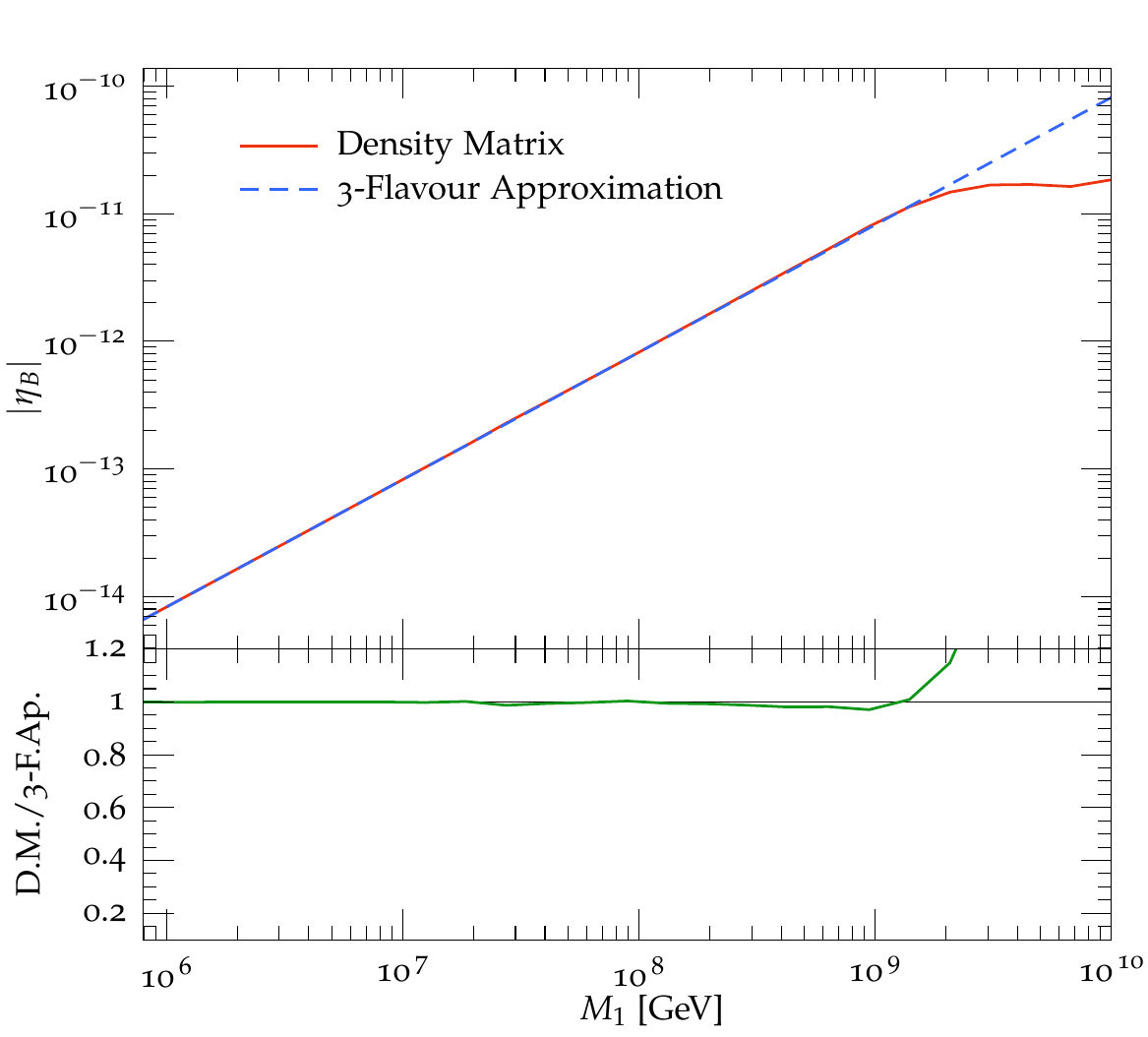}
\includegraphics[width=0.48\textwidth]{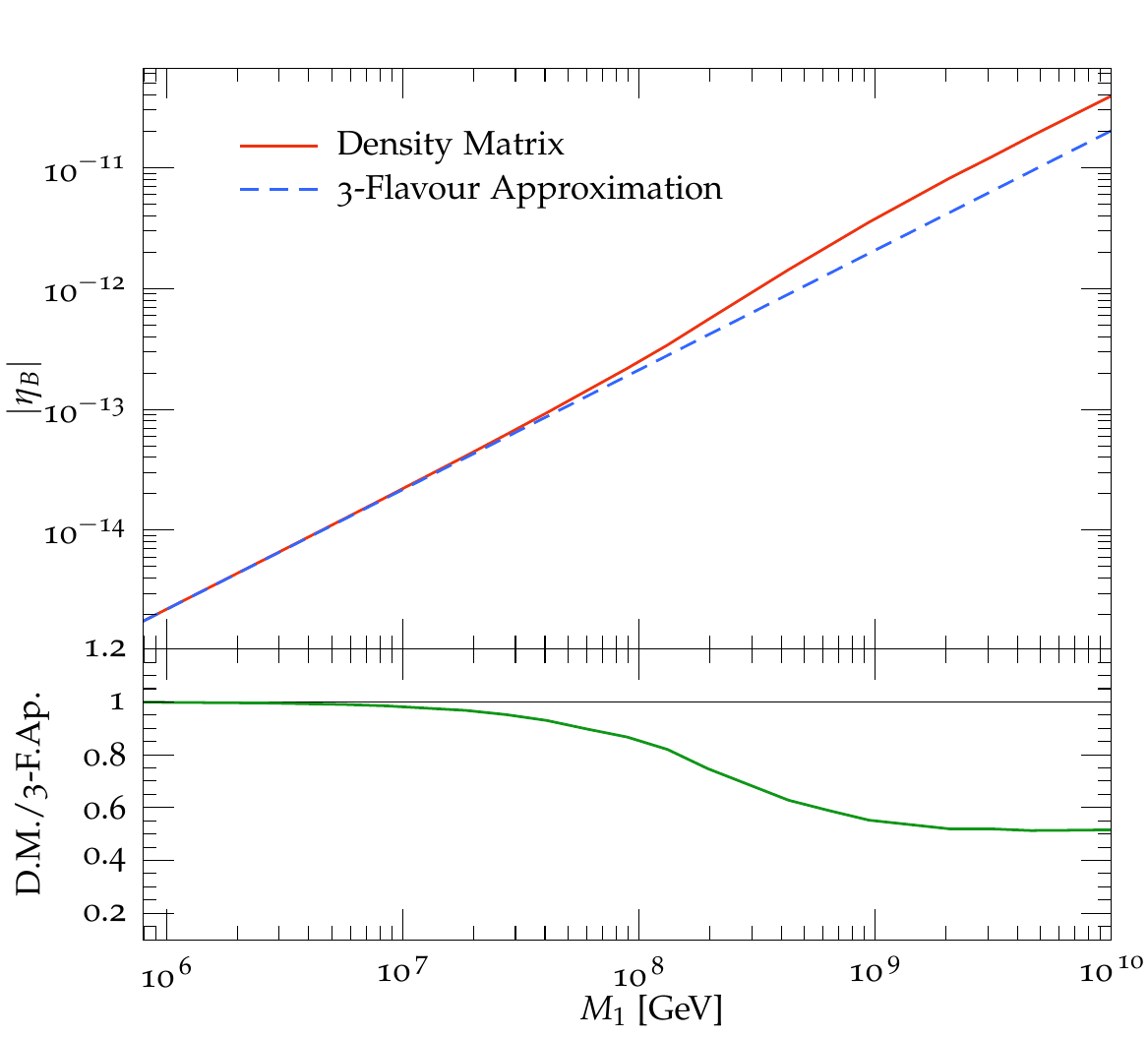}
\includegraphics[width=0.48\textwidth]{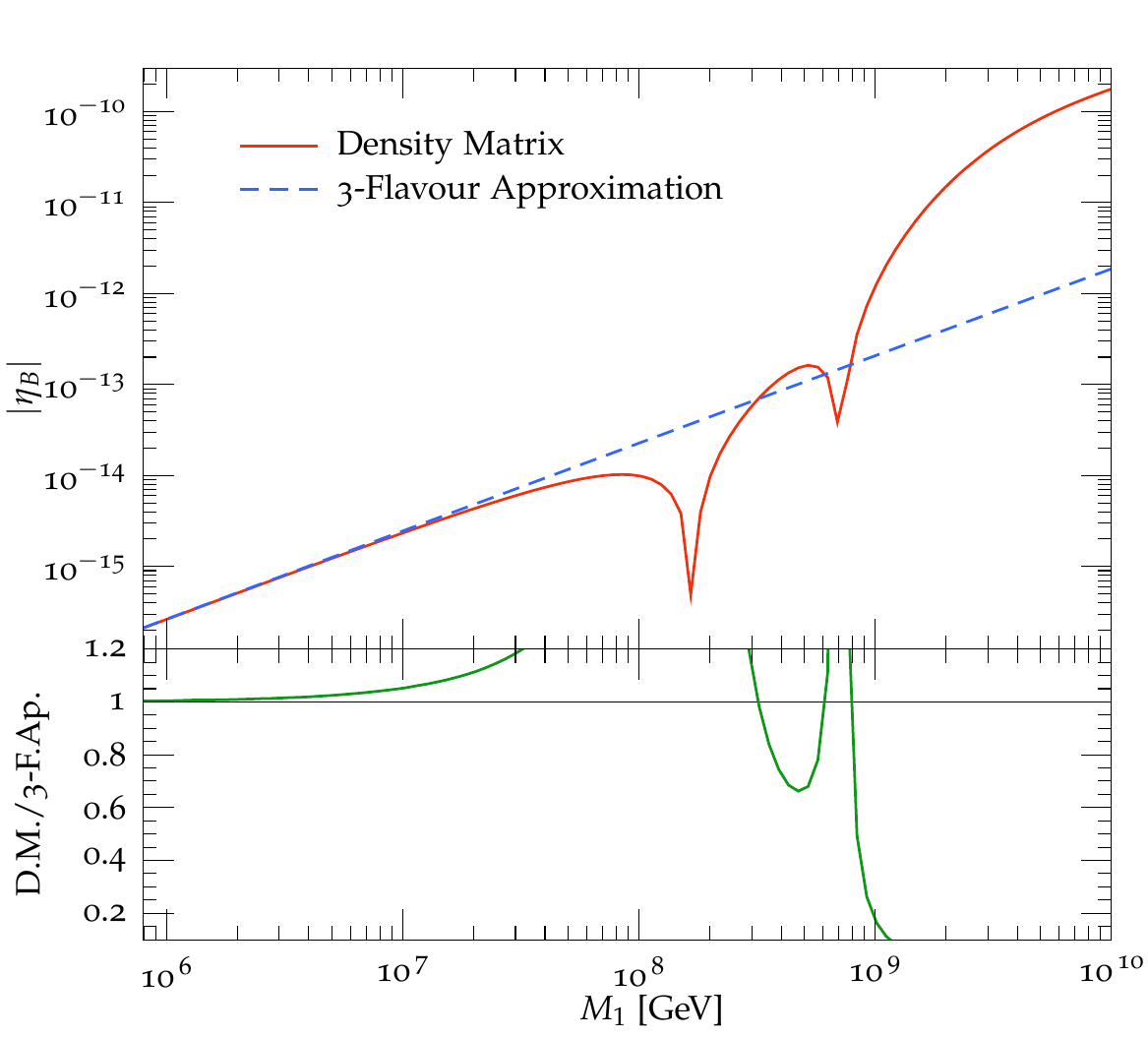}
\includegraphics[width=0.48\textwidth]{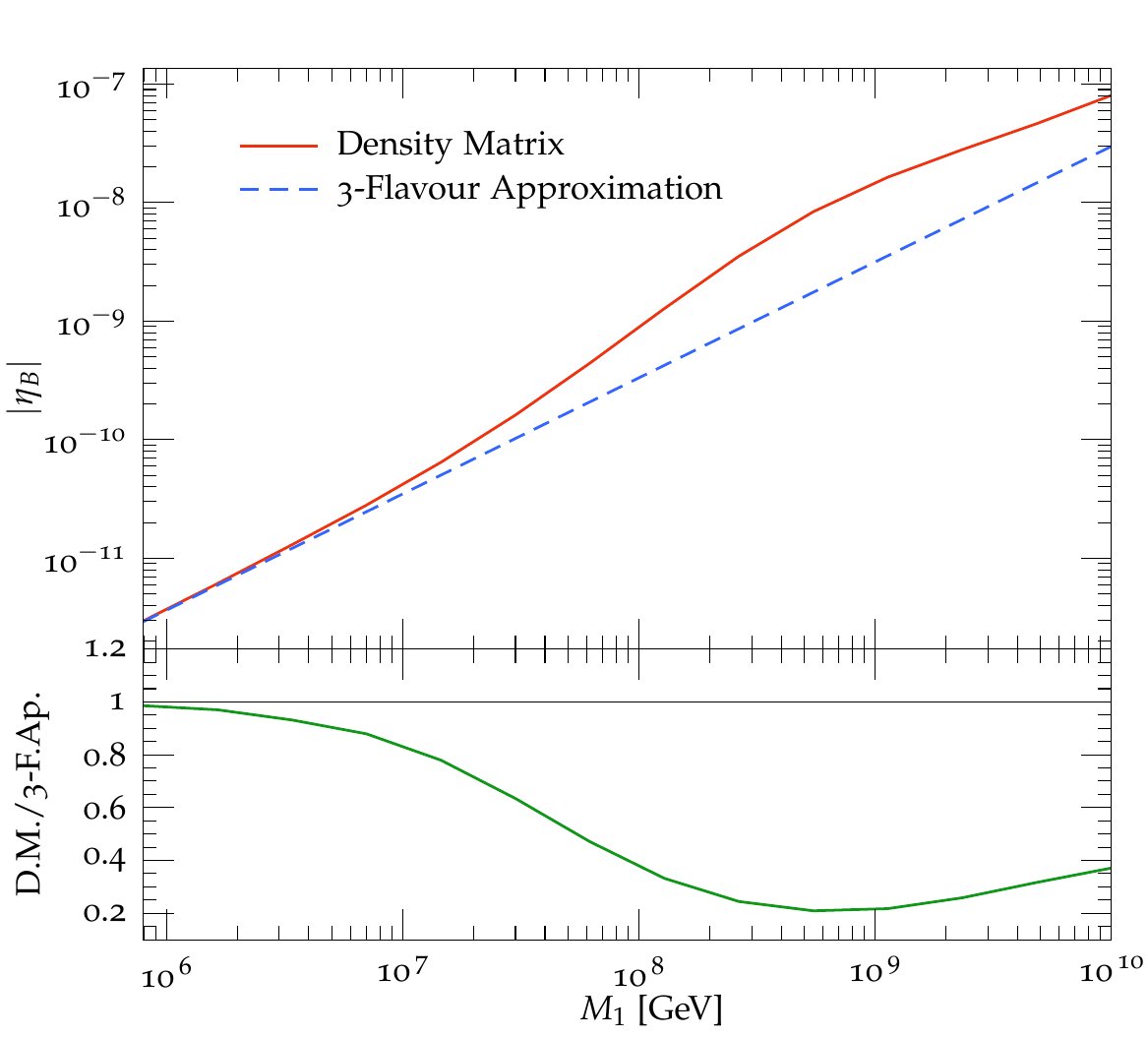}
\caption{$\eta_{B}$ is shown as a function of varying $M_{1}$ for density matrix (red) and fully three-flavoured  (blue, dashed) equations. The solutions for BP1 (BP2) is shown on the top left (right) and BP3 (BP4) on bottom left (right) with $m_{1}=10^{-2}$eV.
 }
\end{figure*}

As can be seen from \figref{fig:DMBE}, the deviation between the two is generally small ($<5\%$) for $M_{1}\sim10^{6}$ GeV. 
In the case of BP1, the Boltzmann equations do not deviate from the density matrix solutions until $M_{1}\approx10^{9}$ GeV and for $M_{1}\approx10^{6}$ GeV, the discrepancy between the two is negligible.
A more pronounced deviation is exhibited in BP2, BP3 and BP4, with an underestimation from the Boltzmann equations particularly evident in BP4 in which the $R$-matrix has relatively large elements.
In this example, the deviation between the solutions even for low masses $M_{1}=10^{7}\,\text{GeV}$ and $M_{1}=10^{6}\,\text{GeV}$ 
is $\sim 20\%$ and $\sim 5\%$ respectively.
The discrepancy  grows as a function of $M_{1}$. 
As can be seen from these benchmark points, the fully three-flavoured equations may not well approximate the 
density matrix equations well even for $M_{1}\ll 10^{9}$ GeV. As we are interested in exploring the parameter space over a range of values of $M_1$, we shall use the more accurate density matrix equations.

Here we summarise the approximations and physical effects that we shall exclude from our calculation  but whose inclusion would increase the accuracy of our calculations. Such effects include lepton number-changing scattering processes, spectator effects \cite{Buchmuller:2001sr,Nardi:2005hs,Schwaller:2014hna}, thermal corrections \cite{Kiessig:2010pr,Giudice:2003jh} and the inclusion of quantum statistical factors  \cite{DeSimone:2007gkc,Beneke:2010dz,Anisimov:2010dk,Beneke:2010wd}.

$\lvert \Delta L\rvert=1$ scattering and related washout processes  occur as a result of Higgs and lepton mediated scattering involving the top quark and gauge bosons. It has been 
demonstrated  that scatterings involving the top quark are most important at relatively low temperatures, $T<M$ \cite{Davidson:2008bu}. Therefore, the effects of $\lvert \Delta L\rvert=1$ scattering on the strong washout regime (where the bulk of lepton asymmetry is produced at $T>M$) are small and have been estimated to affect the final lepton asymmetry to a level less than $\sim \mathcal{O}(10)\%$ \cite{Blanchet:2006be,Frossard:2013bra}. However, for weak washout these corrections are necessary for 
 a correct calculation of the final lepton asymmetry \cite{Nardi:2007jp}. 
Spectator processes cause the  redistribution of  the  asymmetry generated in the leptonic doublets amongst other particle species in the thermal bath. These processes typically  protect the lepton asymmetry from washout and therefore increase the efficiency of leptogenesis \cite{Garbrecht:2014kda}. Although the inclusion of spectator effects could further lower the scale of successful thermal leptogenesis, we relegate the inclusion of these effects for further studies. Besides, the neglect of these effects leads to an overly-conservative estimate of the amount of baryon asymmetry produced.
Quantum kinetic equations are derived from the first principles of NE-QFT  based on the Closed-Time Path (CTP) formalism. This approach resolves unitarity issues and properly accounts for the effect of quantum statistics on the lepton asymmetry. However, it has been shown there is little qualitative difference between the 
density matrix and CTP approach in the strong washout regime \cite{Beneke:2010wd}. This is because in the strong washout regime, where the decays and inverse decays of the heavy Majorana neutrinos occur much faster than the expansion rate of the Universe, the majority of the lepton asymmetry is produced at temperatures smaller than the mass of the decaying heavy Majorana neutrinos. As a consequence, the contributions of the particle distribution functions are heavily Boltzmann-suppressed. 

We  demonstrate in  \appref{sec:decayparam} that the regions of parameter space we explore in this work 
 correspond to  strong washout.  This allows us to make two justifiable simplifications to our kinetic equations which are  more easily implementable for a phenomenological study. Firstly, we  ignore the 
 impact of lepton number-changing scatterings and secondly we  solve kinetic equations using the density matrix formalism rather than equations derived from NE-QFT.

\section{Computational Methods}\label{sec:NT}
The computational core of this work is solving a set of coupled
differential equations as shown in \equaref{eq:full3}.  We use the {\sc
Python} interface for complex differential equations~\cite{odeintw} to the {\sc LSODA} algorithm~\cite{odepack} that is available
in {\sc Scientific Python}~\cite{scipy}.

Our aim is to find regions of the model parameter space, $\mathbf{p}$, that
yield values of $\eta_B(\mathbf{p})$ that are consistent with the measurement
$\eta_{B_\text{CMB}}=(6.10\pm0.04)\times10^{-10}$. In order to do so, we have to
use an efficient sampling method. This is mainly for three reasons. Firstly,
the parameter space has a relatively high dimension. Secondly, the function
$\eta_B(\mathbf{p})$ itself does not vary smoothly with changes of
$\mathbf{p}$. In fact, tiny variations of the input parameters yield 
function values differing in many orders of magnitude and sign.  Thirdly, the
computation of  $\eta_B(\mathbf{p})$ for a single point is relatively expensive
and can take up to the order of seconds. Thus any attempt of a brute-force
parameter scan is doomed to fail. Finally, we are not only interested in a single
best-fit point but also a region of confidence that resembles the measurement
uncertainty.

We found the use of {\sc
Multinest}~\cite{Feroz:2008xx,Feroz:2007kg,2013arXiv1306.2144F} (more precisely,
{\sc pyMultiNest}~\cite{pymultinest}, a wrapper around {\sc Multinest} written
in {\sc Python}) to be particularly well suited to address all the
aforementioned complications associated to this task. The {\sc Multinest} algorithm has seen
wide and very successful application in astronomy and cosmology.  It provides a
nested sampling algorithm that calculates Bayesian posterior distributions
which we will utilise in order to define regions of confidence.

In all our scenarios, {\sc Multinest} uses a flat prior and the  following log-likelihood
as objective function
\begin{equation}\label{eq:Likelihood}
\begin{aligned}
    \log L =  -\frac{1}{2}\left(\frac{\eta_B(\vec{p}) - {\eta_B}_{CMB}}{\Delta{\eta_B}_{CMB}}\right)^2.
\end{aligned}
\end{equation}
Once a {\sc Multinest} run is finished, we use {\sc SuperPlot}~\cite{Fowlie:2016hew} to visualise
the posterior projected onto a two-dimensional plane.

\section{Results}
We present the solutions to the density matrix equations of \equaref{eq:full3}
for the case of one and two decaying heavy Majorana neutrinos in \secref{sec:1DS} and \secref{sec:2DS}, respectively. 
In principle, it is necessary to consider the decay of 
all three heavy Majorana neutrinos, however we first consider the decay of the lightest heavy Majorana neutrino as computationally
this scenario is less expensive than the two and three decaying heavy Majorana neutrinos case.  
In \secref{sec:2DS}, we demonstrate the scale of thermal leptogenesis
involving the decay of two heavy Majorana neutrinos does not change significantly 
from the scenario of one decaying case. These two scenarios are qualitatively and quantitatively similar and so we do not proceed to the case where the third heavy Majorana neutrino contributes to the baryon asymmetry through their decays.

\subsection{Results from $N_{1}$ Decays}\label{sec:1DS}
\begin{figure*}[t]\label{fig:evolve}
\includegraphics[width=0.3\textwidth]{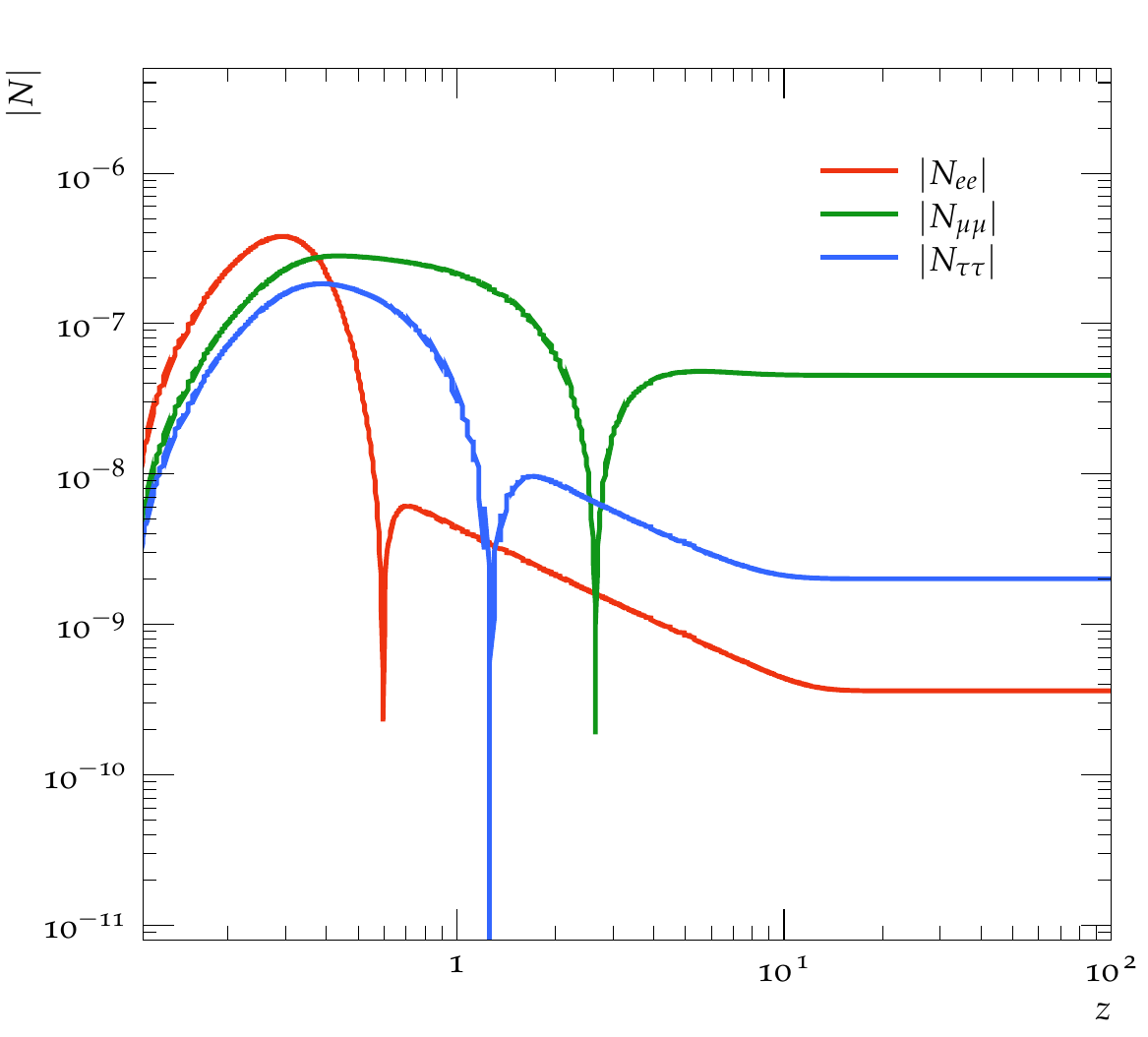}
\includegraphics[width=0.3\textwidth]{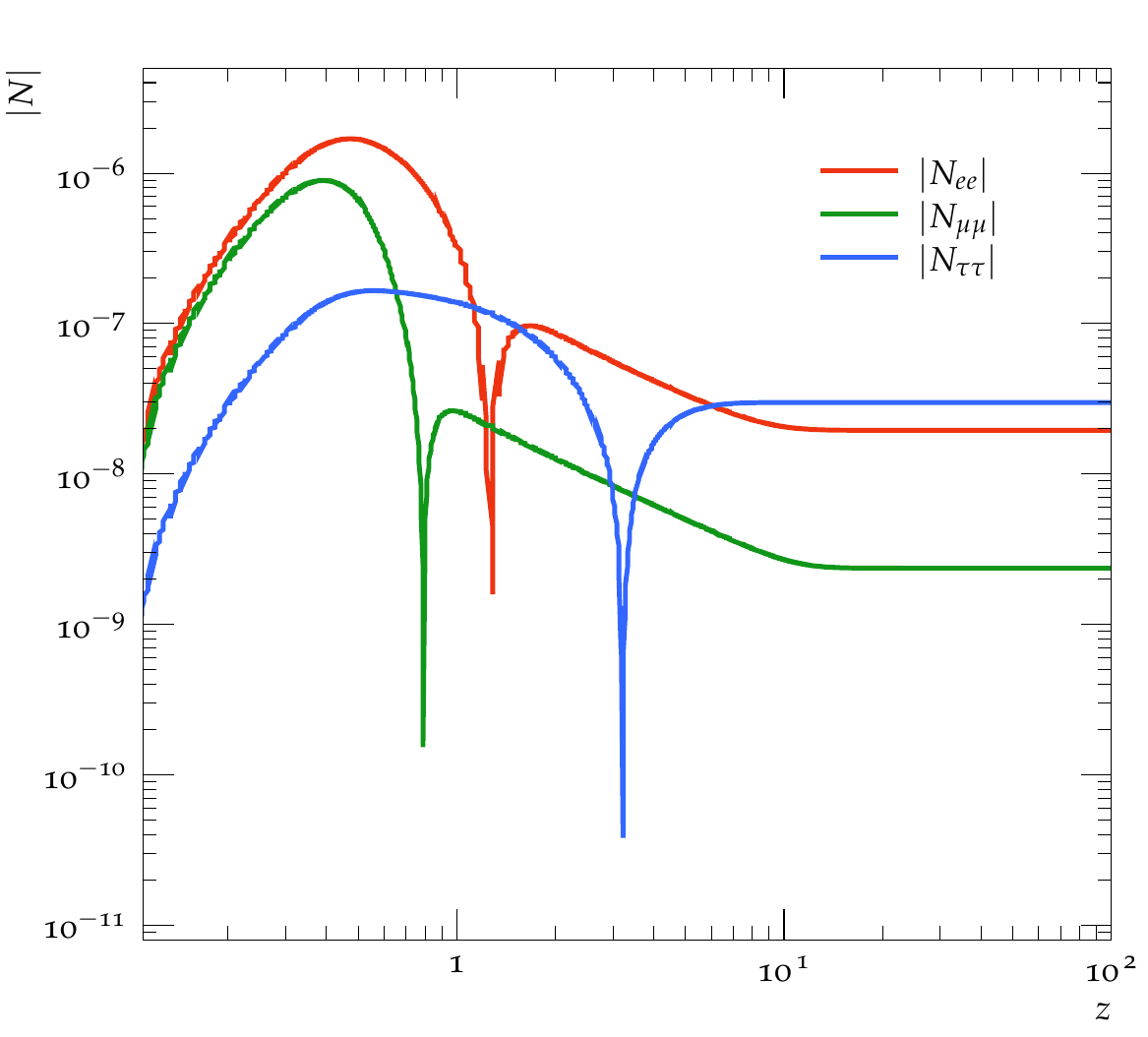}
\includegraphics[width=0.3\textwidth]{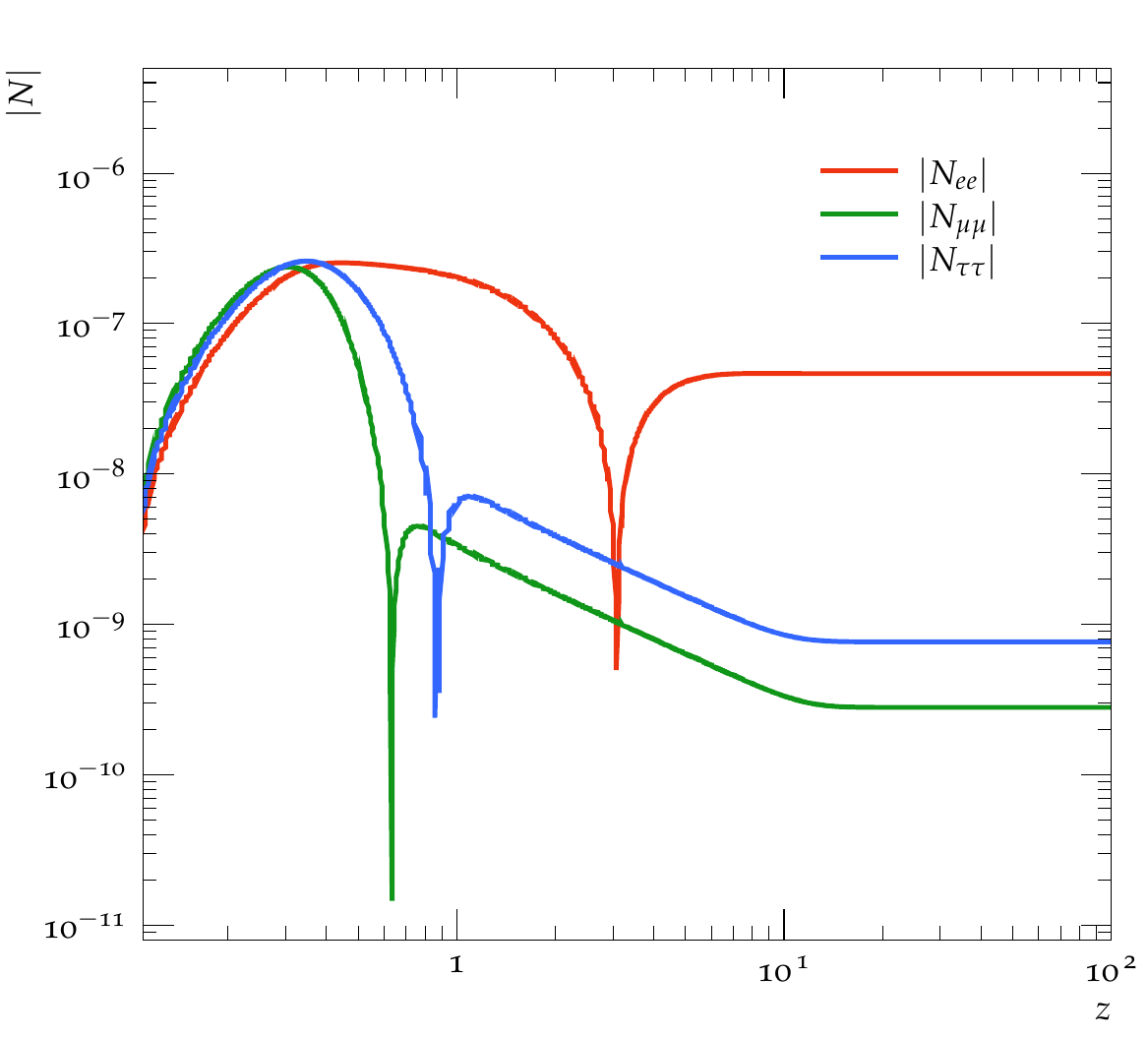}
\includegraphics[width=0.3\textwidth]{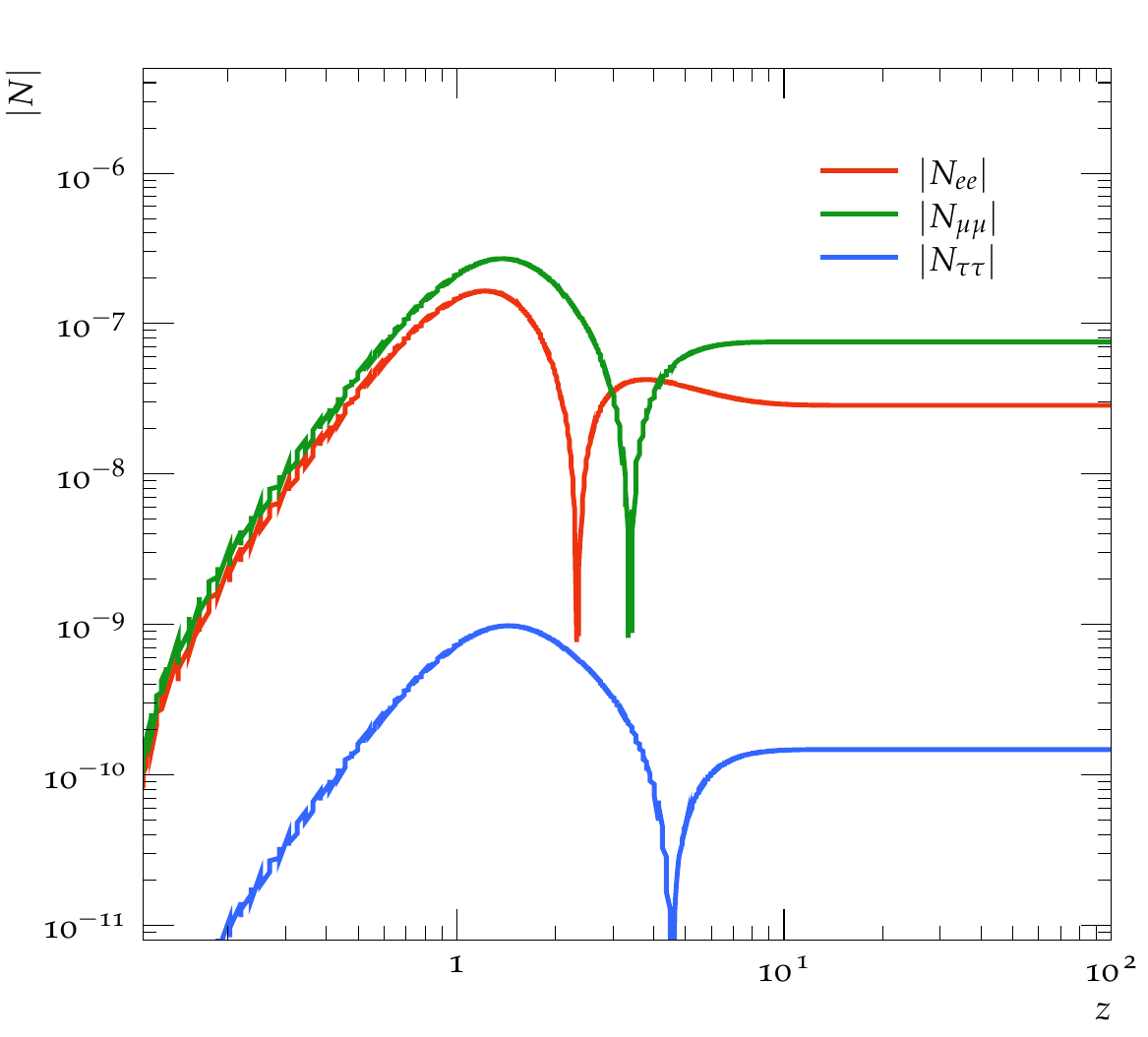}
\includegraphics[width=0.3\textwidth]{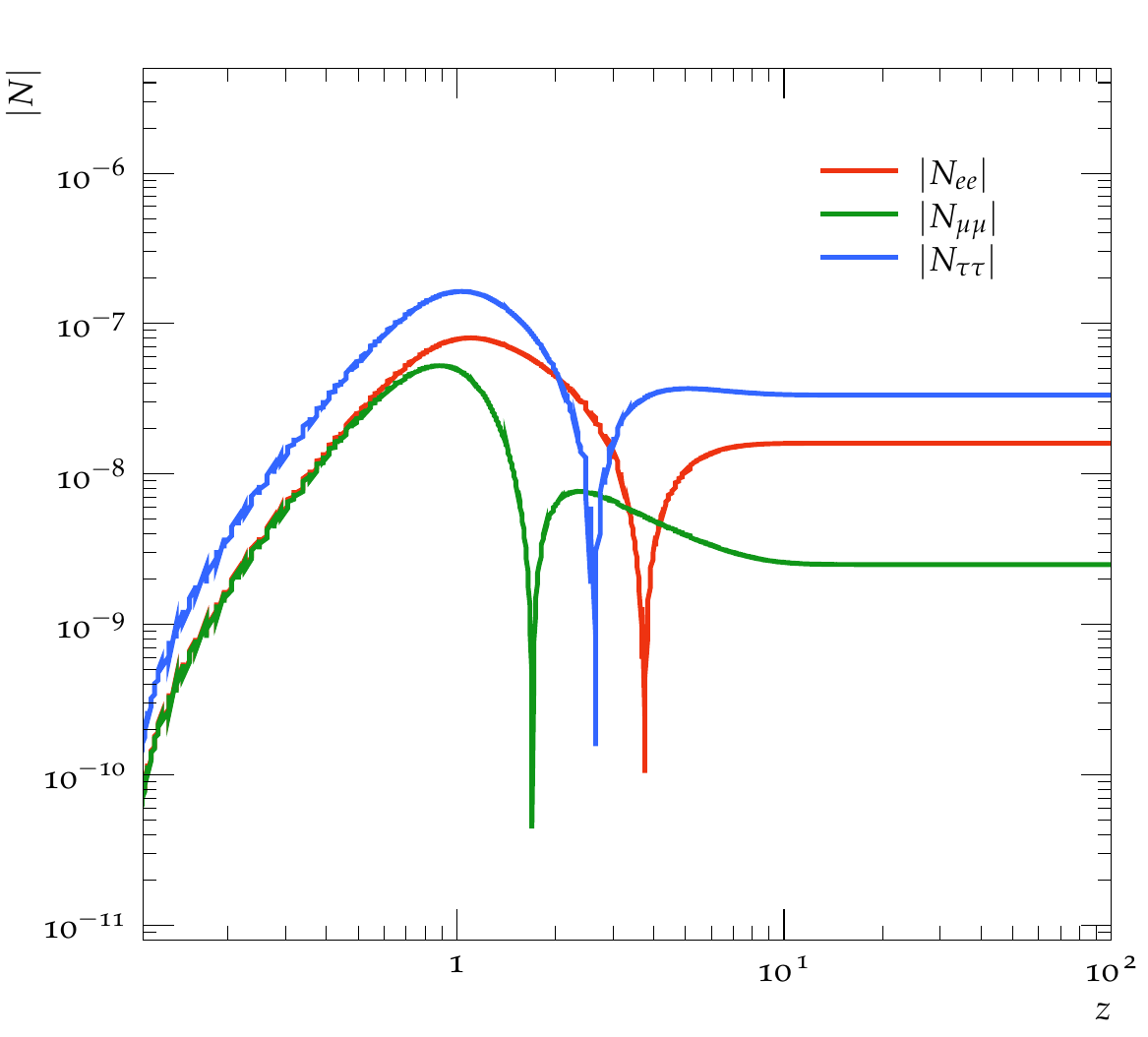}
\includegraphics[width=0.3\textwidth]{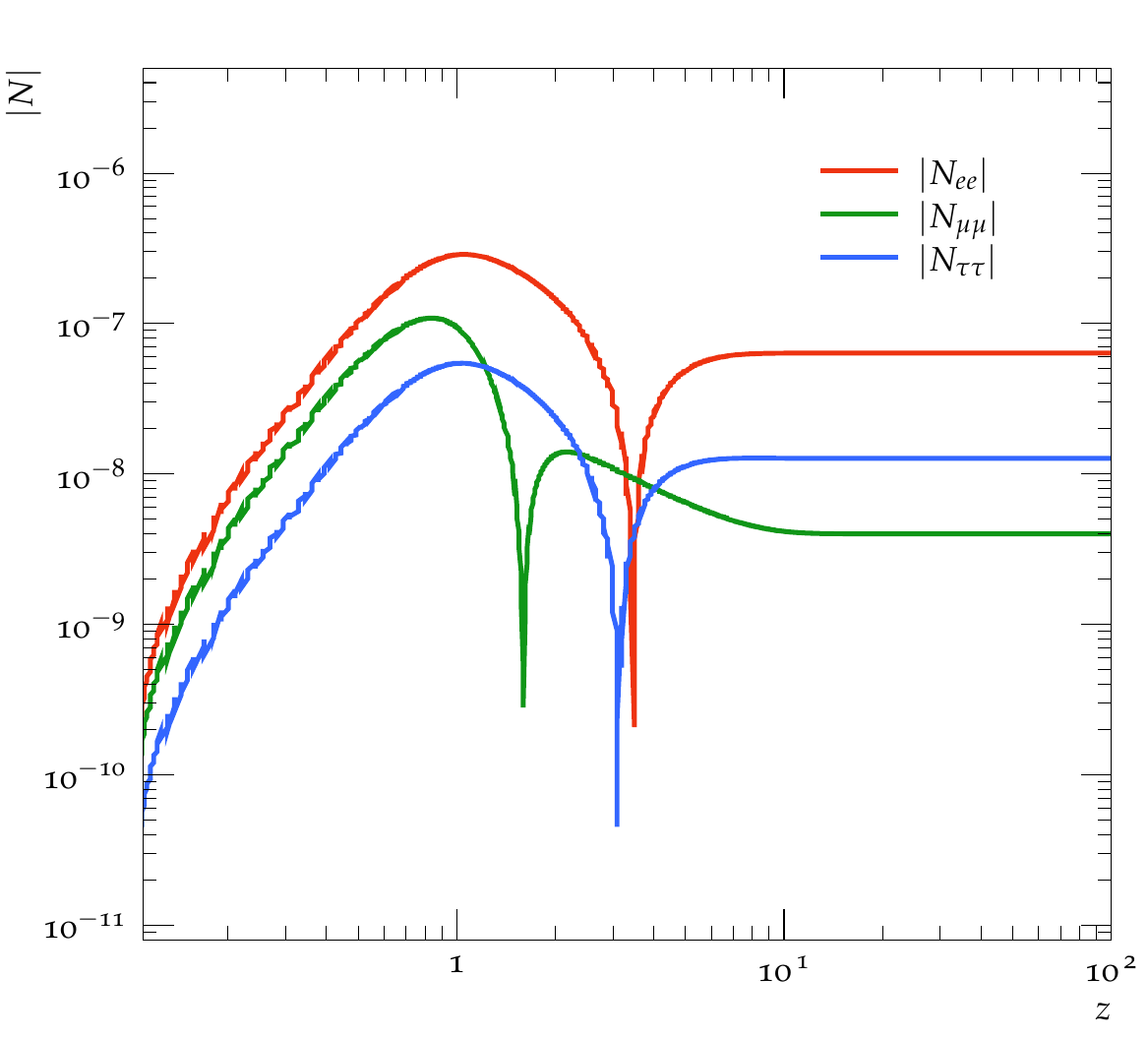}
\caption{The top (bottom) three plots from left to right show evolution of the $B-L$ asymmetry for each flavour  evolved as a function of $z=M_{1}/T$ for the best-fit points of $S_{1}$, $S_{2}$ and $S_{3}$
($\overline{S_{1}}$, $\overline{S_{2}}$ and $\overline{S_{3}}$) respectively. }
\end{figure*}
\begin{figure*}[t]\label{fig:NO1}
\includegraphics[width=0.9\textwidth]{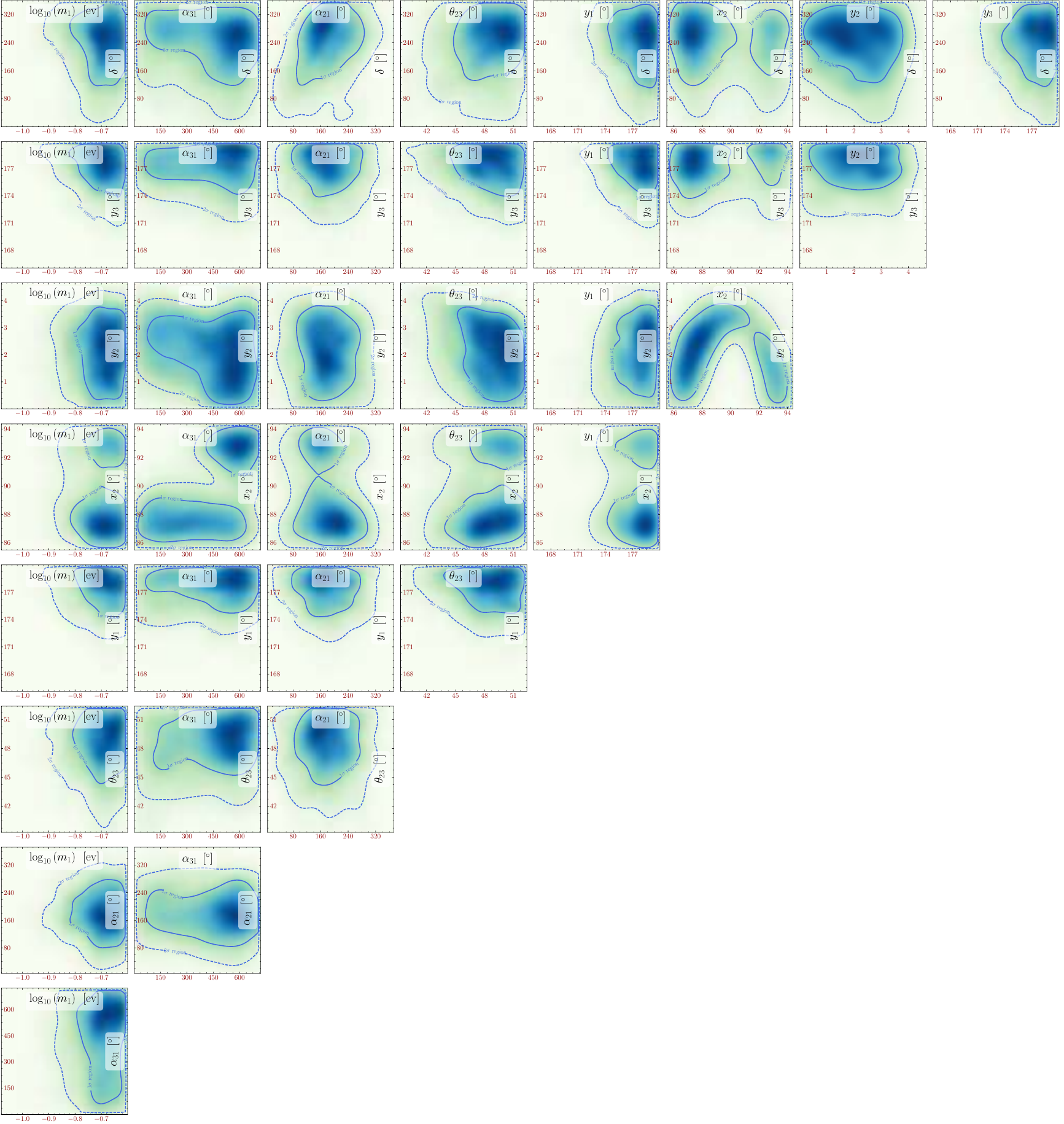}
\caption{\label{fig:NO1DS} $S_{1}$: Triangle plot showing the two-dimensional projection  of the 11-dimensional model parameter space for posterior distributions using  normal ordering with one-decaying heavy Majorana neutrino and 
heavy Majorana neutrino mass spectrum: $M_{1}=10^{6}$ GeV, $M_{2}=3.15\, M_{1}$, $M_{3}=3.15\,M_{2}$. The contours correspond to 68$\%$ and 95$\%$ confidence levels respectively.}\end{figure*}
\begin{figure*}[t]\label{fig:IO1}
\includegraphics[width=0.9\textwidth]{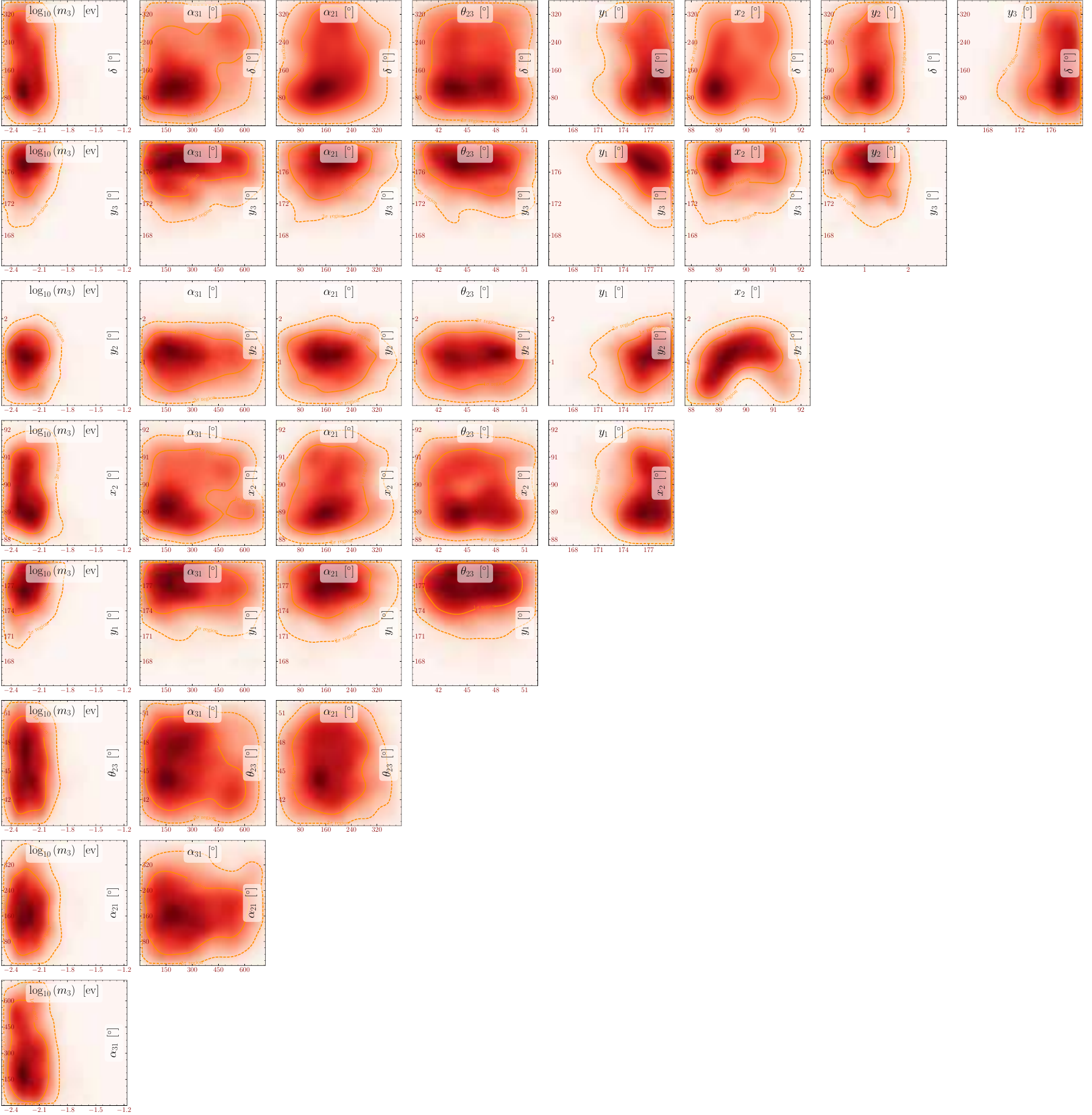}
\caption{\label{fig:IO1DS} $\overline{S_{1}}$: Triangle plot showing the two-dimensional projection   of the 11-dimensional model parameter space for posterior distributions using inverted  ordering and with one-decaying heavy Majorana neutrino mass spectrum:  $M_{1}=10^{6}$ GeV, $M_{2}=3.15\, M_{1}$, $M_{3}=3.15\,M_{2}$. The contours correspond to 68$\%$ and 95$\%$ confidence levels respectively.}
\end{figure*}
\begin{table*}[t]\label{tab:bestfitall}
  \centering
  \begin{tabular}{c|c|c|c|c|c|c|c|c|c|c|c|c|c|c}
    & $\theta_{23}(^\circ)$ & $\delta(^\circ)$    & $\alpha_{21}(^\circ)$ &  $\alpha_{31}(^\circ)$ &  $x_{1}(^\circ)$    & $y_{1}(^\circ)$ &  $x_{2}(^\circ)$ & $y_{2}(^\circ)$    & $x_{3}(^\circ)$ &  $y_{3}(^\circ)$  &
    $m_{1(3)}$ (eV) &       $M_{1}$ (GeV) &    $M_{2}$ (GeV)&    $M_{3}$(GeV)\\
          \hline\hline
$S_{1}$&  $46.24$  & $281.21$  & $181.90$ & $344.71$ & $132.23$ & $179.88$ & $87.81$ & $2.88$ & $-30.25$ & $177.5$ & $0.120$ & $10^{6.0}$ & $10^{6.5}$ & $10^{7.0}$ \\
$S_{2}$& $46.57$ & $88.26$ &  $116.07$ & $420.44$ & $44.36$ & $171.78$ & $86.94$ & $2.96$ & $97.01$ & $174.30$ & $0.079$ & $10^{6.5}$ & $10^{7}$ & $10^{7.5}$ \\
$S_{3}$&   $46.63$ & $31.71$ & $130.95$ & $649.65$ &  $-72.33$ & $170.54$ & $86.96$ & $2.22$ & $-1.86$ & $178.31$
& $0.114$ & $10^{6.5}$ & $10^{7.2}$ & $10^{7.9}$\\
\hline
$\overline{S_{1}}$ & $40.56$ & $158.51$ & $157.48$ & $511.0$ & $-16.23$ & $179.29$ & $90.04$ & $1.29$ & $-107.14$ & $179.22$ & 
$0.0047$ & $10^{6.0}$ & $10^{6.5}$ & $10^{7.0}$ \\
$\overline{S_{2}}$ & $43.67$ & $201.02$ &  $238.77$ & $658.33$ & $-39.88$ & $178.68$ & $88.12$ & $2.46$ & $53.97$ & $158.01$
& $0.0133$ & $10^{6.5}$ & $10^{7.0}$ & $10^{7.5}$ \\
$\overline{S_{3}}$ & $43.64$& $57.28$& $179.87$ & $292.95$ & $86.58$ & $174.40$ & $91.11$ & $1.61$ & $134.48$ & $173.74$ 
& $0.012$ & $10^{6.5}$ & $10^{7.2}$ & $10^{7.9}$\\
\hline
${F.T}^{\text{loop}}$ & $44.59$ & $140.04$ & $537.15$ & $291.89$ & $164.06$ & $-149.85$ &
$178.99$ & $49.15$ & $93.39$ & $-14.50$ & $0.15882$ &   $10^{9.0}$ & $10^{9.5}$ & $10^{10}$\\
${F.T}^{\text{tree}}$ & $43.81$ & $31.59$ & $681.96$ & $276.19$ & $ 271.56$ & $-125.27$ &
$14.95$ & $-11.50$ & $344.87$ & $5.22$ & $0.0041$ &   $10^{9.0}$ & $10^{9.5}$ & $10^{10}$\\
      \end{tabular}
  \caption{The best-fit points for the leptogenesis scenarios in Figs \ref{fig:NO1}-\ref{fig:IO3} are given and are all consistent with $\eta_{B}=\left(6.10\pm 0.04\right)\times 10^{-10}$,
  $\theta_{13}=8.52^{\circ}$ and $\theta_{12}=33.63^{\circ}$. The upper (lower) three rows are the best-fit points for normal (inverted) ordering. The final two rows are the best fit points for normal ordering in the loop and tree-level dominated scenarios.}
\end{table*}
%
 \begin{figure*}[t]\label{1DSFT}
\includegraphics[width=0.3\textwidth]{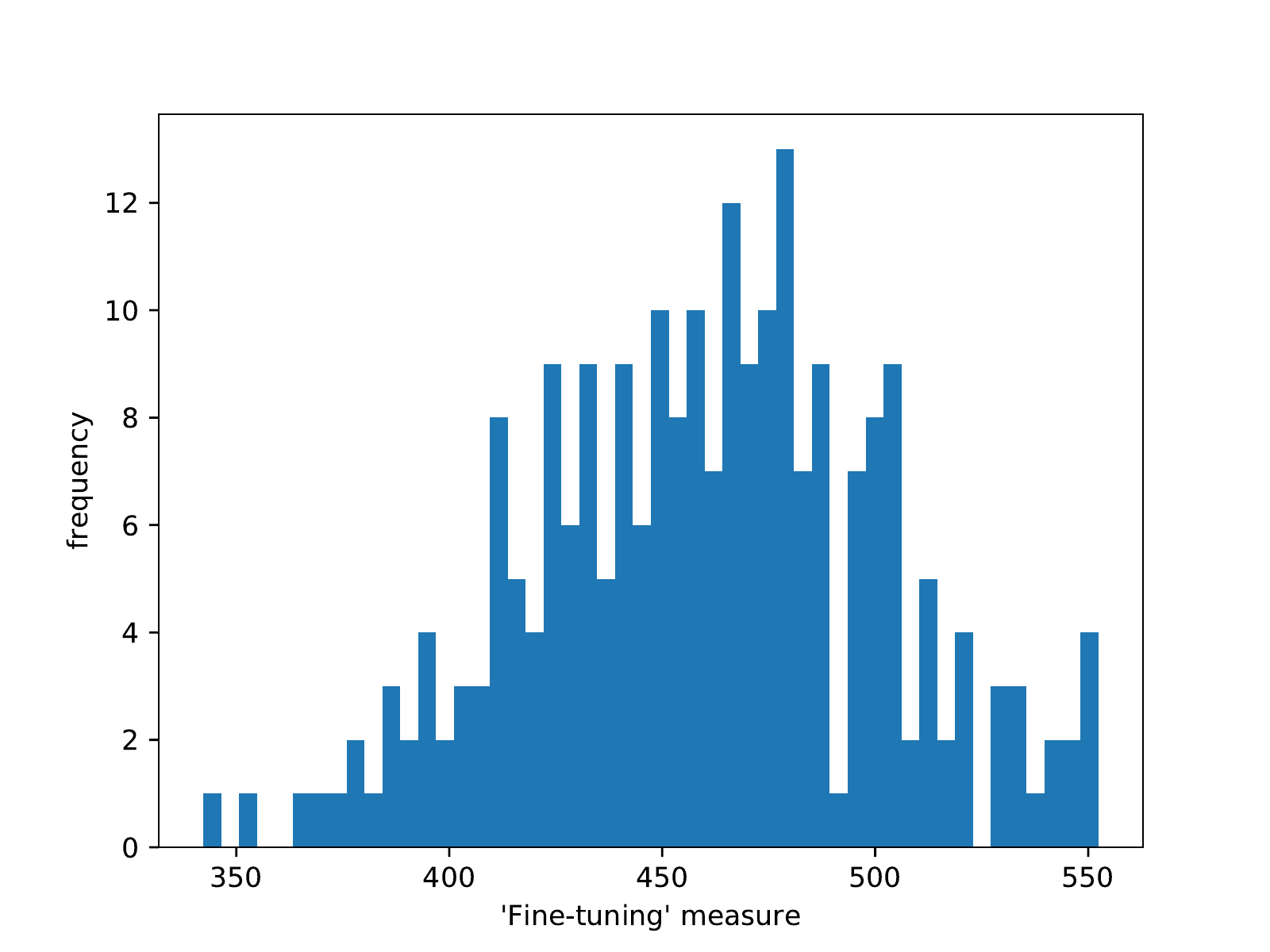}
\includegraphics[width=0.3\textwidth]{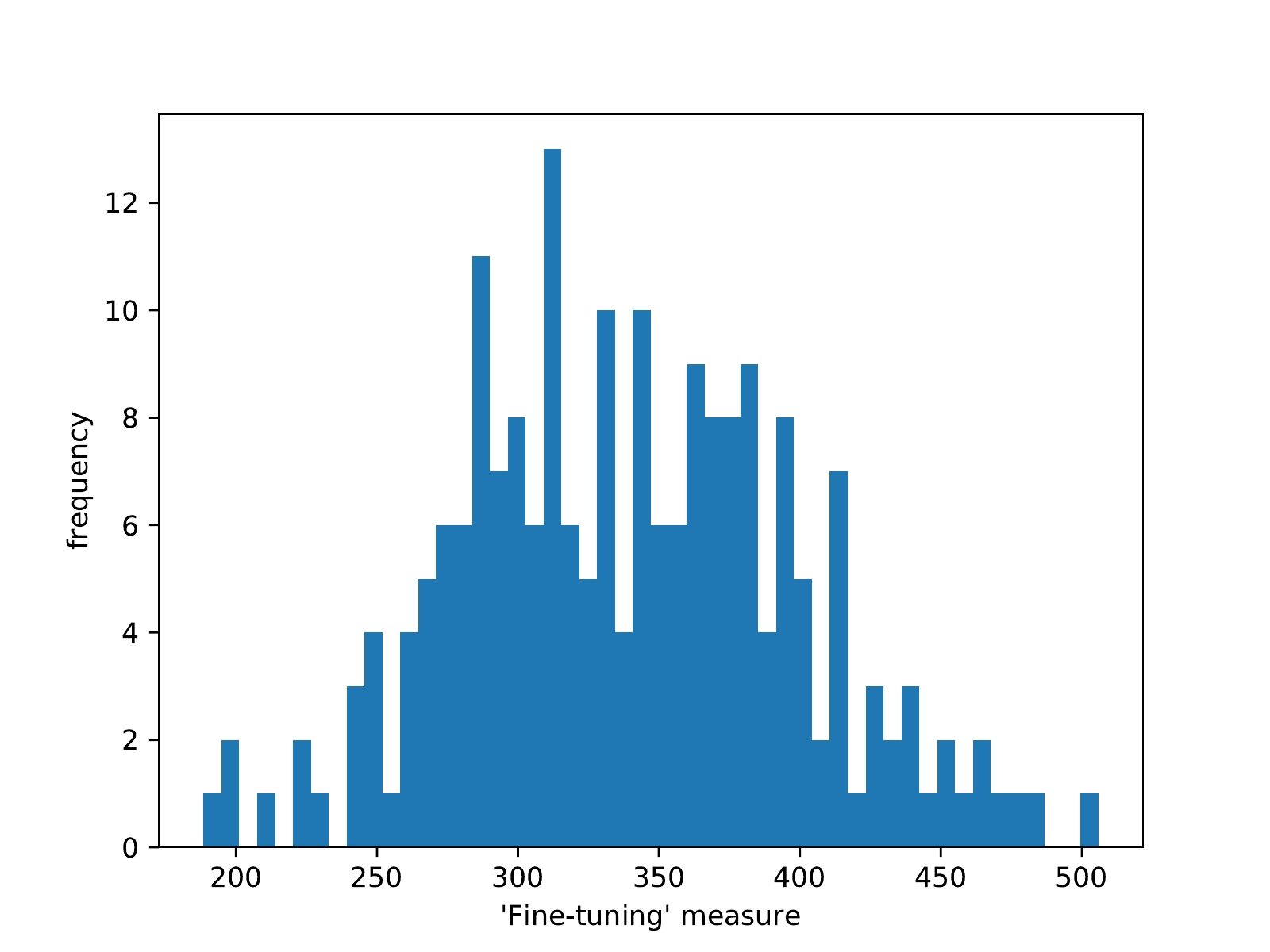}
\includegraphics[width=0.3\textwidth]{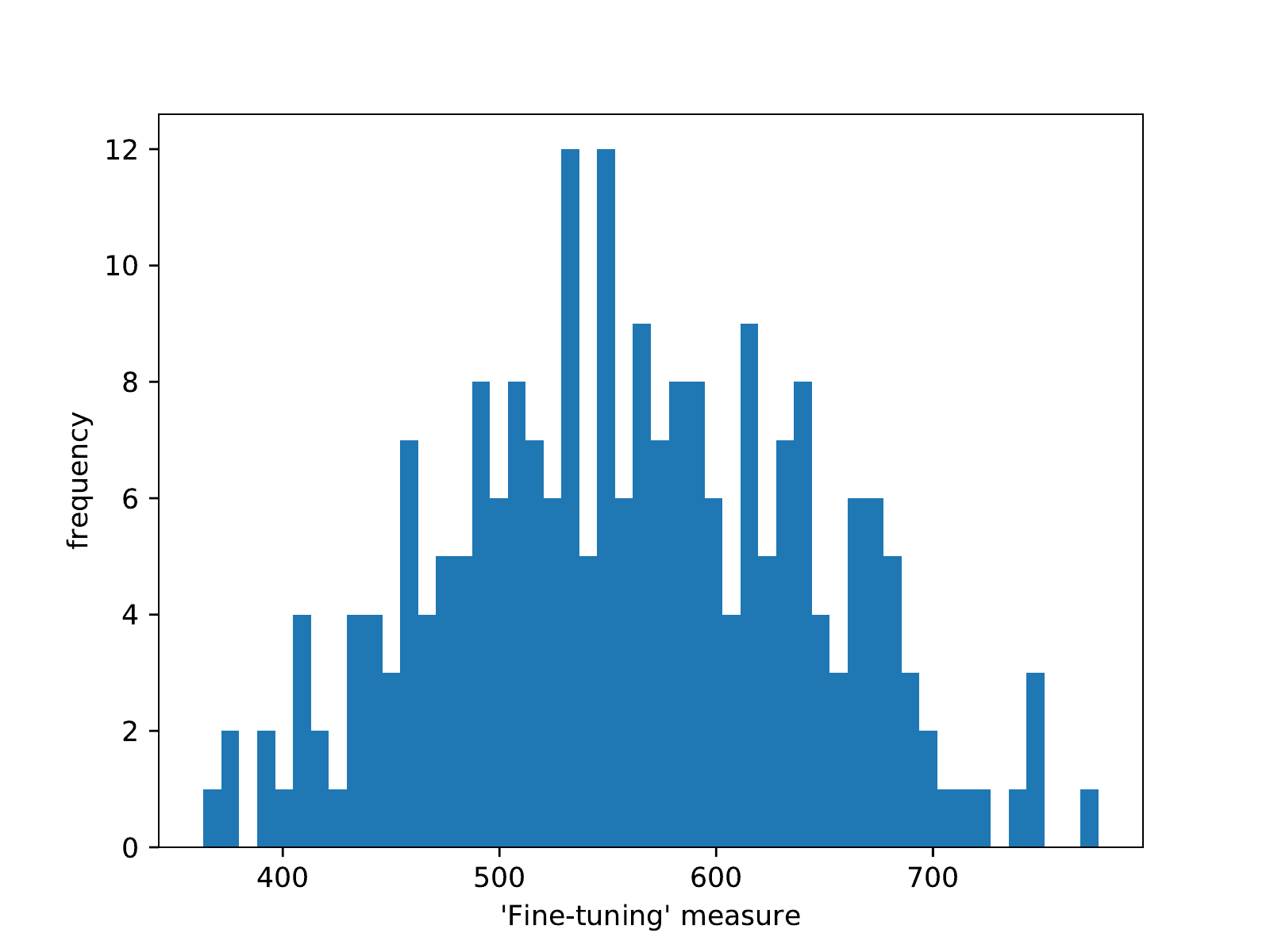}
\includegraphics[width=0.3\textwidth]{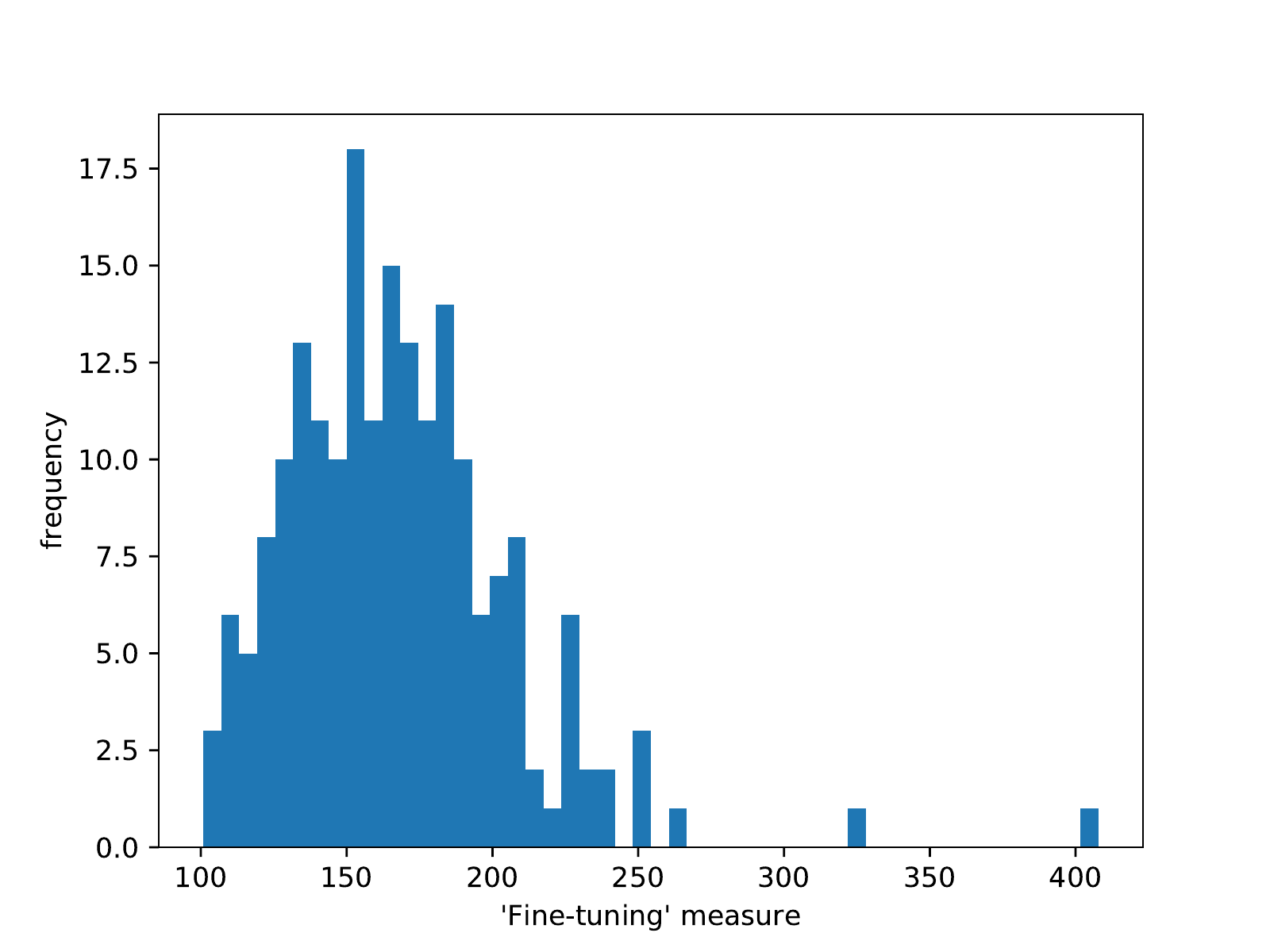}
\includegraphics[width=0.3\textwidth]{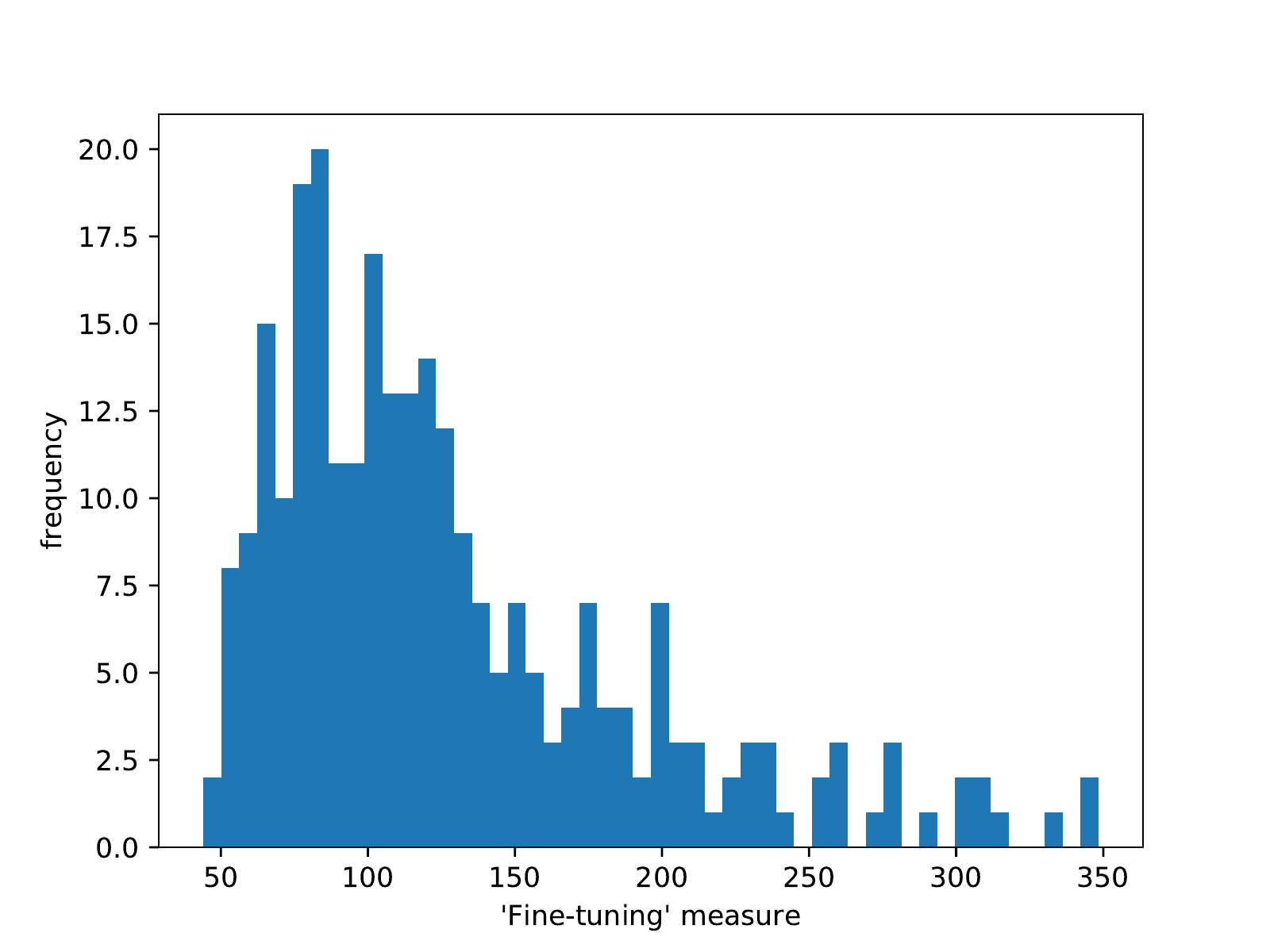}
\includegraphics[width=0.3\textwidth]{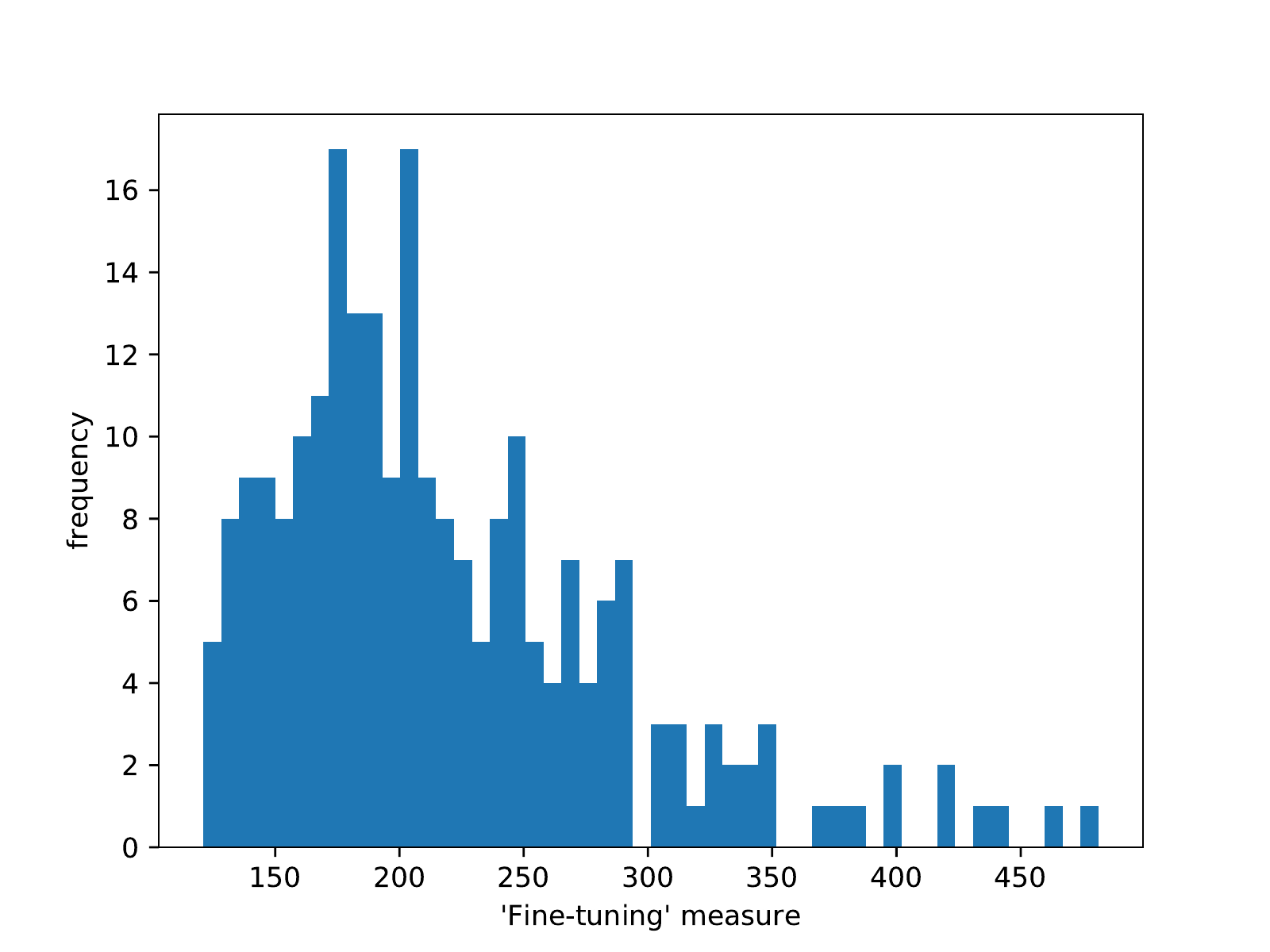}
\caption{The top (bottom) three plots from left to right show  the fine-tuning for regions of the model parameter space within 1$\sigma$ of measured $\eta_B$ for $S_{1}$, $S_{2}$ and $S_{3}$
($\overline{S_{1}}$, $\overline{S_{2}}$ and $\overline{S_{3}}$) respectively. }
\end{figure*}
As detailed in \secref{sec:NT}, solving the density matrix equations of  \equaref{eq:longDM} for an 18-dimensional model parameter space, $\mathbf{p}$, is a challenging numerical task. 
In order to reduce the volume, $\mathbf{p}$, we shall fix certain parameters. Firstly, as the solar and reactor mixing angles 
are relatively precisely measured, we shall use the values for these angles from global fit data \cite{Esteban:2016qun}.
Although we allow the  lightest neutrino mass ($m_{1}$ for NO and $m_{3}$ for IO) to vary within the experimentally allowable region, given by
the sum of neutrino masses, the other two light masses are determined from  the best-fit values of the atmospheric and solar mass squared splittings from global fit data \cite{Esteban:2016qun}. Finally, we fix the heavy Majorana mass spectrum leaving only 11 of the 18 parameters of \textbf{p} to be varied.

In all scenarios we choose a set of initial values for $M_{1}$, $M_{2}$ and $M_{3}$, in which, as mentioned, we ensure $M_3 > 3 M_2$ and $M_2 > 3 M_1$. We explore the parameter space and find the regions consistent with $\eta_{B_{CMB}}$ to a 1 and 2$\sigma$ level. Through the inspection of the fine-tuning of the solutions in the regions of $1\sigma$ agreement, we decide either to lower the scale of $M_1$  or not (while keeping the ratios $M_2/M_1$ and $M_3/M_2$ fixed). The lower the scale, the higher the fine-tuning and thus the greater the impact of higher-order corrections. Thus we do not further lower the scale when either  the two-loop contributions becomes greater than a few percent or when the fine-tuning exceeds $\mathcal{O}(1000)$ (see \appref{sec:2loops}) . If one were to  incorporate the effects of higher radiative orders, the parameter space could be explored at even lower scales where the fine-tuning is greater.

In the case of one decaying heavy Majorana neutrino contributing to the lepton asymmetry, we shall pick out six scenarios in total,  all of which
 satisfy a mild hierarchical spectrum:
 \begin{itemize}
 \item $S_1$:\\
 $M_{1}=10^{6}\,\text{GeV}, \,M_2/M_1 \simeq3.15, \,M_3/M_2\simeq3.15 $;
  \item $S_2$:\\
 $M_{1}=10^{6.5}\,\text{GeV}, \,M_2/M_1 \simeq3.15, \,M_3/M_2\simeq3.15 $;
 \item $S_3$:\\
 $M_{1}=10^{6.5}\,\text{GeV}, \,M_2/M_1 \simeq5, \,M_3/M_2\simeq5 $;
 \end{itemize}
for normally ordered light neutrino masses. In the case of inverted ordering, we shall denote these scenarios as $\overline{S_{1}}, \overline{S_{2}}$ and $\overline{S_{3}}$.
Scenarios $S_{1}$ ($\overline{S_{1}}$) and $S_{2}$ ($\overline{S_{2}}$) have the same mass ratios, with $S_{1}$ ($\overline{S_{1}}$) corresponding to the lowest value of the scale $M_1$ with acceptable fine-tuning values and $S_{2}$ ($\overline{S_{2}}$) presented for comparison. $S_{3}$ ($\overline{S_{3}}$) corresponds to the lowest scale for its given set of mass ratios. In \figref{fig:evolve}, we provide the temperature evolution of the absolute magnitude 
of the lepton asymmetry number densities, $\lvert n_{\alpha\alpha}\rvert, \alpha=e, \mu, \tau$ for the best-fit points of each scenario.

 The parameters of the PMNS matrix
are varied within their allowable or measured  3$\sigma$ range: $\delta\in(0, 360)^{\circ}$, $\theta_{23}\in(38.6, 52.5)^{\circ}$ and  $\alpha_{21}, \alpha_{31}\in(0, 720)^{\circ}$.
 We solve the density matrix equations of \equaref{eq:longDM} assuming a vanishing initial abundance of $N_{1}$ with an end point of the integration, $z\approx100$ after which $\eta_B$ is  constant. 
  In addition, we ensure the Yukawa couplings, $Y_{\alpha\beta}$, are perturbative and the CP-asymmetry does not suffer from resonant effects as detailed in \appref{sec:res}.

The plots in \figref{fig:NO1} show two-dimensional projections of the eleven-dimensional posterior\footnote{As each individual plot of the triangle plots is relatively small, we provide the following link  to view each  individually: \url{https://gitlab.dur.scotgrid.ac.uk/leptogenesis-public/thermal/wikis/home}}. The dark (light) blue contours correspond to the regions of parameter space consistent with 68$\%$ (95$\%$) confidence levels. 
In addition to the two-dimensional posterior plots we provide the best-fit point for each heavy Majorana neutrino mass spectrum scenario as shown in \tabref{tab:bestfitall} where the upper (lower) three rows of the table correspond to normal (inverted) ordering.  

For the two-dimensional posterior plots of scenario $S_{1}$, as shown in \figref{fig:NO1}, the  region of the model parameter space consistent to a 1$\sigma$ level with the observed baryon asymmetry favours larger values of the CP-violating Dirac phase,  $120 \leq\delta(^\circ)\leq 360$. 
The likelihood function appears  to be more sensitive to  $\alpha_{21}$ than $\alpha_{31}$: from \figref{fig:NO1}, we observe $80 \leq \alpha_{21}(^\circ)\leq 270$ while
  $65 \leq\alpha_{31}(^\circ)\leq 720$ is consistent with the measured baryon asymmetry to a 1$\sigma$ level. Although the atmospheric mixing angle may take most value within its 3$\sigma$ range, the likelihood function favours values close to $45^\circ$ and in the upper octant.
 The values of the  lightest neutrino mass  which are consistent with the observed $\eta_{B}$ tend to be close to the upper limit, which for normal ordering is $m_{1} \simeq  3.32\times10^{-1}\,\text{eV}$. This strong dependence of $\eta_B$  on the lightest neutrino mass agrees with work which investigated (two) flavoured thermal leptogenesis  \cite{Molinaro:2007uv}.

In general, the likelihood function is more sensitive to the imaginary than the real 
components of the $R$-matrix. For example, we find that $\eta_{B}$ is relatively insensitive to $x_{1}$ and $x_{3}$: $x_1, x_{3} \in (-180^{\circ}, 180^{\circ})$ is consistent with the measured $\eta_{B}$ to 2$\sigma$ level. On the contrary, the likelihood function is highly sensitive to $x_{2}$ with preferred values of approximately $90^{\circ}$. 
We note that the two-dimensional projections onto parameters $x_{1}$ and $x_{3}$ are not included in the triangle plots as $\eta_{B}\left(\textbf{p}\right)$ exhibits flat directions in
both these parameters and the two-dimensional projection plots show little interesting structure\footnote{However, these plots are included in the aforementioned link.}.
The complex components of the $R$-matrix 
are likely to be  within a small range: $y_{1}\simeq 180^{\circ}$, $y_{2}\simeq 3^{\circ}$ and $y_{3}\simeq 180^{\circ}$ where the explanation for this structure has been detailed
in \secref{sec:ftdiscus}. Given the mass of the decaying heavy Majorana neutrino is relatively light, it would be expected that large phases of the PMNS and $R$-matrix are favoured as these ensure the Yukawa couplings are sufficiently large.

 The triangle plots  for larger masses of $M_{1}$ and more hierarchical heavy Majorana neutrino spectra
  of $S_{2}$ and $S_{3}$  are shown in \figref{fig:NO2} and \figref{fig:NO3} respectively. 
  Unsurprisingly, on comparison of  scenario $S_{1}$ and $S_{2}$ (which share the same mass splitting but different values of $M_{1}$) we observe the scenario with the larger heavy Majorana neutrino masses has a larger region of  the model parameter space consistent with the measured $\eta_{B_{CMB}}$. Moreover, as expected, the constraints on the $R$- and PMNS-matrix parameters in scenario $S_{2}$ are weaker yet qualitatively similar to $S_{1}$. In particular, the $m_{1}$-dependence in $S_{2}$ is less severe than in the scenario of $S_{1}$; for example in 
  \figref{fig:NO1} the 2$\sigma$ allowed region for the  lightest neutrino mass  is  $1.25\times10^{-1}\leq m_{1}(\text{eV})\leq3.32\times10^{-1}$ while in the case of  \figref{fig:NO2}, $3.16\times10^{-2}\leq m_{1}(\text{eV})\leq3.32\times10^{-1}$.  For smaller values of $m_1$, successful leptogenesis is possible for larger values of the heavy Majorana neutrino mass $M_1$. For larger heavy Majorana neutrino mass splitting, we anticipate the model parameter volume consistent with data will be reduced. This is  because the CP-asymmetry becomes increasingly suppressed for larger mass splittings.  This effect is confirmed upon comparison of \figref{fig:NO2} and \figref{fig:NO3} where the former has milder mass splitting.
In contrast to $S_{1}$, in the case of both $S_2$ and $S_3$, the likelihood function favours values of $\theta_{23}$ close to $45^\circ$ and in the lower octant.

The triangle plot showing the two-dimensional posterior distributions of the 11-dimensional model parameter space for  $\overline{S_{1}}$ is shown in \figref{fig:IO1}. The dark (light) red contours correspond to the regions of parameter space consistent with 68$\%$ (95$\%$) confidence levels. As anticipated, the points of the model space consistent with the measurement are different from the normal ordering case and the volume of parameter space $\mathbf{p}$ consistent with the measured $\eta_{B_{CMB}}$ is less constrained. In particular we observe that the  likelihood function is relatively insensitive to changes of $\delta$, $\alpha_{31}$ and $\theta_{23}$.  
However, this scenario displays a similar feature to $S_{1}$,  where the likelihood function favours values of $\alpha_{21}\leq 360^{\circ}$. 

Additionally, the likelihood has a flat direction in the $x_{1}$ and $x_{3}$ parameters of the $R$-matrix (as discussed in \secref{sec:ftdiscus}). We observe that all values of $x_{1}$ and $x_{3}$ are consistent to a 2$\sigma$ level with the measured $\eta_{B}$; however, the likelihood is very sensitive to $x_{2}$ with $x_{2}\simeq 90^{\circ}$. 
 Similarly, to the normal ordering scenario the imaginary phases of R are constrained with $y_{1}\simeq 180^{\circ}$, $y_{2}\simeq 2^{\circ}$ and $y_{3}\simeq 180^{\circ}$.
 The triangle plots for larger masses of $M_{1}$ and more hierarchical spectra of $\overline{S_{2}}$ and $\overline{S_{3}}$ are shown in  \figref{fig:IO2} and \figref{fig:IO3} respectively.  As seen in the case of normal ordering, the scenario with the slightly more hierarchical mass spectrum ($M_{2}=5\,M_{1}$,  $M_{3} =5\, M_{2}$) has a slightly smaller volume of parameter space consistent with the data than the case of 
 the milder hierarchy. 
 
 Although we allow for the possibility there exists  a certain level of cancellation between the tree and one-loop level contributions to the
light neutrino masses, we  avoid regions of the parameter space where the perturbative series no longer converges. We present the 
fine-tuning measure
  defined in  \equaref{eq:FT} for the regions of the model parameter space within 1$\sigma$ of the measured $\eta_{B}$.
To be explicit, the top (bottom) three plots of  \figref{1DSFT} shows the distribution of the fine-tuning measure within the  1$\sigma$ region of $S_1$, $S_2$ and $S_3$ ($\overline{S_1}$, $\overline{S_2}$ and $\overline{S_3}$) shown in \figref{fig:NO1}, \figref{fig:NO2} and \figref{fig:NO3} (\figref{fig:IO1}, \figref{fig:IO2} and \figref{fig:IO3}) respectively.  Moreover increasing the spread from 1$\sigma$ to 5$\sigma$  would  allow for a broader spread of fine-tuning values, both smaller and larger. 
%
%

In general, for normal ordering, the fine-tuning measure for points within 1$\sigma$ is $\mathcal{O}\left(100\right)$. We observe that the minimal fine-tuning value for $S_{1}\approx330$. Somewhat unsurprisingly, the scenario with the large mass of decaying heavy Majorana neutrino, $S_{2}$, has smaller fine-tuning due to the fact the complex phases of the $R$-matrix may attain a broader range of values. We observe the minimum fine-tuning measure in the case of  $S_{2}$ to be $\approx180$. However, in the case of $S_{3}$ (where the decaying heavy Majorana neutrino mass is the same as $S_2$ the mass splitting between the heavy Majorana neutrinos is larger) the fine-tuning values are in general larger due to the increased mass of $N_{3}$.

The fine-tuning present in the case of inverted ordering is, in general, less than in the case of normal ordering. The minimum value of fine-tuning present in $\overline{S_{1}}\simeq 100$. Again, the same pattern emerges as in the case of normal ordering where the fine-tuning in $\overline{S_{2}}$ ($\overline{S_{3}}$) is less (greater) than $\overline{S_{1}}$. In fact, for $\overline{S_{2}}$ the minimum fine-tuning $\approx40$. Again, we emphasise the fine-tuning we present here is for points in $\textbf{p}$ within 1$\sigma$ of
the best fit value of $\eta_{B_{CMB}}$ and allowing for an increase in the spread around the best fit value would allow for smaller (and larger) values of  fine-tuning.

At such scales, $T\ll 10^{9}$ GeV, it is impossible to have successful leptogenesis without some degree of cancellation between the tree and one-loop level contributions. 
However, we did investigate if there existed regions of $\mathbf{p}$ such that thermal leptogenesis was viable (within 1$\sigma$ of the central value of $\eta_{B_{CMB}}$) where  either the tree or one-loop level contribution dominates. In the latter scenario, where the radiative corrections dominate over the tree-level contributions, the fine-tuning measure should be close to unity as $\lvert m^{\text{1-loop}}\rvert/\lvert \left(m^{\text{tree}} + m^{\text{1-loop}}\right)\rvert\approx1$ for $m^{\text{tree}}\ll m^{\text{1-loop}}$. We applied the same numerical procedure to solve the density matrix equations with one decaying heavy Majorana neutrino and vetoed points in $\mathbf{p}$ if the fine-tuning measure was not within the boundary $0.9\leq\text{F.T}\leq1.1$. After scanning a series of differing heavy Majorana neutrino mass spectra, we found the loop-dominated scenario was possible, assuming normal ordering, for $M_{1}=10^{9}$ GeV with $M_{2}=3.15 M_1$ and $M_{3}=3.15M_{2}$. The best-fit point is denoted as ${F.T}^{\text{loop}}$ in \tabref{tab:bestfitall} and the triangle plot of the two-dimensional posterior distributions may be found on the provided webpage. In the former scenario, where the tree-level contributions dominates, the fine-tuning measure will be close to zero. Using {\sc Multinest} to search for regions of $\mathbf{p}$ consist with tree-domination we required the fine-tuning to be within the boundary $0\leq\text{F.T}\leq0.2$. We found no solutions compatible with this condition for $M_1<10^9$ GeV.
However, we did find a single single point consistent with a fine-tuning $\approx0.18$ for a mass spectrum of $M_{1}=10^9$GeV, $M_{2}\approx 3.15M_1$ and $M_{3}\approx 3.15M_2$. Note that a two-dimensional projection of the posterior is not possible and we simply provide the value of this point as ${F.T}^{\text{tree}}$ in  \tabref{tab:bestfitall}.
 For larger values of $M_1$ more points will exist that satisfy the condition and so we regard ${F.T}^{\text{tree}}$ as the solution of lowest $M_1$ in which the tree-level is the dominant contribution. The absolute values of the Yukawa matrix elements are listed for all scenarios for reference in \appref{sec:AbsYukawaMatrices}.

We note that it is possible to reduce the fine-tuning by considering the scenario where $M_2=M_3$. Such a scenario may result from the introduction of a partial symmetry into the type-I seesaw. As, in this section, we only consider the case that $N_1$ decays this does not lead to  resonant leptogenesis. As an example, consider $S_1$ but with $M_2=M_3\approx5.05 \times 10^6 \text{ GeV}$. Such a  point in $\mathbf{p}$ leads to $\eta_B = 6.1 \times 10^{-10}$, which is in good agreement with the experimental value. In this case, $N_2$ and $N_3$ act as two Majorana components of a pseudo-Dirac pair. The contribution of $N_2$ and $N_3$ to the tree-level mass is cancelled (as together they are lepton number conserving) and a dramatic reduction in our fine-tuning measure occurs, resulting in F.T. $\approx2.1$. This is similar to the scenarios considered in~\cite{Racker:2012vw} and will not be further discussed in this paper.

In summary, foregoing fine-tuning of the light neutrino masses
 $\gtrsim\mathcal{O}(10)$, it is possible to lower the scale of 
 non-resonant thermal leptogenesis to $T\sim 10^{6}$ GeV with a mildly hierarchical
 heavy Majorana neutrino mass spectrum. 
 At such intermediate scales, interactions mediated by the SM
 charged lepton Yukawa couplings are greater than the Hubble rate. We have properly
 accounted for such effects as we calculated the lepton asymmetry from three-flavoured density 
 matrix equations.
 In the case of normally ordered light neutrinos, larger values of the 
 $\delta$  are favoured in conjunction with an atmospheric mixing angle close to $\theta_{23} = 45^{\circ}$ (slightly above or below depending on the scenario, see \tabref{tab:bestfitall}). We observe that larger masses of $m_{1}$ are favoured as this  compensates for decreasing $M_{1}$. In the scenario of an inverted ordered mass spectrum, the likelihood function shows little sensitivity to changes in the low-energy neutrino parameters. On the other hand, the $R$-matrix is comparatively highly constrained. In addition, we present the distribution of the fine-tuning measure within 1$\sigma$ of the measured $\eta_{B}$ and  found the fine-tuning  was in general smaller for inverted than normal ordering and usually took values $\sim\mathcal{O}\left(100\right)$. 
 We find that the minimum observed value of the fine-tuning measure in the vicinity of the best-fit  is $\sim$40. However, \emph{at} the most likely point, the F.T. assumes values $\sim\mathcal{O}(100)$.
%

\subsection{Results from $N_{2}$ Decays}\label{sec:2DS}
 \begin{figure*}[t]\label{2DSFT}
\includegraphics[width=0.4\textwidth]{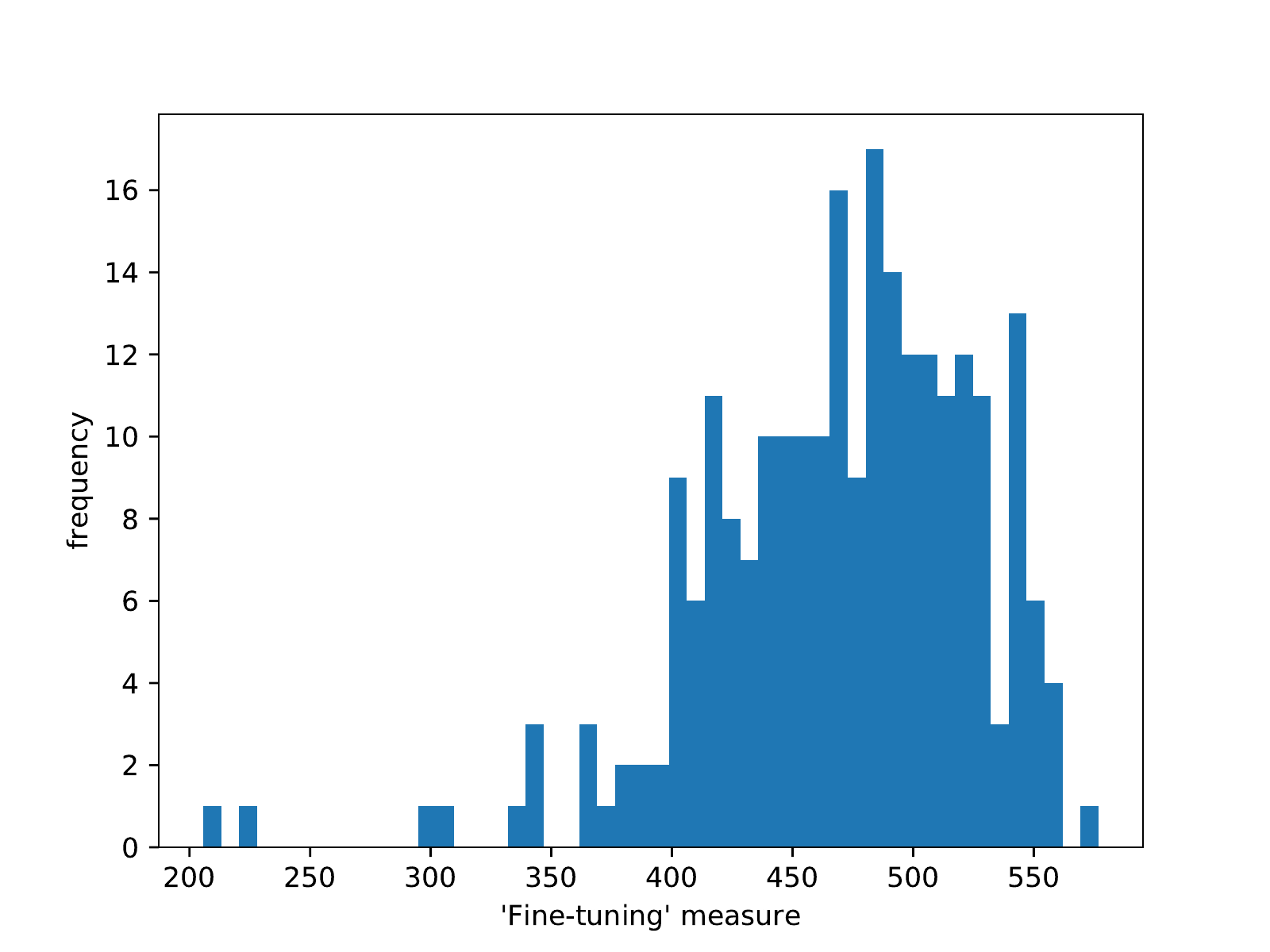}
\includegraphics[width=0.4\textwidth]{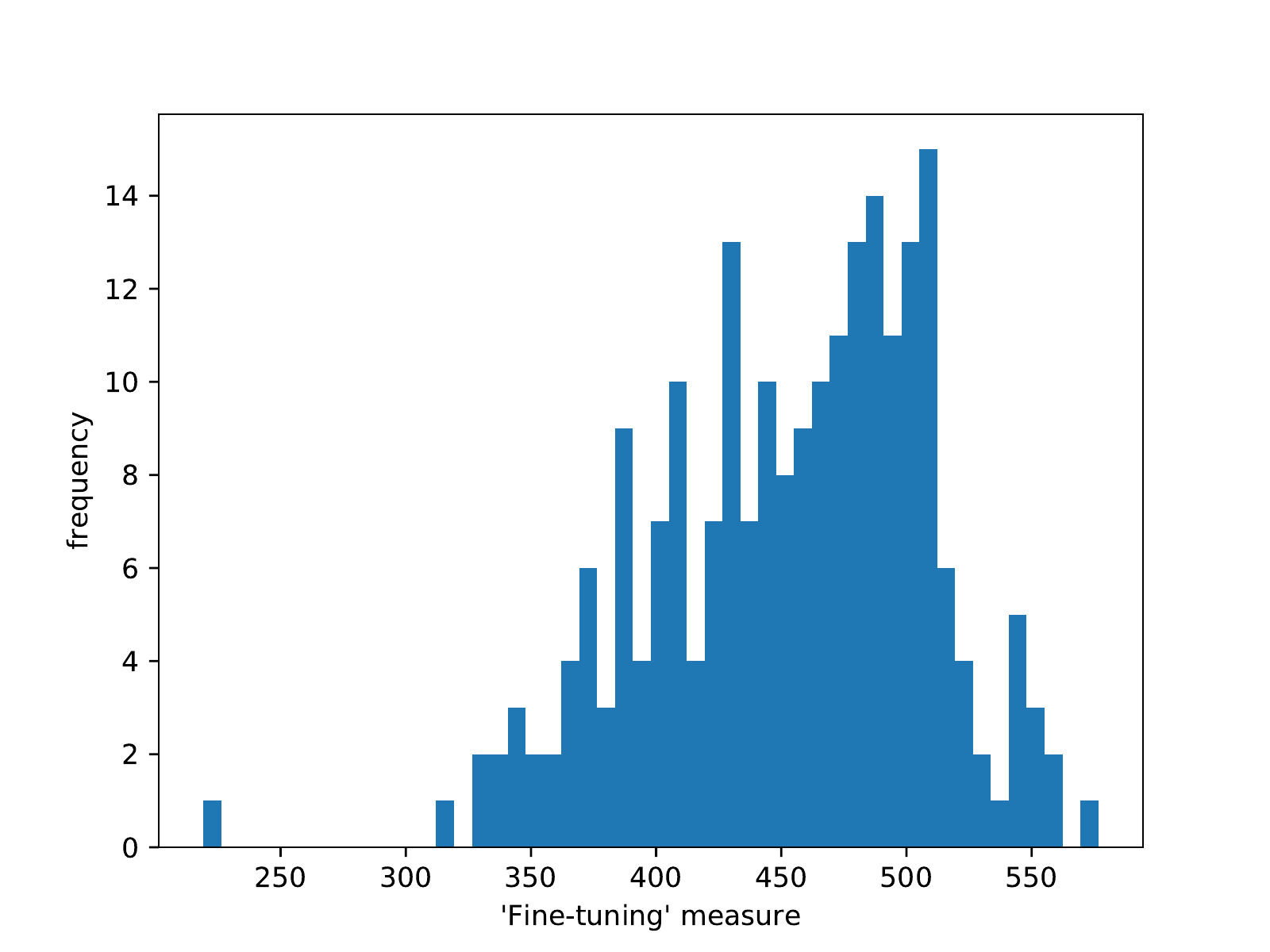}
\caption{The left (right) plot shows the fine-tuning for regions of the model parameter space within 1$\sigma$ of measured $\eta_B$ for $S_{4}$ ($\overline{S_{4}}$).}
\end{figure*}
\begin{table*}[t]\label{tab:bestfit2DS}
  \centering
  \begin{tabular}{c|c|c|c|c|c|c|c|c|c|c|c|c|c|c}
    & $\theta_{23}(^\circ)$ & $\delta(^\circ)$    & $\alpha_{21}(^\circ)$ &  $\alpha_{31}(^\circ)$ &  $x_{1}(^\circ)$    & $y_{1}(^\circ)$ &  $x_{2}(^\circ)$ & $y_{2}(^\circ)$    & $x_{3}(^\circ)$ &  $y_{3}(^\circ)$  &
    $m_{1(3)}$ (eV) &       $M_{1}$ (GeV) &    $M_{2}$ (GeV)&    $M_{3}$(GeV)\\
          \hline\hline
$S_{4}$&  $47.85$  & $105.65$  & $133.40$ & $367.99$ & $-99.50$ & $178.77$ & $94.22$ & $0.12$ & $-9.59$ & $172.53$ & $0.208$ & $10^{6.7}$ & $10^{7.5}$ & $10^{8.1}$ \\
$\overline{S_{4}}$& $44.11$ & $243.0$ &  $347.54$ & $437.04$ & $14.94$ & $167.76$ & $90.79$ & $1.42$ & $132.12$ & $178.29$ & $0.0084$ & $10^{6.7}$ & $10^{7.5}$ & $10^{8.1}$ \\
      \end{tabular}
  \caption{The best-fit points for the leptogenesis scenarios in Figs \ref{fig:NODS21}-\ref{fig:IODS21} are given and are all consistent with $\eta_{B}=\left(6.10\pm 0.04\right)\times 10^{-10}$,
  $\theta_{13}=8.52^{\circ}$ and $\theta_{12}=33.63^{\circ}$. The upper (lower) row is the best-fit points for normal (inverted) ordering.}
\end{table*}
%
%
\begin{figure*}[t]\label{fig:NODS21}
\includegraphics[width=0.99\textwidth]{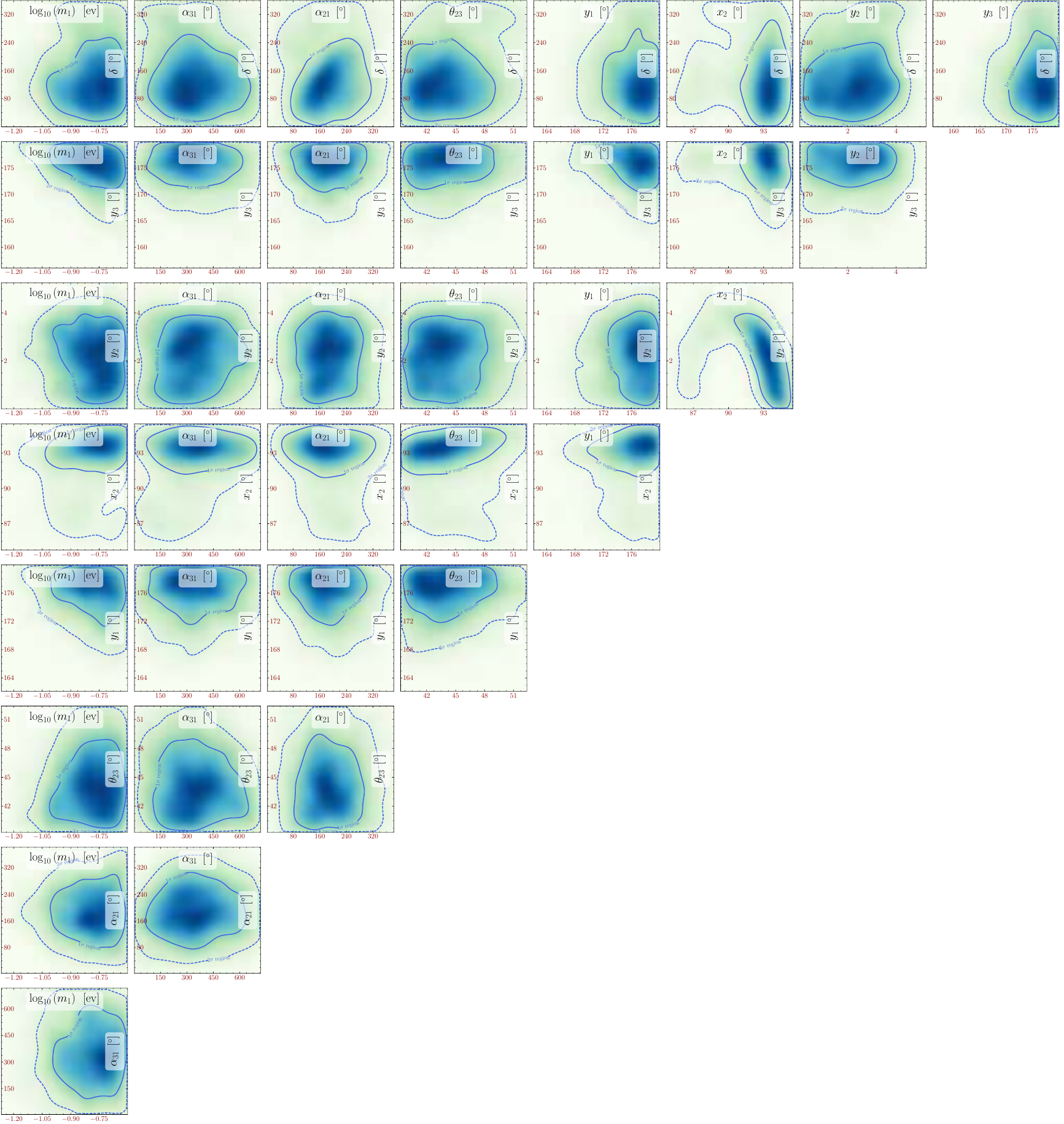}
\caption{$S_{4}$: Triangle plot showing the two-dimensional projection   of the 11-dimensional model parameter space for posterior distributions using  normal ordering, with two-decaying 
steriles neutrinos and mass spectrum: $M_{1}=10^{6.7}$ GeV, $M_{2}=5.0\, M_{1}$, $M_{3}=5.0\,M_{2}$. The contours correspond to 68$\%$ and 95$\%$ confidence levels respectively.}
\end{figure*}
\begin{figure*}[t]\label{fig:IODS21}
\includegraphics[width=0.99\textwidth]{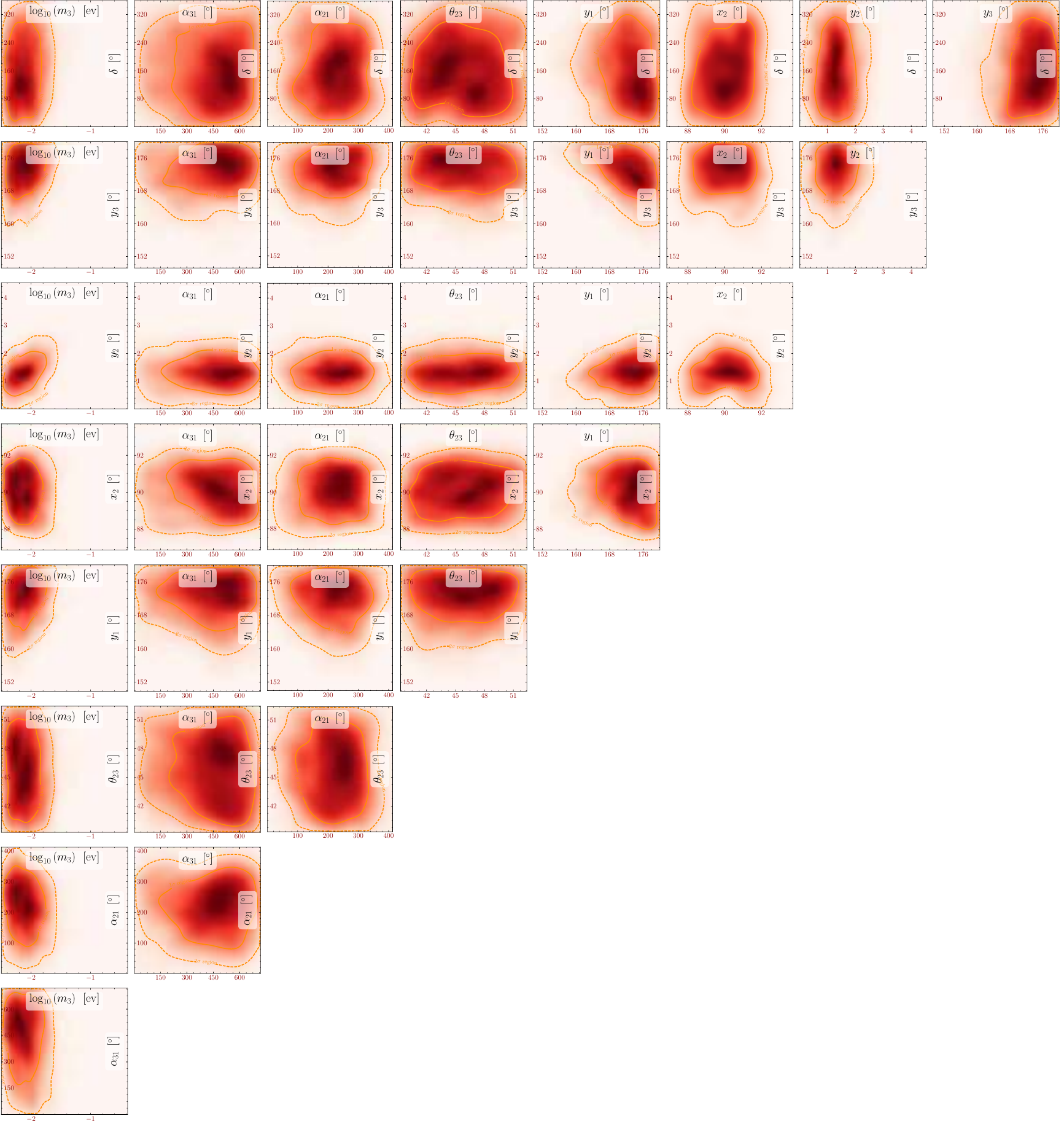}
\caption{\label{fig:IO1DS} $\overline{S_{4}}$: Triangle plot showing the two-dimensional projection of the 11-dimensional model parameter space for posterior distributions using inverted  ordering, with two-decaying 
steriles and mass spectrum:  $M_{1}=10^{6.7}$ GeV, $M_{2}=5.0\, M_{1}$, $M_{3}=5.0\,M_{2}$. The contours correspond to 68$\%$ and 95$\%$ confidence levels respectively.}\end{figure*}
In this section, we explore the possibility that the decay of two heavy Majorana neutrinos contributes to the baryon asymmetry. In this setup, the density matrix equations 
follow rather straightforwardly  from \equaref{eq:full3} and the numerical procedure to find the two-dimensional posterior plots is the same as discussed in \secref{sec:1DS}. The qualitative difference between this case and the former as discussed in \secref{sec:NT} is that now $N_{2}$ may decay in addition to $N_{1}$. As $M_{2}>M_{1}$, $N_{2}$ will decay before $N_{1}$ with the average time between the two decays determined by the hierarchy of their masses. 

In \cite{Antusch:2011nz} the authors explored thermal leptogenesis using the decay of two heavy Majorana neutrinos in the limit the third is decoupled from the theory. Using analytic estimates, they found
 the minimal mass of the lightest heavy Majorana neutrino, for successful leptogenesis, to be $M_{1}\sim 1.3\times10^{11}$ GeV assuming a mildly hierarchical mass spectrum.
In this scenario, we explored a number of heavy Majorana neutrino mass scenarios and found the lowest mass of $N_{1}$ which allowed  for successful leptogenesis was $M_{1}=10^{6.7}$ GeV with 
$M_{2}\approx 6.3 M_{1}$ and $M_{3}\approx4 M_{2}$. We denote these two scenarios as $S_4$ and $\overline{S_4}$ for normal and inverted ordering respectively and the best-fit point and corresponding triangle plots are shown in \figref{fig:NODS21} and \figref{fig:IODS21}.

Naively, one would think that the decay of two heavy Majorana neutrinos would further lower the scale of leptogenesis; however, this is not the case as there is
non-trivial interplay between the decays and washout processes of $N_{2}$ and $N_{1}$. 
We note that contribution of the third heavy Majorana neutrino to the lepton asymmetry in these scenarios is negligible as the CP-asymmetry $\epsilon^{(3)}_{\alpha \beta}$ is several orders of magnitude lower than that of the other two and its washout term $W_3$ decays far faster.

Unlike in the previous section, we find the two-dimensional posterior projections in this case
for both orderings do not appear to be too dissimilar.  In both cases, the likelihood function is  insensitive to $\delta$. In addition, the atmospheric mixing angle can be in the lower or upper octant and there is strong dependence on large values of $m_{1}$ ($m_{3}$) in $S_{4}$ ($\overline{S_{4}}$).  The dependence of the likelihood on the $R$-matrix parameters is similar to the cases discussed in \secref{sec:1DS}; we find
 $x_{1}$ and $x_{3}$ may take any values while $x_{2}\simeq 90^{\circ}$. Likewise, two of the imaginary components of the $R$-matrix are constrained to be large  $y_{1}, y_{3} \simeq 180^{\circ}$ while the other is nearly vanishing $y_{2}\simeq 2.5^{\circ}$.  For reference, the corresponding absolute value Yukawa matrices are given in \secref{sec:AbsYukawaMatrices}.
In a similar fashion to \secref{sec:1DS}, we present the fine-tuning measure for the regions of the model parameter space within 1$\sigma$ of the measured $\eta_{B}$. We observe for normal and inverted ordering the fine-tuning $\sim \mathcal{O}\left(100\right)$. 
\section{Discussion of Fine-tuned Results}\label{sec:ftdiscus}

We may gain an understanding of why fine-tuned solutions were found by the numerical machinery through inspection of  the structure of the Yukawa matrix at the best-fit points. Looking at the solutions for one and two decaying heavy Majorana neutrino scenarios, we observe that generically $|y_1| \approx 180^\circ$, $y_2 \approx 0^\circ$, $|y_3| \approx 180^\circ$ and $\lvert x_2 \rvert \approx 90^{\circ}$. 
Consider as a typical example $S_1$, for which the orthogonal $R$-matrix assumes the following form
\[
R \approx 
\left(
\begin{array}{ccc}
 -\frac{i}{2} e^{y_3} \cos x_2 & \frac{1}{2} e^{y_3} \cos x_2 &
   \sin x_2 \\
 \frac{i}{2} e^{y_1+y_3} & -\frac{1}{2} e^{y_1+y_3}
    & \frac{1}{2} e^{y_1} \cos x_2 \\
 \frac{1}{2} e^{y_1+y_3} & \frac{i}{2} e^{y_1+y_3}
    & -\frac{i}{2} e^{y_1} \cos x_2 \\
\end{array}
\right),
\]
which has the structure
\begin{equation}\label{RStructure}
R \approx
\left(
\begin{array}{ccc}
 R_{11} & R_{12} & R_{13} \\
 -i R_{22} & R_{22} & R_{23} \\
 -R_{22} & -i R_{22} & -i R_{23} \\
\end{array}
\right)\text{.}
\end{equation}
The appearance of $y_{1}$ and $y_{3}$ in the exponentials, and the proximity of $x_{2}$ to $90^\circ$, result in $\lvert R_{13} \rvert \sim 1$,  $\lvert R_{1i} \rvert \ll \lvert R_{22} \rvert$ and $\lvert R_{i3} \rvert \ll \lvert R_{22} \rvert$.

In the case of the asymmetries $\epsilon^{(1)}_{\alpha\alpha}$, 
generated in the $N_1$ decays,
and for the best-fit values of the parameters listed in Table \ref{tab:bestfitall}, the 
leading term in the
expansion of the function $f_1(x_j/x_1)$ in powers of $x_j/x_1 = M^2_1/M^2_j \ll 
1$, $j=2,3$, as can be shown,
gives a sub-dominant contribution. The dominant contribution is 
generated by the next-to-leading
term in the expansion of $f_1(x_j/x_1)$ as well as by the leading term in 
the expansion of  the self-energy function $f_2(x_j/x_1)$ in powers of $x_j/x_1 = 
M^2_1/M^2_j \ll 1$.
Under the approximation $m_1=m_2$, the part of the asymmetry proportional to $f_1$ (which we call $\epsilon^{(1)}_{\alpha \alpha} \left( f_1 \right)$) is
\[
\begin{aligned}
& \epsilon^{(1)}_{\alpha \alpha}\left(f_1\right) = \frac{3}{16 \pi \left(Y^{\dagger} Y\right)_{11}} \frac{M_1^2}{v^4} \frac{5}{9} \frac{M_1^2}{M_2^2} \times \\
& \left(m_1^2 |U_{\alpha 1}+i U_{\alpha 2}|^2 \Im\left[\left(R_{11}^* R_{21} \right)^2 \right] \right. \\
& \left. +  m_1 \sqrt{m_1 m_3} \Im \left[R_{11}^* R_{21}^2 U_{l3}^* R_{13}^* \left(U_{\alpha 1}+iU_{\alpha 2}\right) \right] \right).
\end{aligned}
\]
and
\[
\begin{aligned}
\epsilon^{(1)}_{\alpha \alpha}\left(f_2\right) = \frac{3}{16 \pi \left(Y^{\dagger} Y\right)_{11}} \frac{M_1^2}{v^4} \frac{2}{3} m_1^{\frac{3}{2}} \sqrt{m_3} |R_{21}|^2 \times \\ 
\left( \sum_{j=2,3} \frac{M_1}{M_j} \right) \Im \left[R_{11} R_{13}^* U_{l3}^* \left(U_{\alpha 1}+iU_{\alpha 2}\right) \right] .
\end{aligned}
\]
Numerical estimates at the best-fit values of Table~\ref{tab:bestfitall}
show that this second contribution (the resonance function  
contribution) is somewhat larger than the first one, although,
the baryon asymmetry in the cases studied by us is produced in the  
non-resonance regime.

In the density matrix equations, the CP-asymmetry parameters enter in the combinations
\begin{align}
\nonumber &\epsilon^{(1)}_{ee}(f_1) +\epsilon^{(1)}_{ee}(f_2),\\
\nonumber &\epsilon^{(1)}_{\mu\mu}(f_1) + \epsilon^{(1)}_{\mu\mu}(f_2),\\
\nonumber &\epsilon^{(1)}_{\tau\tau}(f_1) + \epsilon^{(1)}_{\tau\tau}(f_2),
\end{align}
in the three-flavour regime.

Thus, although for our best-fit scenarios $\epsilon^{(1)}_{ee}(f_2) +
\epsilon^{(1)}_{\mu\mu}(f_2)
+ \epsilon^{(1)}_{\tau\tau}(f_2)$ may be $\mathcal{O}\left(10^{-22}\right)$, this does not mean that the
$\epsilon^{(1)}_{\alpha \alpha}(f_2)$ give a negligible contribution
in the generation of the lepton (baryon) asymmetry.

We note that there is a factor $(Y^{\dagger} Y)^{-1}_{11}$ in the diagonal CP-asymmetries $\epsilon^{(1)}_{\alpha \alpha}$ (\equaref{eq:CPoff}) for the lightest heavy Majorana neutrino and a factor $(Y^{\dagger} Y)_{11}$ (\equaref{eq:washout}) appears in the washout term $W_{1}$. Thus, we naively expect that in order achieve successful leptogenesis, by reducing the washout, $(Y^{\dagger} Y)_{11}$ should be made small. Expanding this quantity, in terms of the $R$-matrix elements and the remaining CI parameters, we find
\[
\left( Y^{\dagger} Y \right)_{11} =  \frac{M_{1}}{v^2} \left( m_1 |R_{11}|^2 + m_2 |R_{12}|^2 + m_3 |R_{13}|^2 \right).
\]
Thus, with the assumption that this quantity should be small, the relative smallness of the elements $R_{1i}$ is explained and with it the values of $x_2$ and $y_2$.

Similarly, given the dependence on $|R_{21}|$ in $\epsilon^{(1)}_{\alpha \alpha}(f_2)$, it may be expected that we should maximise the values of $y_1$ and $y_3$. With these imaginary parts of $\omega_{1}$ and $\omega_{3}$ large, the values of the corresponding  real parts $x_{1}$ and $x_{3}$ is immaterial. This is reflected in the relative flatness of their directions in the parameter space plots. The dependence on $m_1$ in $(Y^{\dagger} Y)_{11}$ may initially lead one to expect $m_1$ to be minimised. That this is not the case is due to the factors $m_1^{2}$ or $m_1^{3/2} \sqrt{m_3}$ appearing in the expressions for $\epsilon^{(1)}_{\alpha \alpha}$. In order to maximise these CP-asymmetries, one would expect $m_1$ to be found at its largest allowed value (determined by the constraint on the sum of the neutrino masses).

Let us now examine how these choices of parameters affect the expressions for the tree- and one-loop light neutrino masses. We may estimate the light masses using the largest value of the Yukawa matrix ($\sim 10^{-2}$ in the case of $S_1$, see \appref{sec:AbsYukawaMatrices}) and the smallest heavy mass $M_1=10^6 \text{ GeV}$:
\[
m^{\text{tree}} \sim v^2 \frac{Y^2}{M_1} \sim \mathcal{O}\left( 10^{-6} \text{ GeV} \right).
\]
This mass is too large from the point of view of the experimental bound and yet the numerical machinery is enforcing neutrino masses which sum to $< 1 \text{ eV}$. Let us investigate why this estimate fails.
This structure of the $R$-matrix leads to the following structure for the Dirac mass matrix:
\[
m_D \sqrt{f} =\left(
\begin{array}{ccc}
 \delta_1, & u, &  -i u +\delta_2 \\
\end{array}
\right),
\]
in which $|\delta_2| \ll |\delta_1| \ll |u|$ where each of $\delta_1$, $\delta_2$ and $u$ are $3$-component complex vectors.
We may rewrite the tree and one-loop masses in terms of this relatively simple matrix $m_D \sqrt{f}$
\[
m^{\text{tree}} = \left(m_D \sqrt{f} \right) M^{-1} f^{-1} \left(m_D \sqrt{f} \right)^T,
\]
where  the commutativity of the diagonal matrices $M$ and $f$ has been exploited.
For the one-loop contribution we find
\[
m^{\text{1-loop}} = \left(m_D \sqrt{f} \right) \left(f-M^{-1}\right) f^{-1} \left(m_D \sqrt{f} \right)^T.
\]
This ensures that their sum is
\[
\begin{aligned}
m_{\nu} & = m_D \sqrt{f} \left(m_D \sqrt{f}\right)^T \\
& = \delta_1 \delta_1^T + u \delta_2^T +\delta_2 u^T + \delta_2 \delta_2^T.
\end{aligned}
\]
Due to the relative smallness of the elements of  $\delta_i$, the light neutrino mass matrix may be considerably smaller than would be expected from a naive estimate based on the size of $u$. Neglecting terms containing a $\delta_i$, we find that
\[
m^{\text{tree}} = - m^{\text{1-loop}}.
\]
This is the mechanism by which the fine-tuned mass matrices are arrived at.

Although in this analysis, the results of $S_1$ were used, the other solutions differ essentially only in the sign used for $y_i$. This introduces a different pattern of minus signs in the matrix of Eq. \ref{RStructure} (and hence also in the expression for $m_D \sqrt{f}$) which does not affect the overall argument. Note that this argument is true even for the solutions of the two-decaying heavy Majorana neutrinos equations.\\

\section{Summary and Conclusions}\label{sec:discus}
In this work we have explored the viable model parameter space of
thermal leptogenesis associated to a type-I seesaw mechanism.
To do so, we numerically solved the three-flavoured density matrix equations \cite{Blanchet:2011xq} for
 one and two-decaying heavy Majorana neutrinos. 
Of the eighteen dimensional model parameter space, seven parameters were fixed from
neutrino oscillation data, cosmological constraints and consideration of a mildly hierarchical heavy Majorana neutrino mass
spectrum.

To find the regions of parameter space consistent with the measured baryon-to-photon ratio we 
used {\sc pyMultiNest} which implements a nested sampling algorithm to calculate Bayesian posterior 
distributions which are utilised to find regions of confidence. In addition, we ensured the Yukawa matrix entries respected perturbativity and 
we  protected against resonance effects 
by assuming a mildly hierarchical heavy Majorana neutrino mass spectrum. In the case of one decaying heavy Majorana neutrino, we found the lightest heavy Majorana neutrino mass that could successfully generate the baryon asymmetry, with our choice of upper bound on $R$-matrix components, to be
$M_{1}\simeq 10^{6}$ GeV.
This is 
possible as regions of the parameter space which have levels of fine-tuning in the light neutrino mass matrix  $>\mathcal{O}(10)$ were explored.
In conjunction,  eleven parameters  were allowed to vary thus compensating for the smaller heavy Majorana neutrino masses.
%
Moreover, with normal ordering, maximally CP-violating values of $\delta$ and 
$\theta_{23}$ close to $45^{\circ}$ (in most cases slightly larger than $45^\circ$, see \tabref{tab:bestfitall}) is preferred. In addition, there was strong dependence on the mass of the lightest
neutrino. On the other hand, we found in the case of  inverted ordering there were no strong constraints on low
energy neutrino parameters.  For this scenario, the level of fine-tuning was $\sim \mathcal{O}\left(100\right)$. 
In the case of one decaying heavy Majorana neutrino, we found the scenario with the smallest fine-tuning, at intermediate scales,
was $\overline{S_{2}}$, (F.T $\sim40$) with a heavy Majorana neutrino spectrum $M_{1}=10^{6.5}$ GeV, $M_{2}\approx3.15M_1$
and  $M_{3}\approx3.15M_2$. We showed also that fine tuning would not be necessary at all if $M_2 
= M_3$, when the one loop contribution to the light Majorana neutrino mass matrix is strongly suppressed. We also explored the possibility that either the tree or one-loop radiative corrections dominate the neutrino mass matrix. 
We  found the lowest scale
possible for this scenario, assuming a mildly hierarchical spectrum, was $M_{1}=10^9$ GeV.
As discussed, a motivation for exploring
leptogenesis at intermediate scales is to avoid large corrections to the Higgs mass.
Although, we found regions of the parameter space of three-flavoured thermal leptogenesis
consistent with the observed baryon asymmetry, we did not seek to minimise $\delta\mu^2$
and relegate this to a future study.

Finally, we investigated the case of two decaying heavy Majorana neutrinos. We found the lowest scale for both normal and
inverted ordering to be $M_{1}=10^{6.7}$ GeV. This scale is higher than in the one decaying heavy Majorana neutrino case because the scale of the washout is larger for $N_2$ and its CP-asymmetry is small in comparison with $N_1$. Although the washout for $N_2$ decays much more quickly than for $N_1$, it still has an appreciable effect on the final lepton asymmetry and so one must raise the scale of the heavy Majorana neutrino masses to achieve successful leptogenesis. In this paper, we did not include spectator effects which could potentially further lower the scale of thermal leptogenesis
and may be investigated in future work. 

\begin{acknowledgements}
We would like to thank Pasquale di Bari, Bjoern Garbrecht and Andrea de Simone for useful discussions and advice regarding this work. We are grateful to Alexis Plascencia, Carlos Tamarit and Ye-Ling Zhou for lively discussions on leptogenesis. J.T. would like to thank Bogdan Dobrescu and Pedro Machado for helpful advice regarding fine-tuning. K.M. and S.P. acknowledge the (partial) support from the European Research Council under the European Union Seventh Framework Programme (FP/2007-2013) / ERC Grant NuMass agreement n. [617143].  S.P. would like to acknowledge partial support from the Wolfson Foundation and the Royal Society, and also thanks SISSA for support and hospitality during part of this work. S.P. and S.T.P. acknowledge partial support from the European Unions Horizon 2020 research and innovation programme under the Marie Sklodowska Curie grant agreements No 690575 (RISE InvisiblesPlus) and No 674896 (ITN ELUSIVE). The work of S.T.P. was supported in part by the INFN program on Theoretical Astroparticle Physics (TASP) and by the World Premier International Research Center Initiative (WPI Initiative), MEXT, Japan. J.T. would like to express a special thanks to the Mainz Institute for Theoretical Physics (MITP) and International School for Advanced Studies (SISSA) for their hospitality and support where part of this work was completed. This manuscript has been authored by Fermi Research Alliance, LLC under Contract No. DE-AC02-07CH11359 with the U.S. Department of Energy, Office of Science, Office of High Energy Physics. This material is based upon work supported by the U.S. Department of Energy, Office of Science, Office of Advanced Scientific Computing Research, Scientific Discovery through Advanced Computing (SciDAC) program.
\end{acknowledgements}
\appendix 
\section{Thermal Width}\label{sec:thermalwidth}
The terms proportional to the thermal width in \equaref{eq:full3} are explicitly given below
\begin{equation}\label{LT1}
\begin{aligned}
\quad H(z)\sim 1.66 \sqrt{g_{*}}\frac{M^2_{1}}{M_{P}}\frac{1}{z^2}\implies zH(z)=1.66 g_{*}\frac{M^2_{1}}{M_{P}}\frac{1}{z},
\end{aligned}
\end{equation}
where  $z=\frac{M_{1}}{T}$ and $M_{P}$ is the Planck mass. From \cite{Blanchet:2011xq}
\begin{equation}\label{LT2}
\Im(\Lambda_{\alpha})=8\times 10^{-3}f^2_{\alpha}T.
\end{equation}
Using \equaref{LT1} and \equaref{LT2} we find
\begin{equation}\label{eq:Lamb}
\frac{\Im(\Lambda_{\alpha})}{Hz}=\frac{8\times 10^{-3} f^2_{\alpha}M_{P}}{1.66 \sqrt{g_{*}}M_{1}},
\end{equation}
in which $M_{P}=1.22\times 10^{19}$ GeV and $g_{*}=106.75$. Note that $f_{\tau}$ is the $\tau$ charged lepton Yukawa coupling
\begin{equation}
\begin{aligned}
m_{\tau}&=f_{\tau} v\implies f_{\tau}=\frac{m_{\tau}}{v}=\frac{1.776}{174}\sim 1.02\times 10^{-2},\\
m_{\mu}&=f_{\mu} v\implies f_{\mu}=\frac{m_{\mu}}{v}=\frac{0.105}{174}\sim 6.03\times 10^{-4},
\end{aligned}
\end{equation}
where all the units above are in GeV. We can rewrite \equaref{eq:Lamb}
\begin{equation}
\begin{aligned}
\frac{\Im(\Lambda_{\tau})}{Hz} & =4.66\times 10^{-8}\frac{M_{P}}{M_{1}},\\
\frac{\Im(\Lambda_{\mu})}{Hz} & = 1.69 \times 10^{-10} \frac{M_{P}}{M_{1}}.
\end{aligned}
\end{equation}
%
\section{Strong Washout}\label{sec:decayparam}
In this paper we have assumed that the density matrix approximation is appropriate (as opposed to the more accurate NE-QFT approaches). We also assume that the baryon asymmetry is insensitive to the initial values of the particle abundances. These assumptions are justified if we are working in the strong washout regime defined by $K_{1} \gg 1$ where
\begin{equation}
K_{1}\equiv\frac{\tilde{\Gamma}_{1}}{Hz},
\end{equation}
with 
\begin{equation}
\tilde{\Gamma}_1 = \frac{M_1 \left(m_D^{\dagger} m_D\right)_{11}}{8 \pi v^2},
\end{equation}
the decay rate of $N_1$ into leptons and anti-leptons at zero temperature.
In this appendix we provide justification for assuming $K_{1} \gg 1$ is generally satisfied.

Employing the tree-level appropriate CI parametrisation, we find
\begin{equation}
\begin{aligned}
K_{1}  & =\frac{(m^{\dagger}_{D}m_{D})_{11}}{M_{1} }\frac{1}{10^{-3} \text{ eV}},\\
&=\frac{(Y^{\dagger}Y)_{11}v^2}{M_{1}}\frac{1}{10^{-12}\text{ GeV}},\\
& = \frac{m_{1}\left| R_{11} \right|^2 + m_{2}\left| R_{12} \right|^2 + m_{3}\left| R_{13} \right|^2}{10^{-12}\text{ GeV}}.
\end{aligned}
\end{equation}

For the normally ordered mass  spectrum, following experimental constraints on the masses, $m_2$ and $m_3$ are increasing functions of $m_1$. Thus, if the elements of $R$ are fixed, $K_1$ is smallest when $m_1=0$. A random scan over the angles of $R$ (allowing $x_i$ in $[0, 360] ^\circ$ and $y_i$ in $[0, 180]^\circ$) for $10^6$ points leads to the conclusion that $>99.9\%$ of points in the parameter space lead to $K_1>1$ and  $\sim 99.7\%$ of points lead to $K_1 > 10$.

In IO, a random scan of $10^6$ points found none for which $K_1 < 1$ only $9$ points for which $\mathcal{K}_1 < 10$. Thus, the experimental constraints in both the IO and NO case greatly favour strong washout.

In conclusion, very few points in the parameter space satisfy  the experimental constraints on the neutrino mass-squared differences and mass-sum whilst simultaneously achieving weak washout. Thus it is a safe assumption that the washout is strong and our numerical methods are accurate.
\section{The resonance region}\label{sec:res}

The analytical expressions used CP asymmetry parameters have been calculated under the assumption that the the heavy Majorana neutrinos have well-separated masses such that the usual Feynman rules may be used in perturbation theory. The meaning of well-separated here is such that the mass differences are significantly larger than their decay rates. In this appendix we investigate this assumption.

\begin{figure*}[h!]\label{fig:NO2}
\includegraphics[width=0.9\textwidth]{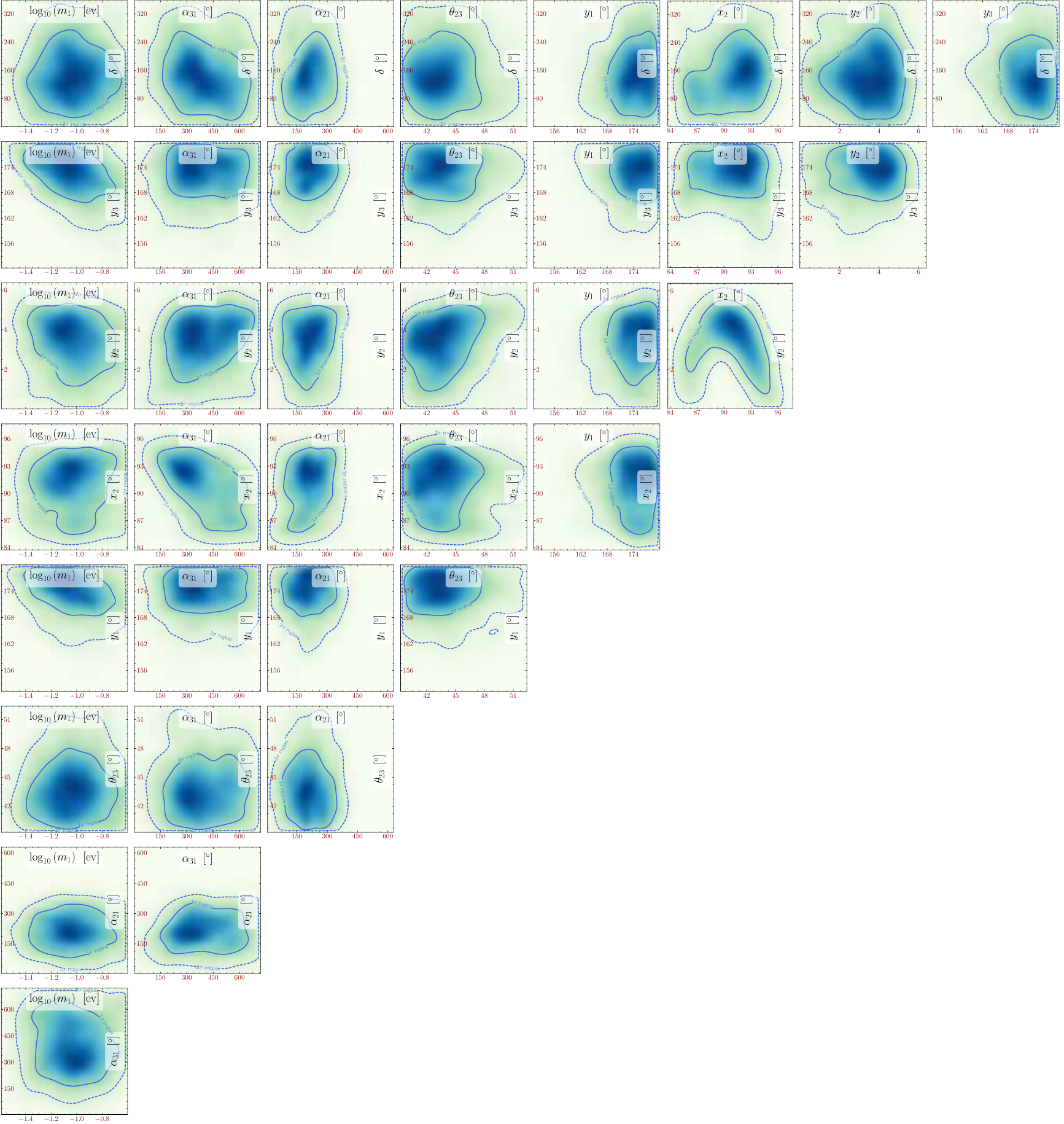}
\caption{ $S_{2}$: Triangle plot showing the two-dimensional projection   of the 11-dimensional model parameter space for posterior distributions using  normal ordering with one-decaying heavy Majorana neutrino and 
heavy Majorana neutrino mass spectrum: $M_{1}=10^{6.5}$ GeV, $M_{2}=3.15\, M_{1}$, $M_{3}=3.15\,M_{2}$. The contours correspond to 68$\%$ and 95$\%$ confidence levels respectively.}
\end{figure*}

\begin{figure*}[h!]\label{fig:NO3}
\includegraphics[width=0.9\textwidth]{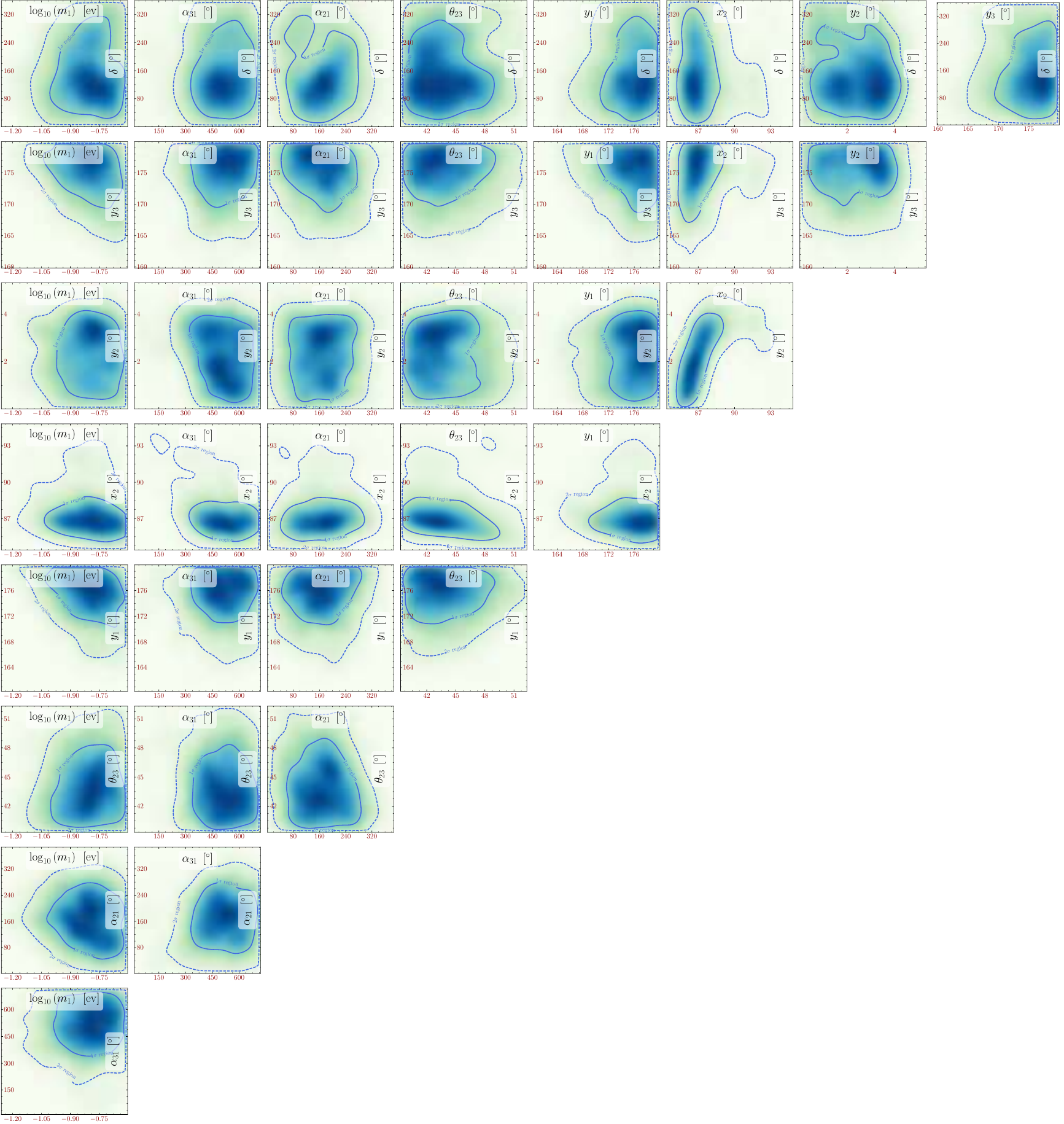}
\caption{ $S_{3}$:  Triangle plot showing the two-dimensional projection   of the 11-dimensional model parameter space for posterior distributions using normal ordering with one-decaying heavy Majorana neutrino and 
heavy Majorana neutrino mass spectrum: $M_{1}=10^{6.5}$ GeV, $M_{2}=5\, M_{1}$, $M_{3}=5\,M_{2}$. The contours correspond to 68$\%$ and 95$\%$ confidence levels respectively.}
\end{figure*}

\begin{figure*}[h!]\label{fig:IO2}
\includegraphics[width=0.9\textwidth]{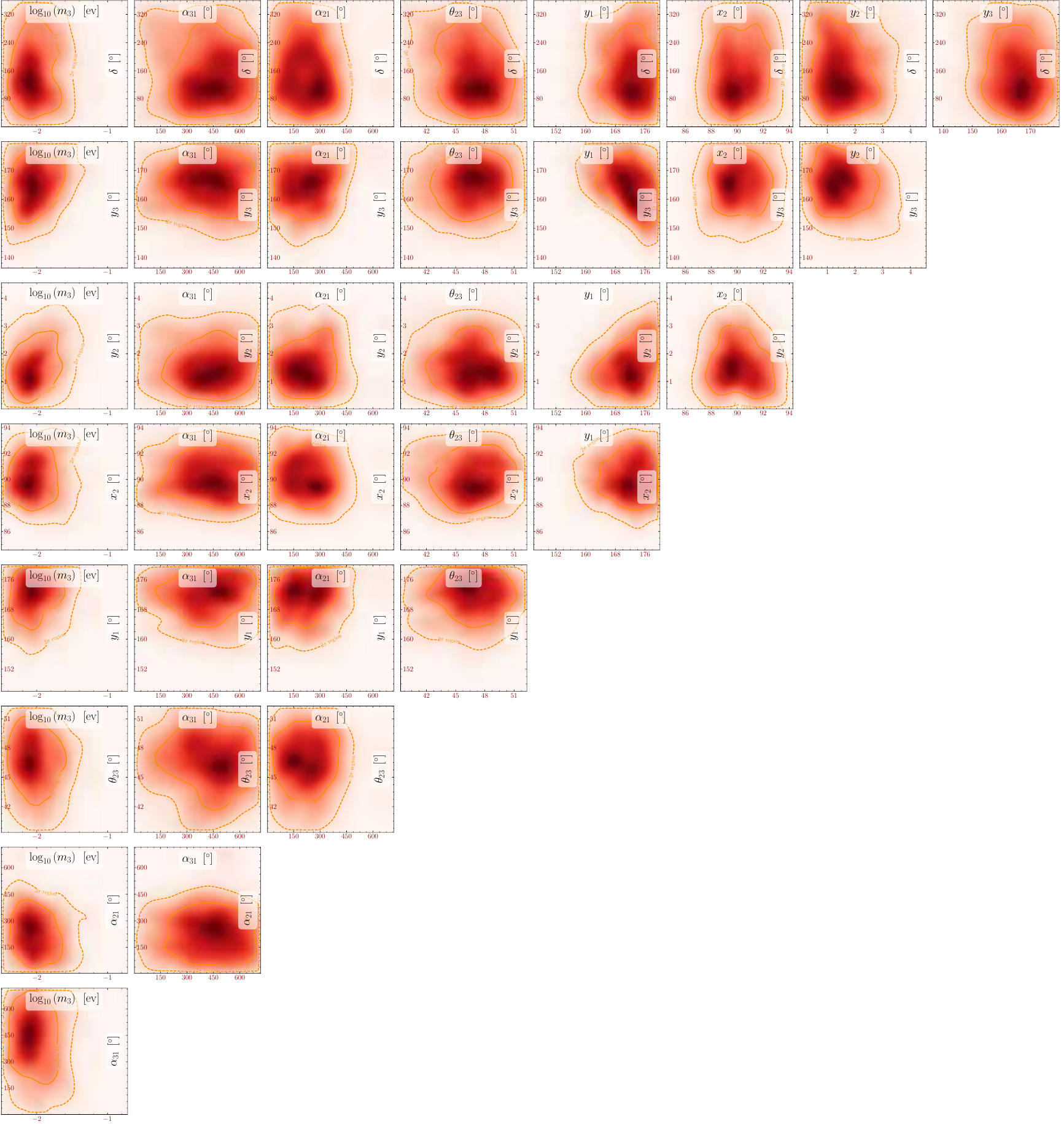}
\caption{ $\overline{S_{2}}$:  Triangle plot showing the two-dimensional projection   of the 11-dimensional model parameter space for posterior distributions using inverted ordering with one-decaying heavy Majorana neutrino and 
heavy Majorana neutrino mass spectrum: $M_{1}=10^{6.5}$ GeV, $M_{2}=3.15\, M_{1}$, $M_{3}=3.15\,M_{2}$. The contours correspond to 68$\%$ and 95$\%$ confidence levels respectively.}
\end{figure*}

\begin{figure*}[h!]\label{fig:IO3}
\includegraphics[width=0.9\textwidth]{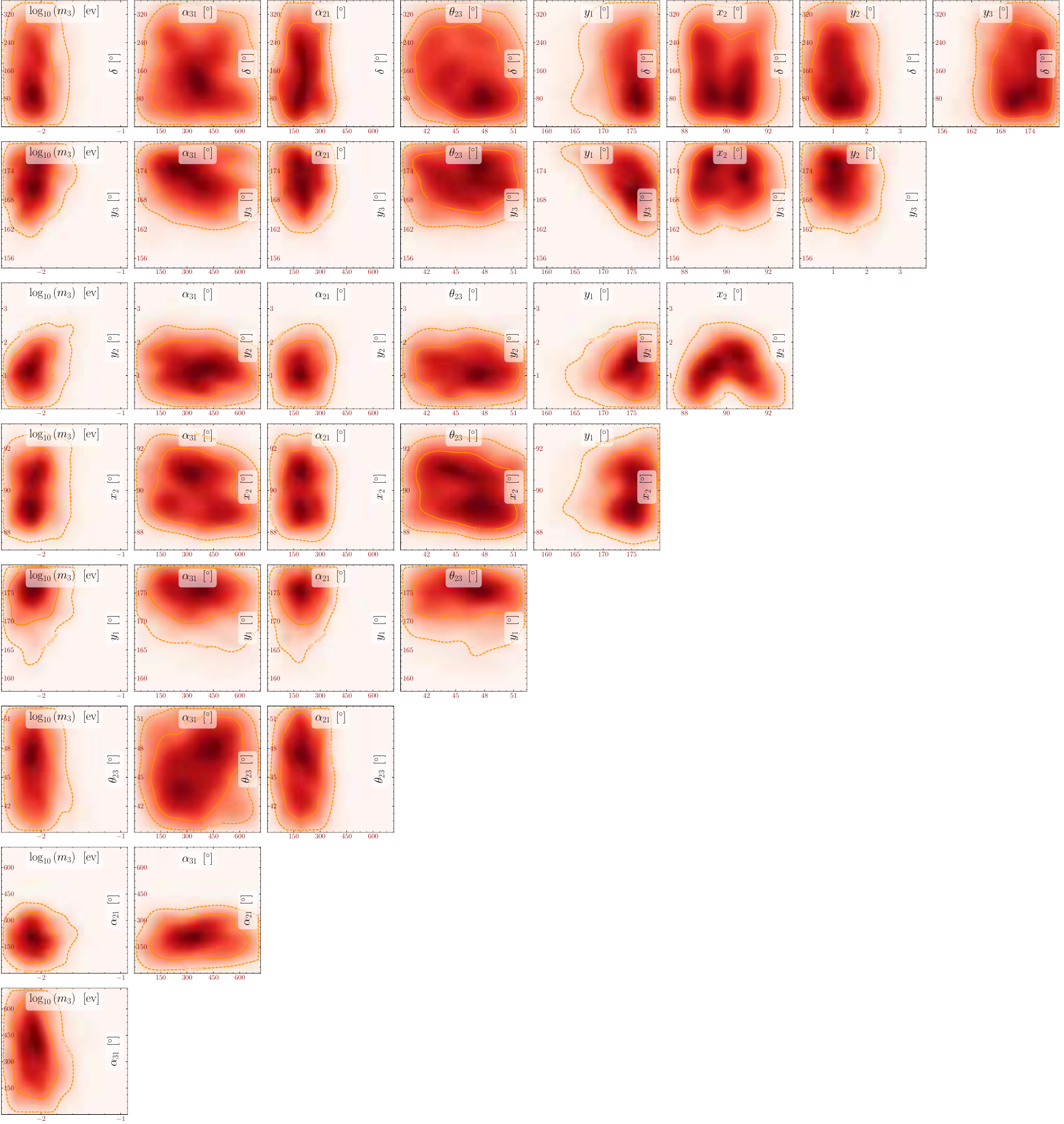}
\caption{ $\overline{S_{3}}$: Triangle plot showing the two-dimensional projection   of the 11-dimensional model parameter space for posterior distributions using inverted ordering with one-decaying heavy Majorana neutrino and 
heavy Majorana neutrino mass spectrum: $M_{1}=10^{6.5}$ GeV, $M_{2}=5\, M_{1}$, $M_{3}=5\,M_{2}$. The contours correspond to 68$\%$ and 95$\%$ confidence levels respectively.}
\end{figure*}

The total CP asymmetry parameter is defined in terms of $\Gamma_1$, the decay rate for $N_1 \rightarrow \phi^{\dagger} l$ and $\bar{\Gamma}_{1}$, the rate for CP conjugate process $N_1 \rightarrow \phi l^{\dagger}$, as
\begin{equation}
\epsilon^{(1)} = \sum_{\alpha} \epsilon^{(1)}_{\alpha \alpha}\equiv \frac{\Gamma_1-\bar{\Gamma}_1}{\Gamma_1+\bar{\Gamma}_1},
\end{equation}
and the decay terms are
\begin{equation}
D_1\left(z\right) \equiv \frac{\Gamma_1+\bar{\Gamma}_1}{Hz}.
\end{equation}
As analytical expressions for these are well-known we may put them to use in finding the decay rate. We have
\begin{equation}
\Gamma_1 = \frac{Hz}{2} \left( \epsilon^{(1)} +1 \right) D_1 \left(z\right).
\end{equation}
The Hubble parameter $H$ in a radiation-dominated Universe is, from the Friedmann equation,
\begin{equation}
H = -\frac{\dot{T}}{T} = \sqrt{\frac{8 \pi G}{3}} \sqrt{\frac{\pi^2 g_{*}}{30}} T^2,
\end{equation}
which may be expressed in terms of $M_1$, $z$ and the Planck mass $M_P$ with
\begin{equation}
H = \frac{M_1^2}{M_P} \sqrt{\frac{2 \pi}{3}} \sqrt{\frac{\pi^2 g_{*}}{30}} \frac{1}{z^2}.
\end{equation}
Thus, the decay rate may be written in terms of the functions that are typically written in the Boltzman equations by
\begin{align}
\Gamma_1 & = \frac{M_1^2}{M_P}  \sqrt{\frac{2 \pi}{3}} \sqrt{\frac{\pi^2 g_{*}}{30}} \left(1+\epsilon^{(1)}\right) \frac{D_1\left(z\right)}{z}
\\ & \approx \left(7.03 \times 10^{-19} \text{ GeV}^{-1}\right) \times M_1^2  \left(1+\epsilon^{(1)}\right) \frac{D_1\left(z\right)}{z}.
\end{align}
In order to avoid the resonance region we require that
\begin{equation}
\Gamma_1 \ll M_2-M_1.
\end{equation}
To test this, the PMNS angles; $x_1, x_2, x_3$; and $M_1,M_2,M_3$ were fixed according to the best-fit points for NO (\figref{fig:NO1},  \figref{fig:NO2} and \figref{fig:NO3}) and also for IO (\figref{fig:IO1},  \figref{fig:IO2} and \figref{fig:IO3}) and a random scan over the remaining parameters for $10^5$ points was performed with the criterion
\begin{equation}
\frac{\Gamma_1}{M_2-M_1} > 0.01.
\end{equation}
We  found there were no points which verified this condition and thus the assumption of non-resonance is justified.

\section{Higher-order radiative corrections}\label{sec:2loops}
We have been careful to include the one-loop radiative corrections to the light neutrino masses. In doing so we have expanded the region of the parameter space in which we may accurately explore leptogenesis. Of course, there may also be regions in which the higher-order corrections are important. We may ask the question \textit{how can we be sure that the neglect of two-loops, three-loops etc. was legitimate?}

A pragmatic approach is to perform an order-of-magnitude estimate of the effects of the higher-order corrections for those points in the parameter space of most significance to our result: the best-fit points for the scenarios $S_1$ to $\overline{S_4}$ and $F.T^{\text{loop}}$, $F.T^{\text{tree}}$. If, in these scenarios, the higher-order corrections appear small then our main conclusions are left untouched.

Our estimate of the two-loop effect (which we shall assume generically dominates three or more loops) will be given by two extra factors of the Yukawa couplings and the conventional loop factor $\left(4 \pi \right)^{-2}$ to the one-loop effect. Let us use
\begin{equation}
m^{\text{2-loop}} = \frac{1}{\left(4 \pi \right)^2} |Y_{\text{max.}}|^2 m^{\text{1-loop}},
\end{equation}
with $|Y_{\text{max.}}|$ the largest element of the matrix of absolute values of the Yukawas, as a conservative estimate (over-estimate) of the second-order radiative correction to neutrino masses. (This is similar to the estimate used in~\cite{Lopez-Pavon:2015cga}.)

\begin{table}[t]\label{tab:mtm1lm2l}
\centering
\begin{tabular}{ c | c | c | c  }
   &  $\sum_i \left(m^{\text{tree}}+m^{\text{1-loop}}\right)_i \text{ (eV)}$ & $\sum_i m^{\text{2-loop}}_i \text{ (eV)}$ \\
 \hline \hline
$S_1$ &  $3.70\times 10^{-1}$ & $1.69\times 10^{-3}$\\
$S_2$ &  $2.52\times 10^{-1}$ & $1.12\times 10^{-3}$\\
$S_3$ &  $3.53\times 10^{-1}$ & $4.25\times 10^{-3}$\\
$S_4$ &  $6.30\times 10^{-1}$ & $5.81\times 10^{-2}$\\
$\overline{S_1}$ & $1.13\times 10^{-1}$ & $1.98\times 10^{-4}$ \\
$\overline{S_2}$ & $1.16\times 10^{-1}$ & $2.22\times 10^{-4}$ \\
$\overline{S_3}$ & $1.14\times 10^{-1}$ & $1.95\times 10^{-3}$\\
$\overline{S_4}$ & $1.09\times 10^{-1}$ & $1.91\times 10^{-3}$\\
$F.T^{\text{loop}}$ & $8.65\times 10^{-2}$ & $1.07\times 10^{-6}$\\
$F.T^{\text{tree}}$ & $6.39\times 10^{-2}$ & $7.58\times 10^{-8}$\\
\end{tabular}\caption{Comparisons of the (sum of singular values of the) tree plus one-loop correct light mass matrix to the two-loop estimate.}
\end{table}

From table \ref{tab:mtm1lm2l}, we see that the two-loop contributions generally provide small corrections and therefore that corrections beyond one-loop order are safely neglected at these points.

\clearpage
\section{Yukawa matrices}\label{sec:AbsYukawaMatrices}

Here we provide a table of the absolute values of the Yukawa matrices ($\lvert Y \rvert$) for the best-fit points of each scenario considered in \tabref{tab:bestfitall} and \tabref{tab:bestfit2DS}.

\begin{table}[h]\label{tab:mtm1lm2l}
\centering
\begin{tabular}{ c | c | c}
   &  $\lvert Y \rvert$\\
 \hline \hline
$S_1$ &  $\left(
\begin{array}{ccc}
 1.20501\times 10^{-5} & 5.84226\times 10^{-3} & 1.04449\times 10^{-2} \\
 6.50743\times 10^{-5} & 2.0441\times 10^{-2} & 3.65463\times 10^{-2} \\
 7.26332\times 10^{-6} & 2.11503\times 10^{-2} & 3.78139\times 10^{-2} \\
\end{array}
\right)$\\
\hline\\
$S_2$ &  $\left(
\begin{array}{ccc}
 1.78047\times 10^{-5} & 1.16361\times 10^{-2} & 2.08046\times 10^{-2} \\
 1.00881\times 10^{-5} & 2.21656\times 10^{-2} & 3.96322\times 10^{-2} \\
 1.02069\times 10^{-4} & 2.55\times 10^{-2} & 4.55925\times 10^{-2} \\
\end{array}
\right)$\\
\hline \\
$S_3$ &  $\left(
\begin{array}{ccc}
 3.07775\times 10^{-5} & 1.59166\times 10^{-2} & 2.84583\times 10^{-2} \\
 1.23975\times 10^{-5} & 3.77326\times 10^{-2} & 6.74663\times 10^{-2} \\
 1.14533\times 10^{-4} & 3.93327\times 10^{-2} & 7.03289\times 10^{-2} \\
\end{array}
\right)$\\
\hline \\
$S_4$ &  $\left(
\begin{array}{ccc}
 2.54075\times 10^{-5} & 3.09962\times 10^{-2} & 6.2255\times 10^{-2} \\
 1.52369\times 10^{-5} & 7.01974\times 10^{-2} & 1.40989\times 10^{-1} \\
 1.99141\times 10^{-4} & 8.33171\times 10^{-2} & 1.67344\times 10^{-1} \\
\end{array}
\right)$\\
\hline \\
$\overline{S_1}$ & $\left(
\begin{array}{ccc}
 5.37412\times 10^{-6} & 3.81344\times 10^{-3} & 6.81765\times 10^{-3} \\
 4.68081\times 10^{-6} & 1.03898\times 10^{-2} & 1.85756\times 10^{-2} \\
 2.33498\times 10^{-5} & 1.28236\times 10^{-2} & 2.29271\times 10^{-2} \\
\end{array}
\right)$\\
\hline \\
$\overline{S_2}$ & $\left(
\begin{array}{ccc}
 5.37412\times 10^{-6} & 3.81344\times 10^{-3} & 6.81765\times 10^{-3} \\
 4.68081\times 10^{-6} & 1.03898\times 10^{-2} & 1.85756\times 10^{-2} \\
 2.33498\times 10^{-5} & 1.28236\times 10^{-2} & 2.29271\times 10^{-2} \\
\end{array}
\right)$\\
\hline \\
$\overline{S_3}$ & $\left(
\begin{array}{ccc}
 5.37412\times 10^{-6} & 3.81344\times 10^{-3} & 6.81765\times 10^{-3} \\
 4.68081\times 10^{-6} & 1.03898\times 10^{-2} & 1.85756\times 10^{-2} \\
 2.33498\times 10^{-5} & 1.28236\times 10^{-2} & 2.29271\times 10^{-2} \\
\end{array}
\right)$\\
\hline \\
$\overline{S_4}$ & $\left(
\begin{array}{ccc}
 5.37412\times 10^{-6} & 3.81344\times 10^{-3} & 6.81765\times 10^{-3} \\
 4.68081\times 10^{-6} & 1.03898\times 10^{-2} & 1.85756\times 10^{-2} \\
 2.33498\times 10^{-5} & 1.28236\times 10^{-2} & 2.29271\times 10^{-2} \\
\end{array}
\right)$\\
\hline \\
$F.T^{\text{loop}}$ & $\left(
\begin{array}{ccc}
 6.27292\times 10^{-4} & 1.68158\times 10^{-2} & 2.98125\times 10^{-2} \\
 2.86893\times 10^{-3} & 3.06908\times 10^{-2} & 5.56779\times 10^{-2} \\
 9.98924\times 10^{-4} & 2.62581\times 10^{-2} & 4.69331\times 10^{-2} \\
\end{array}
\right)$
\\
\hline \\
$F.T^{\text{tree}}$ & $\left(
\begin{array}{ccc}
 2.08179\times 10^{-4} & 3.44059\times 10^{-3} & 6.19056\times 10^{-3} \\
 3.20671\times 10^{-4} & 5.48821\times 10^{-3} & 9.63727\times 10^{-3} \\
 2.05748\times 10^{-4} & 5.38578\times 10^{-3} & 9.37847\times 10^{-3} \\
\end{array}
\right)$\\
\end{tabular}\caption{Absolute values of the Yukawas for each scenario listed in \tabref{tab:bestfitall} and \tabref{tab:bestfit2DS}.}
\end{table}

\clearpage

\bibliographystyle{apsrev4-1}
\bibliography{lepbib}{}

\begin{thebibliography}{113}%
\makeatletter
\providecommand \@ifxundefined [1]{%
 \@ifx{#1\undefined}
}%
\providecommand \@ifnum [1]{%
 \ifnum #1\expandafter \@firstoftwo
 \else \expandafter \@secondoftwo
 \fi
}%
\providecommand \@ifx [1]{%
 \ifx #1\expandafter \@firstoftwo
 \else \expandafter \@secondoftwo
 \fi
}%
\providecommand \natexlab [1]{#1}%
\providecommand \enquote  [1]{``#1''}%
\providecommand \bibnamefont  [1]{#1}%
\providecommand \bibfnamefont [1]{#1}%
\providecommand \citenamefont [1]{#1}%
\providecommand \href@noop [0]{\@secondoftwo}%
\providecommand \href [0]{\begingroup \@sanitize@url \@href}%
\providecommand \@href[1]{\@@startlink{#1}\@@href}%
\providecommand \@@href[1]{\endgroup#1\@@endlink}%
\providecommand \@sanitize@url [0]{\catcode `\\12\catcode `\$12\catcode
  `\&12\catcode `\#12\catcode `\^12\catcode `\_12\catcode `\%12\relax}%
\providecommand \@@startlink[1]{}%
\providecommand \@@endlink[0]{}%
\providecommand \url  [0]{\begingroup\@sanitize@url \@url }%
\providecommand \@url [1]{\endgroup\@href {#1}{\urlprefix }}%
\providecommand \urlprefix  [0]{URL }%
\providecommand \Eprint [0]{\href }%
\providecommand \doibase [0]{http://dx.doi.org/}%
\providecommand \selectlanguage [0]{\@gobble}%
\providecommand \bibinfo  [0]{\@secondoftwo}%
\providecommand \bibfield  [0]{\@secondoftwo}%
\providecommand \translation [1]{[#1]}%
\providecommand \BibitemOpen [0]{}%
\providecommand \bibitemStop [0]{}%
\providecommand \bibitemNoStop [0]{.\EOS\space}%
\providecommand \EOS [0]{\spacefactor3000\relax}%
\providecommand \BibitemShut  [1]{\csname bibitem#1\endcsname}%
\let\auto@bib@innerbib\@empty
\bibitem [{\citenamefont {Patrignani}\ \emph {et~al.}(2016)\citenamefont
  {Patrignani} \emph {et~al.}}]{Patrignani:2016xqp}%
  \BibitemOpen
  \bibfield  {author} {\bibinfo {author} {\bibfnamefont {C.}~\bibnamefont
  {Patrignani}} \emph {et~al.} (\bibinfo {collaboration} {Particle Data
  Group}),\ }\href {\doibase 10.1088/1674-1137/40/10/100001} {\bibfield
  {journal} {\bibinfo  {journal} {Chin. Phys.}\ }\textbf {\bibinfo {volume}
  {C40}},\ \bibinfo {pages} {100001} (\bibinfo {year} {2016})}\BibitemShut
  {NoStop}%
\bibitem [{\citenamefont {Ade}\ \emph {et~al.}(2016)\citenamefont {Ade} \emph
  {et~al.}}]{Ade:2015xua}%
  \BibitemOpen
  \bibfield  {author} {\bibinfo {author} {\bibfnamefont {P.~A.~R.}\
  \bibnamefont {Ade}} \emph {et~al.} (\bibinfo {collaboration} {Planck}),\
  }\href {\doibase 10.1051/0004-6361/201525830} {\bibfield  {journal} {\bibinfo
   {journal} {Astron. Astrophys.}\ }\textbf {\bibinfo {volume} {594}},\
  \bibinfo {pages} {A13} (\bibinfo {year} {2016})},\ \Eprint
  {http://arxiv.org/abs/1502.01589} {arXiv:1502.01589 [astro-ph.CO]}
  \BibitemShut {NoStop}%
\bibitem [{\citenamefont {Sakharov}(1967)}]{Sakharov:1967dj}%
  \BibitemOpen
  \bibfield  {author} {\bibinfo {author} {\bibfnamefont {A.~D.}\ \bibnamefont
  {Sakharov}},\ }\href {\doibase 10.1070/PU1991v034n05ABEH002497} {\bibfield
  {journal} {\bibinfo  {journal} {Pisma Zh. Eksp. Teor. Fiz.}\ }\textbf
  {\bibinfo {volume} {5}},\ \bibinfo {pages} {32} (\bibinfo {year} {1967})},\
  \bibinfo {note} {[Usp. Fiz. Nauk161,61(1991)]}\BibitemShut {NoStop}%
\bibitem [{\citenamefont {Fukugita}\ and\ \citenamefont
  {Yanagida}(1986)}]{Fukugita:1986hr}%
  \BibitemOpen
  \bibfield  {author} {\bibinfo {author} {\bibfnamefont {M.}~\bibnamefont
  {Fukugita}}\ and\ \bibinfo {author} {\bibfnamefont {T.}~\bibnamefont
  {Yanagida}},\ }\href {\doibase 10.1016/0370-2693(86)91126-3} {\bibfield
  {journal} {\bibinfo  {journal} {Phys. Lett.}\ }\textbf {\bibinfo {volume}
  {B174}},\ \bibinfo {pages} {45} (\bibinfo {year} {1986})}\BibitemShut
  {NoStop}%
\bibitem [{\citenamefont {Khlebnikov}\ and\ \citenamefont
  {Shaposhnikov}(1988)}]{Khlebnikov:1988sr}%
  \BibitemOpen
  \bibfield  {author} {\bibinfo {author} {\bibfnamefont {S.~{\relax Yu}.}\
  \bibnamefont {Khlebnikov}}\ and\ \bibinfo {author} {\bibfnamefont {M.~E.}\
  \bibnamefont {Shaposhnikov}},\ }\href {\doibase 10.1016/0550-3213(88)90133-2}
  {\bibfield  {journal} {\bibinfo  {journal} {Nucl. Phys.}\ }\textbf {\bibinfo
  {volume} {B308}},\ \bibinfo {pages} {885} (\bibinfo {year}
  {1988})}\BibitemShut {NoStop}%
\bibitem [{\citenamefont {Covi}\ \emph {et~al.}(1996)\citenamefont {Covi},
  \citenamefont {Roulet},\ and\ \citenamefont {Vissani}}]{Covi:1996wh}%
  \BibitemOpen
  \bibfield  {author} {\bibinfo {author} {\bibfnamefont {L.}~\bibnamefont
  {Covi}}, \bibinfo {author} {\bibfnamefont {E.}~\bibnamefont {Roulet}}, \ and\
  \bibinfo {author} {\bibfnamefont {F.}~\bibnamefont {Vissani}},\ }\href
  {\doibase 10.1016/0370-2693(96)00817-9} {\bibfield  {journal} {\bibinfo
  {journal} {Phys. Lett.}\ }\textbf {\bibinfo {volume} {B384}},\ \bibinfo
  {pages} {169} (\bibinfo {year} {1996})},\ \Eprint
  {http://arxiv.org/abs/hep-ph/9605319} {arXiv:hep-ph/9605319 [hep-ph]}
  \BibitemShut {NoStop}%
\bibitem [{\citenamefont {Covi}\ and\ \citenamefont
  {Roulet}(1997)}]{Covi:1996fm}%
  \BibitemOpen
  \bibfield  {author} {\bibinfo {author} {\bibfnamefont {L.}~\bibnamefont
  {Covi}}\ and\ \bibinfo {author} {\bibfnamefont {E.}~\bibnamefont {Roulet}},\
  }\href {\doibase 10.1016/S0370-2693(97)00287-6} {\bibfield  {journal}
  {\bibinfo  {journal} {Phys. Lett.}\ }\textbf {\bibinfo {volume} {B399}},\
  \bibinfo {pages} {113} (\bibinfo {year} {1997})},\ \Eprint
  {http://arxiv.org/abs/hep-ph/9611425} {arXiv:hep-ph/9611425 [hep-ph]}
  \BibitemShut {NoStop}%
\bibitem [{\citenamefont {Pilaftsis}(1997)}]{Pilaftsis:1997jf}%
  \BibitemOpen
  \bibfield  {author} {\bibinfo {author} {\bibfnamefont {A.}~\bibnamefont
  {Pilaftsis}},\ }\href {\doibase 10.1103/PhysRevD.56.5431} {\bibfield
  {journal} {\bibinfo  {journal} {Phys. Rev.}\ }\textbf {\bibinfo {volume}
  {D56}},\ \bibinfo {pages} {5431} (\bibinfo {year} {1997})},\ \Eprint
  {http://arxiv.org/abs/hep-ph/9707235} {arXiv:hep-ph/9707235 [hep-ph]}
  \BibitemShut {NoStop}%
\bibitem [{\citenamefont {Buchmuller}\ and\ \citenamefont
  {Plumacher}(1998)}]{Buchmuller:1997yu}%
  \BibitemOpen
  \bibfield  {author} {\bibinfo {author} {\bibfnamefont {W.}~\bibnamefont
  {Buchmuller}}\ and\ \bibinfo {author} {\bibfnamefont {M.}~\bibnamefont
  {Plumacher}},\ }\href {\doibase 10.1016/S0370-2693(97)01548-7} {\bibfield
  {journal} {\bibinfo  {journal} {Phys. Lett.}\ }\textbf {\bibinfo {volume}
  {B431}},\ \bibinfo {pages} {354} (\bibinfo {year} {1998})},\ \Eprint
  {http://arxiv.org/abs/hep-ph/9710460} {arXiv:hep-ph/9710460 [hep-ph]}
  \BibitemShut {NoStop}%
\bibitem [{\citenamefont {Davidson}\ and\ \citenamefont
  {Ibarra}(2002)}]{Davidson:2002qv}%
  \BibitemOpen
  \bibfield  {author} {\bibinfo {author} {\bibfnamefont {S.}~\bibnamefont
  {Davidson}}\ and\ \bibinfo {author} {\bibfnamefont {A.}~\bibnamefont
  {Ibarra}},\ }\href {\doibase 10.1016/S0370-2693(02)01735-5} {\bibfield
  {journal} {\bibinfo  {journal} {Phys. Lett.}\ }\textbf {\bibinfo {volume}
  {B535}},\ \bibinfo {pages} {25} (\bibinfo {year} {2002})},\ \Eprint
  {http://arxiv.org/abs/hep-ph/0202239} {arXiv:hep-ph/0202239 [hep-ph]}
  \BibitemShut {NoStop}%
\bibitem [{\citenamefont {Buchmuller}\ \emph {et~al.}(2002)\citenamefont
  {Buchmuller}, \citenamefont {Di~Bari},\ and\ \citenamefont
  {Plumacher}}]{Buchmuller:2002rq}%
  \BibitemOpen
  \bibfield  {author} {\bibinfo {author} {\bibfnamefont {W.}~\bibnamefont
  {Buchmuller}}, \bibinfo {author} {\bibfnamefont {P.}~\bibnamefont {Di~Bari}},
  \ and\ \bibinfo {author} {\bibfnamefont {M.}~\bibnamefont {Plumacher}},\
  }\href {\doibase 10.1016/S0550-3213(02)00737-X,
  10.1016/j.nuclphysb.2007.11.030} {\bibfield  {journal} {\bibinfo  {journal}
  {Nucl. Phys.}\ }\textbf {\bibinfo {volume} {B643}},\ \bibinfo {pages} {367}
  (\bibinfo {year} {2002})},\ \bibinfo {note} {[Erratum: Nucl.
  Phys.B793,362(2008)]},\ \Eprint {http://arxiv.org/abs/hep-ph/0205349}
  {arXiv:hep-ph/0205349 [hep-ph]} \BibitemShut {NoStop}%
\bibitem [{\citenamefont {Ellis}\ and\ \citenamefont
  {Raidal}(2002)}]{Ellis:2002xg}%
  \BibitemOpen
  \bibfield  {author} {\bibinfo {author} {\bibfnamefont {J.~R.}\ \bibnamefont
  {Ellis}}\ and\ \bibinfo {author} {\bibfnamefont {M.}~\bibnamefont {Raidal}},\
  }\href {\doibase 10.1016/S0550-3213(02)00753-8} {\bibfield  {journal}
  {\bibinfo  {journal} {Nucl. Phys.}\ }\textbf {\bibinfo {volume} {B643}},\
  \bibinfo {pages} {229} (\bibinfo {year} {2002})},\ \Eprint
  {http://arxiv.org/abs/hep-ph/0206174} {arXiv:hep-ph/0206174 [hep-ph]}
  \BibitemShut {NoStop}%
\bibitem [{\citenamefont {Buchmuller}\ \emph {et~al.}(2003)\citenamefont
  {Buchmuller}, \citenamefont {Di~Bari},\ and\ \citenamefont
  {Plumacher}}]{Buchmuller:2003gz}%
  \BibitemOpen
  \bibfield  {author} {\bibinfo {author} {\bibfnamefont {W.}~\bibnamefont
  {Buchmuller}}, \bibinfo {author} {\bibfnamefont {P.}~\bibnamefont {Di~Bari}},
  \ and\ \bibinfo {author} {\bibfnamefont {M.}~\bibnamefont {Plumacher}},\
  }\href {\doibase 10.1016/S0550-3213(03)00449-8} {\bibfield  {journal}
  {\bibinfo  {journal} {Nucl. Phys.}\ }\textbf {\bibinfo {volume} {B665}},\
  \bibinfo {pages} {445} (\bibinfo {year} {2003})},\ \Eprint
  {http://arxiv.org/abs/hep-ph/0302092} {arXiv:hep-ph/0302092 [hep-ph]}
  \BibitemShut {NoStop}%
\bibitem [{\citenamefont {Buchmuller}\ \emph {et~al.}(2005)\citenamefont
  {Buchmuller}, \citenamefont {Di~Bari},\ and\ \citenamefont
  {Plumacher}}]{Buchmuller:2004nz}%
  \BibitemOpen
  \bibfield  {author} {\bibinfo {author} {\bibfnamefont {W.}~\bibnamefont
  {Buchmuller}}, \bibinfo {author} {\bibfnamefont {P.}~\bibnamefont {Di~Bari}},
  \ and\ \bibinfo {author} {\bibfnamefont {M.}~\bibnamefont {Plumacher}},\
  }\href {\doibase 10.1016/j.aop.2004.02.003} {\bibfield  {journal} {\bibinfo
  {journal} {Annals Phys.}\ }\textbf {\bibinfo {volume} {315}},\ \bibinfo
  {pages} {305} (\bibinfo {year} {2005})},\ \Eprint
  {http://arxiv.org/abs/hep-ph/0401240} {arXiv:hep-ph/0401240 [hep-ph]}
  \BibitemShut {NoStop}%
\bibitem [{\citenamefont {Vissani}(1998)}]{Vissani:1997ys}%
  \BibitemOpen
  \bibfield  {author} {\bibinfo {author} {\bibfnamefont {F.}~\bibnamefont
  {Vissani}},\ }\href {\doibase 10.1103/PhysRevD.57.7027} {\bibfield  {journal}
  {\bibinfo  {journal} {Phys. Rev.}\ }\textbf {\bibinfo {volume} {D57}},\
  \bibinfo {pages} {7027} (\bibinfo {year} {1998})},\ \Eprint
  {http://arxiv.org/abs/hep-ph/9709409} {arXiv:hep-ph/9709409 [hep-ph]}
  \BibitemShut {NoStop}%
\bibitem [{\citenamefont {Clarke}\ \emph {et~al.}(2015)\citenamefont {Clarke},
  \citenamefont {Foot},\ and\ \citenamefont {Volkas}}]{Clarke:2015gwa}%
  \BibitemOpen
  \bibfield  {author} {\bibinfo {author} {\bibfnamefont {J.~D.}\ \bibnamefont
  {Clarke}}, \bibinfo {author} {\bibfnamefont {R.}~\bibnamefont {Foot}}, \ and\
  \bibinfo {author} {\bibfnamefont {R.~R.}\ \bibnamefont {Volkas}},\ }\href
  {\doibase 10.1103/PhysRevD.91.073009} {\bibfield  {journal} {\bibinfo
  {journal} {Phys. Rev.}\ }\textbf {\bibinfo {volume} {D91}},\ \bibinfo {pages}
  {073009} (\bibinfo {year} {2015})},\ \Eprint
  {http://arxiv.org/abs/1502.01352} {arXiv:1502.01352 [hep-ph]} \BibitemShut
  {NoStop}%
\bibitem [{\citenamefont {Khlopov}\ and\ \citenamefont
  {Linde}(1984)}]{Khlopov:1984pf}%
  \BibitemOpen
  \bibfield  {author} {\bibinfo {author} {\bibfnamefont {M.~{\relax Yu}.}\
  \bibnamefont {Khlopov}}\ and\ \bibinfo {author} {\bibfnamefont {A.~D.}\
  \bibnamefont {Linde}},\ }\href {\doibase 10.1016/0370-2693(84)91656-3}
  {\bibfield  {journal} {\bibinfo  {journal} {Phys. Lett.}\ }\textbf {\bibinfo
  {volume} {138B}},\ \bibinfo {pages} {265} (\bibinfo {year}
  {1984})}\BibitemShut {NoStop}%
\bibitem [{\citenamefont {Ellis}\ \emph {et~al.}(1984)\citenamefont {Ellis},
  \citenamefont {Kim},\ and\ \citenamefont {Nanopoulos}}]{Ellis:1984eq}%
  \BibitemOpen
  \bibfield  {author} {\bibinfo {author} {\bibfnamefont {J.~R.}\ \bibnamefont
  {Ellis}}, \bibinfo {author} {\bibfnamefont {J.~E.}\ \bibnamefont {Kim}}, \
  and\ \bibinfo {author} {\bibfnamefont {D.~V.}\ \bibnamefont {Nanopoulos}},\
  }\href {\doibase 10.1016/0370-2693(84)90334-4} {\bibfield  {journal}
  {\bibinfo  {journal} {Phys. Lett.}\ }\textbf {\bibinfo {volume} {145B}},\
  \bibinfo {pages} {181} (\bibinfo {year} {1984})}\BibitemShut {NoStop}%
\bibitem [{\citenamefont {Kawasaki}\ \emph {et~al.}(2008)\citenamefont
  {Kawasaki}, \citenamefont {Kohri}, \citenamefont {Moroi},\ and\ \citenamefont
  {Yotsuyanagi}}]{Kawasaki:2008qe}%
  \BibitemOpen
  \bibfield  {author} {\bibinfo {author} {\bibfnamefont {M.}~\bibnamefont
  {Kawasaki}}, \bibinfo {author} {\bibfnamefont {K.}~\bibnamefont {Kohri}},
  \bibinfo {author} {\bibfnamefont {T.}~\bibnamefont {Moroi}}, \ and\ \bibinfo
  {author} {\bibfnamefont {A.}~\bibnamefont {Yotsuyanagi}},\ }\href {\doibase
  10.1103/PhysRevD.78.065011} {\bibfield  {journal} {\bibinfo  {journal} {Phys.
  Rev.}\ }\textbf {\bibinfo {volume} {D78}},\ \bibinfo {pages} {065011}
  (\bibinfo {year} {2008})},\ \Eprint {http://arxiv.org/abs/0804.3745}
  {arXiv:0804.3745 [hep-ph]} \BibitemShut {NoStop}%
\bibitem [{\citenamefont {Racker}\ \emph {et~al.}(2012)\citenamefont {Racker},
  \citenamefont {Pena},\ and\ \citenamefont {Rius}}]{Racker:2012vw}%
  \BibitemOpen
  \bibfield  {author} {\bibinfo {author} {\bibfnamefont {J.}~\bibnamefont
  {Racker}}, \bibinfo {author} {\bibfnamefont {M.}~\bibnamefont {Pena}}, \ and\
  \bibinfo {author} {\bibfnamefont {N.}~\bibnamefont {Rius}},\ }\href {\doibase
  10.1088/1475-7516/2012/07/030} {\bibfield  {journal} {\bibinfo  {journal}
  {JCAP}\ }\textbf {\bibinfo {volume} {1207}},\ \bibinfo {pages} {030}
  (\bibinfo {year} {2012})},\ \Eprint {http://arxiv.org/abs/1205.1948}
  {arXiv:1205.1948 [hep-ph]} \BibitemShut {NoStop}%
\bibitem [{\citenamefont {Raidal}\ \emph {et~al.}(2005)\citenamefont {Raidal},
  \citenamefont {Strumia},\ and\ \citenamefont {Turzynski}}]{Raidal:2004vt}%
  \BibitemOpen
  \bibfield  {author} {\bibinfo {author} {\bibfnamefont {M.}~\bibnamefont
  {Raidal}}, \bibinfo {author} {\bibfnamefont {A.}~\bibnamefont {Strumia}}, \
  and\ \bibinfo {author} {\bibfnamefont {K.}~\bibnamefont {Turzynski}},\ }\href
  {\doibase 10.1016/j.physletb.2005.11.021, 10.1016/j.physletb.2005.01.040}
  {\bibfield  {journal} {\bibinfo  {journal} {Phys. Lett.}\ }\textbf {\bibinfo
  {volume} {B609}},\ \bibinfo {pages} {351} (\bibinfo {year} {2005})},\
  \bibinfo {note} {[Addendum: Phys. Lett.B632,752(2006)]},\ \Eprint
  {http://arxiv.org/abs/hep-ph/0408015} {arXiv:hep-ph/0408015 [hep-ph]}
  \BibitemShut {NoStop}%
\bibitem [{\citenamefont {Mohapatra}\ and\ \citenamefont
  {Senjanovic}(1980)}]{Mohapatra:1979ia}%
  \BibitemOpen
  \bibfield  {author} {\bibinfo {author} {\bibfnamefont {R.~N.}\ \bibnamefont
  {Mohapatra}}\ and\ \bibinfo {author} {\bibfnamefont {G.}~\bibnamefont
  {Senjanovic}},\ }\href {\doibase 10.1103/PhysRevLett.44.912} {\bibfield
  {journal} {\bibinfo  {journal} {Phys. Rev. Lett.}\ }\textbf {\bibinfo
  {volume} {44}},\ \bibinfo {pages} {912} (\bibinfo {year} {1980})}\BibitemShut
  {NoStop}%
\bibitem [{\citenamefont {Gell-Mann}\ \emph {et~al.}(1979)\citenamefont
  {Gell-Mann}, \citenamefont {Ramond},\ and\ \citenamefont
  {Slansky}}]{GellMann:1980vs}%
  \BibitemOpen
  \bibfield  {author} {\bibinfo {author} {\bibfnamefont {M.}~\bibnamefont
  {Gell-Mann}}, \bibinfo {author} {\bibfnamefont {P.}~\bibnamefont {Ramond}}, \
  and\ \bibinfo {author} {\bibfnamefont {R.}~\bibnamefont {Slansky}},\
  }\bibfield  {booktitle} {\emph {\bibinfo {booktitle} {{Supergravity Workshop
  Stony Brook, New York, September 27-28, 1979}}},\ }\href@noop {} {\bibfield
  {journal} {\bibinfo  {journal} {Conf. Proc.}\ }\textbf {\bibinfo {volume}
  {C790927}},\ \bibinfo {pages} {315} (\bibinfo {year} {1979})},\ \Eprint
  {http://arxiv.org/abs/1306.4669} {arXiv:1306.4669 [hep-th]} \BibitemShut
  {NoStop}%
\bibitem [{\citenamefont {Yanagida}(1979)}]{Yanagida:1979as}%
  \BibitemOpen
  \bibfield  {author} {\bibinfo {author} {\bibfnamefont {T.}~\bibnamefont
  {Yanagida}},\ }\bibfield  {booktitle} {\emph {\bibinfo {booktitle}
  {{Proceedings: Workshop on the Unified Theories and the Baryon Number in the
  Universe: Tsukuba, Japan, February 13-14, 1979}}},\ }\href@noop {} {\bibfield
   {journal} {\bibinfo  {journal} {Conf. Proc.}\ }\textbf {\bibinfo {volume}
  {C7902131}},\ \bibinfo {pages} {95} (\bibinfo {year} {1979})}\BibitemShut
  {NoStop}%
\bibitem [{\citenamefont {Minkowski}(1977)}]{Minkowski:1977sc}%
  \BibitemOpen
  \bibfield  {author} {\bibinfo {author} {\bibfnamefont {P.}~\bibnamefont
  {Minkowski}},\ }\href {\doibase 10.1016/0370-2693(77)90435-X} {\bibfield
  {journal} {\bibinfo  {journal} {Phys. Lett.}\ }\textbf {\bibinfo {volume}
  {B67}},\ \bibinfo {pages} {421} (\bibinfo {year} {1977})}\BibitemShut
  {NoStop}%
\bibitem [{\citenamefont {Mohapatra}\ and\ \citenamefont
  {Senjanovic}(1981)}]{Mohapatra:1980yp}%
  \BibitemOpen
  \bibfield  {author} {\bibinfo {author} {\bibfnamefont {R.~N.}\ \bibnamefont
  {Mohapatra}}\ and\ \bibinfo {author} {\bibfnamefont {G.}~\bibnamefont
  {Senjanovic}},\ }\href {\doibase 10.1103/PhysRevD.23.165} {\bibfield
  {journal} {\bibinfo  {journal} {Phys. Rev.}\ }\textbf {\bibinfo {volume}
  {D23}},\ \bibinfo {pages} {165} (\bibinfo {year} {1981})}\BibitemShut
  {NoStop}%
\bibitem [{\citenamefont {Magg}\ and\ \citenamefont
  {Wetterich}(1980)}]{Magg:1980ut}%
  \BibitemOpen
  \bibfield  {author} {\bibinfo {author} {\bibfnamefont {M.}~\bibnamefont
  {Magg}}\ and\ \bibinfo {author} {\bibfnamefont {C.}~\bibnamefont
  {Wetterich}},\ }\href {\doibase 10.1016/0370-2693(80)90825-4} {\bibfield
  {journal} {\bibinfo  {journal} {Phys. Lett.}\ }\textbf {\bibinfo {volume}
  {B94}},\ \bibinfo {pages} {61} (\bibinfo {year} {1980})}\BibitemShut
  {NoStop}%
\bibitem [{\citenamefont {Lazarides}\ \emph {et~al.}(1981)\citenamefont
  {Lazarides}, \citenamefont {Shafi},\ and\ \citenamefont
  {Wetterich}}]{Lazarides:1980nt}%
  \BibitemOpen
  \bibfield  {author} {\bibinfo {author} {\bibfnamefont {G.}~\bibnamefont
  {Lazarides}}, \bibinfo {author} {\bibfnamefont {Q.}~\bibnamefont {Shafi}}, \
  and\ \bibinfo {author} {\bibfnamefont {C.}~\bibnamefont {Wetterich}},\ }\href
  {\doibase 10.1016/0550-3213(81)90354-0} {\bibfield  {journal} {\bibinfo
  {journal} {Nucl. Phys.}\ }\textbf {\bibinfo {volume} {B181}},\ \bibinfo
  {pages} {287} (\bibinfo {year} {1981})}\BibitemShut {NoStop}%
\bibitem [{\citenamefont {Wetterich}(1981)}]{Wetterich:1981bx}%
  \BibitemOpen
  \bibfield  {author} {\bibinfo {author} {\bibfnamefont {C.}~\bibnamefont
  {Wetterich}},\ }\href {\doibase 10.1016/0550-3213(81)90279-0} {\bibfield
  {journal} {\bibinfo  {journal} {Nucl. Phys.}\ }\textbf {\bibinfo {volume}
  {B187}},\ \bibinfo {pages} {343} (\bibinfo {year} {1981})}\BibitemShut
  {NoStop}%
\bibitem [{\citenamefont {Foot}\ \emph {et~al.}(1989)\citenamefont {Foot},
  \citenamefont {Lew}, \citenamefont {He},\ and\ \citenamefont
  {Joshi}}]{Foot:1988aq}%
  \BibitemOpen
  \bibfield  {author} {\bibinfo {author} {\bibfnamefont {R.}~\bibnamefont
  {Foot}}, \bibinfo {author} {\bibfnamefont {H.}~\bibnamefont {Lew}}, \bibinfo
  {author} {\bibfnamefont {X.~G.}\ \bibnamefont {He}}, \ and\ \bibinfo {author}
  {\bibfnamefont {G.~C.}\ \bibnamefont {Joshi}},\ }\href {\doibase
  10.1007/BF01415558} {\bibfield  {journal} {\bibinfo  {journal} {Z. Phys.}\
  }\textbf {\bibinfo {volume} {C44}},\ \bibinfo {pages} {441} (\bibinfo {year}
  {1989})}\BibitemShut {NoStop}%
\bibitem [{\citenamefont {Ma}(1998)}]{Ma:1998dn}%
  \BibitemOpen
  \bibfield  {author} {\bibinfo {author} {\bibfnamefont {E.}~\bibnamefont
  {Ma}},\ }\href {\doibase 10.1103/PhysRevLett.81.1171} {\bibfield  {journal}
  {\bibinfo  {journal} {Phys. Rev. Lett.}\ }\textbf {\bibinfo {volume} {81}},\
  \bibinfo {pages} {1171} (\bibinfo {year} {1998})},\ \Eprint
  {http://arxiv.org/abs/hep-ph/9805219} {arXiv:hep-ph/9805219 [hep-ph]}
  \BibitemShut {NoStop}%
\bibitem [{\citenamefont {Akhmedov}\ \emph {et~al.}(1998)\citenamefont
  {Akhmedov}, \citenamefont {Rubakov},\ and\ \citenamefont
  {Smirnov}}]{Akhmedov:1998qx}%
  \BibitemOpen
  \bibfield  {author} {\bibinfo {author} {\bibfnamefont {E.~K.}\ \bibnamefont
  {Akhmedov}}, \bibinfo {author} {\bibfnamefont {V.~A.}\ \bibnamefont
  {Rubakov}}, \ and\ \bibinfo {author} {\bibfnamefont {A.~{\relax Yu}.}\
  \bibnamefont {Smirnov}},\ }\href {\doibase 10.1103/PhysRevLett.81.1359}
  {\bibfield  {journal} {\bibinfo  {journal} {Phys. Rev. Lett.}\ }\textbf
  {\bibinfo {volume} {81}},\ \bibinfo {pages} {1359} (\bibinfo {year}
  {1998})},\ \Eprint {http://arxiv.org/abs/hep-ph/9803255}
  {arXiv:hep-ph/9803255 [hep-ph]} \BibitemShut {NoStop}%
\bibitem [{\citenamefont {Drewes}\ \emph {et~al.}(2017)\citenamefont {Drewes},
  \citenamefont {Garbrecht}, \citenamefont {Hernandez}, \citenamefont {Kekic},
  \citenamefont {Lopez-Pavon}, \citenamefont {Racker}, \citenamefont {Rius},
  \citenamefont {Salvado},\ and\ \citenamefont {Teresi}}]{Drewes:2017zyw}%
  \BibitemOpen
  \bibfield  {author} {\bibinfo {author} {\bibfnamefont {M.}~\bibnamefont
  {Drewes}}, \bibinfo {author} {\bibfnamefont {B.}~\bibnamefont {Garbrecht}},
  \bibinfo {author} {\bibfnamefont {P.}~\bibnamefont {Hernandez}}, \bibinfo
  {author} {\bibfnamefont {M.}~\bibnamefont {Kekic}}, \bibinfo {author}
  {\bibfnamefont {J.}~\bibnamefont {Lopez-Pavon}}, \bibinfo {author}
  {\bibfnamefont {J.}~\bibnamefont {Racker}}, \bibinfo {author} {\bibfnamefont
  {N.}~\bibnamefont {Rius}}, \bibinfo {author} {\bibfnamefont {J.}~\bibnamefont
  {Salvado}}, \ and\ \bibinfo {author} {\bibfnamefont {D.}~\bibnamefont
  {Teresi}},\ }\href@noop {} {\bibfield  {journal} {\bibinfo  {journal} {n/a}\
  } (\bibinfo {year} {2017})},\ \Eprint {http://arxiv.org/abs/1711.02862}
  {arXiv:1711.02862 [hep-ph]} \BibitemShut {NoStop}%
\bibitem [{\citenamefont {Hernandez}\ \emph {et~al.}(2016)\citenamefont
  {Hernandez}, \citenamefont {Kekic}, \citenamefont {Lopez-Pavon},
  \citenamefont {Racker},\ and\ \citenamefont {Salvado}}]{Hernandez:2016kel}%
  \BibitemOpen
  \bibfield  {author} {\bibinfo {author} {\bibfnamefont {P.}~\bibnamefont
  {Hernandez}}, \bibinfo {author} {\bibfnamefont {M.}~\bibnamefont {Kekic}},
  \bibinfo {author} {\bibfnamefont {J.}~\bibnamefont {Lopez-Pavon}}, \bibinfo
  {author} {\bibfnamefont {J.}~\bibnamefont {Racker}}, \ and\ \bibinfo {author}
  {\bibfnamefont {J.}~\bibnamefont {Salvado}},\ }\href {\doibase
  10.1007/JHEP08(2016)157} {\bibfield  {journal} {\bibinfo  {journal} {JHEP}\
  }\textbf {\bibinfo {volume} {08}},\ \bibinfo {pages} {157} (\bibinfo {year}
  {2016})},\ \Eprint {http://arxiv.org/abs/1606.06719} {arXiv:1606.06719
  [hep-ph]} \BibitemShut {NoStop}%
\bibitem [{\citenamefont {Dev}\ \emph {et~al.}(2017)\citenamefont {Dev},
  \citenamefont {Di~Bari}, \citenamefont {Garbrecht}, \citenamefont {Lavignac},
  \citenamefont {Millington},\ and\ \citenamefont {Teresi}}]{Dev:2017trv}%
  \BibitemOpen
  \bibfield  {author} {\bibinfo {author} {\bibfnamefont {P.~S.~B.}\
  \bibnamefont {Dev}}, \bibinfo {author} {\bibfnamefont {P.}~\bibnamefont
  {Di~Bari}}, \bibinfo {author} {\bibfnamefont {B.}~\bibnamefont {Garbrecht}},
  \bibinfo {author} {\bibfnamefont {S.}~\bibnamefont {Lavignac}}, \bibinfo
  {author} {\bibfnamefont {P.}~\bibnamefont {Millington}}, \ and\ \bibinfo
  {author} {\bibfnamefont {D.}~\bibnamefont {Teresi}},\ }\href@noop {}
  {\bibfield  {journal} {\bibinfo  {journal} {n/a}\ } (\bibinfo {year}
  {2017})},\ \Eprint {http://arxiv.org/abs/1711.02861} {arXiv:1711.02861
  [hep-ph]} \BibitemShut {NoStop}%
\bibitem [{\citenamefont {Antusch}\ \emph {et~al.}(2017)\citenamefont
  {Antusch}, \citenamefont {Cazzato}, \citenamefont {Drewes}, \citenamefont
  {Fischer}, \citenamefont {Garbrecht}, \citenamefont {Gueter},\ and\
  \citenamefont {Klaric}}]{Antusch:2017pkq}%
  \BibitemOpen
  \bibfield  {author} {\bibinfo {author} {\bibfnamefont {S.}~\bibnamefont
  {Antusch}}, \bibinfo {author} {\bibfnamefont {E.}~\bibnamefont {Cazzato}},
  \bibinfo {author} {\bibfnamefont {M.}~\bibnamefont {Drewes}}, \bibinfo
  {author} {\bibfnamefont {O.}~\bibnamefont {Fischer}}, \bibinfo {author}
  {\bibfnamefont {B.}~\bibnamefont {Garbrecht}}, \bibinfo {author}
  {\bibfnamefont {D.}~\bibnamefont {Gueter}}, \ and\ \bibinfo {author}
  {\bibfnamefont {J.}~\bibnamefont {Klaric}},\ }\href@noop {} {\bibfield
  {journal} {\bibinfo  {journal} {n/a}\ } (\bibinfo {year} {2017})},\ \Eprint
  {http://arxiv.org/abs/1710.03744} {arXiv:1710.03744 [hep-ph]} \BibitemShut
  {NoStop}%
\bibitem [{\citenamefont {Antusch}\ and\ \citenamefont
  {Fischer}(2015)}]{Antusch:2015mia}%
  \BibitemOpen
  \bibfield  {author} {\bibinfo {author} {\bibfnamefont {S.}~\bibnamefont
  {Antusch}}\ and\ \bibinfo {author} {\bibfnamefont {O.}~\bibnamefont
  {Fischer}},\ }\href {\doibase 10.1007/JHEP05(2015)053} {\bibfield  {journal}
  {\bibinfo  {journal} {JHEP}\ }\textbf {\bibinfo {volume} {05}},\ \bibinfo
  {pages} {053} (\bibinfo {year} {2015})},\ \Eprint
  {http://arxiv.org/abs/1502.05915} {arXiv:1502.05915 [hep-ph]} \BibitemShut
  {NoStop}%
\bibitem [{\citenamefont {Milanes}\ \emph {et~al.}(2016)\citenamefont
  {Milanes}, \citenamefont {Quintero},\ and\ \citenamefont
  {Vera}}]{Milanes:2016rzr}%
  \BibitemOpen
  \bibfield  {author} {\bibinfo {author} {\bibfnamefont {D.}~\bibnamefont
  {Milanes}}, \bibinfo {author} {\bibfnamefont {N.}~\bibnamefont {Quintero}}, \
  and\ \bibinfo {author} {\bibfnamefont {C.~E.}\ \bibnamefont {Vera}},\ }\href
  {\doibase 10.1103/PhysRevD.93.094026} {\bibfield  {journal} {\bibinfo
  {journal} {Phys. Rev.}\ }\textbf {\bibinfo {volume} {D93}},\ \bibinfo {pages}
  {094026} (\bibinfo {year} {2016})},\ \Eprint
  {http://arxiv.org/abs/1604.03177} {arXiv:1604.03177 [hep-ph]} \BibitemShut
  {NoStop}%
\bibitem [{\citenamefont {Canetti}\ \emph {et~al.}(2014)\citenamefont
  {Canetti}, \citenamefont {Drewes},\ and\ \citenamefont
  {Garbrecht}}]{Canetti:2014dka}%
  \BibitemOpen
  \bibfield  {author} {\bibinfo {author} {\bibfnamefont {L.}~\bibnamefont
  {Canetti}}, \bibinfo {author} {\bibfnamefont {M.}~\bibnamefont {Drewes}}, \
  and\ \bibinfo {author} {\bibfnamefont {B.}~\bibnamefont {Garbrecht}},\ }\href
  {\doibase 10.1103/PhysRevD.90.125005} {\bibfield  {journal} {\bibinfo
  {journal} {Phys. Rev.}\ }\textbf {\bibinfo {volume} {D90}},\ \bibinfo {pages}
  {125005} (\bibinfo {year} {2014})},\ \Eprint {http://arxiv.org/abs/1404.7114}
  {arXiv:1404.7114 [hep-ph]} \BibitemShut {NoStop}%
\bibitem [{\citenamefont {Anelli}\ \emph {et~al.}(2015)\citenamefont {Anelli}
  \emph {et~al.}}]{Anelli:2015pba}%
  \BibitemOpen
  \bibfield  {author} {\bibinfo {author} {\bibfnamefont {M.}~\bibnamefont
  {Anelli}} \emph {et~al.} (\bibinfo {collaboration} {SHiP}),\ }\href@noop {}
  {\bibfield  {journal} {\bibinfo  {journal} {n/a}\ } (\bibinfo {year}
  {2015})},\ \Eprint {http://arxiv.org/abs/1504.04956} {arXiv:1504.04956
  [physics.ins-det]} \BibitemShut {NoStop}%
\bibitem [{\citenamefont {Alekhin}\ \emph {et~al.}(2016)\citenamefont {Alekhin}
  \emph {et~al.}}]{Alekhin:2015byh}%
  \BibitemOpen
  \bibfield  {author} {\bibinfo {author} {\bibfnamefont {S.}~\bibnamefont
  {Alekhin}} \emph {et~al.},\ }\href {\doibase 10.1088/0034-4885/79/12/124201}
  {\bibfield  {journal} {\bibinfo  {journal} {Rept. Prog. Phys.}\ }\textbf
  {\bibinfo {volume} {79}},\ \bibinfo {pages} {124201} (\bibinfo {year}
  {2016})},\ \Eprint {http://arxiv.org/abs/1504.04855} {arXiv:1504.04855
  [hep-ph]} \BibitemShut {NoStop}%
\bibitem [{\citenamefont {Graverini}\ \emph {et~al.}(2015)\citenamefont
  {Graverini}, \citenamefont {Serra},\ and\ \citenamefont
  {Storaci}}]{Graverini:2015dka}%
  \BibitemOpen
  \bibfield  {author} {\bibinfo {author} {\bibfnamefont {E.}~\bibnamefont
  {Graverini}}, \bibinfo {author} {\bibfnamefont {N.}~\bibnamefont {Serra}}, \
  and\ \bibinfo {author} {\bibfnamefont {B.}~\bibnamefont {Storaci}} (\bibinfo
  {collaboration} {SHiP}),\ }\bibfield  {booktitle} {\emph {\bibinfo
  {booktitle} {{Proceedings, 2nd International Summer School on INtelligent
  Signal Processing for FrontIEr Research and Industry (INFIERI 2014): Paris,
  France, July 15-25, 2014}}},\ }\href {\doibase
  10.1088/1748-0221/10/07/C07007} {\bibfield  {journal} {\bibinfo  {journal}
  {JINST}\ }\textbf {\bibinfo {volume} {10}},\ \bibinfo {pages} {C07007}
  (\bibinfo {year} {2015})},\ \Eprint {http://arxiv.org/abs/1503.08624}
  {arXiv:1503.08624 [hep-ex]} \BibitemShut {NoStop}%
\bibitem [{\citenamefont {Shuve}\ and\ \citenamefont
  {Yavin}(2014)}]{Shuve:2014zua}%
  \BibitemOpen
  \bibfield  {author} {\bibinfo {author} {\bibfnamefont {B.}~\bibnamefont
  {Shuve}}\ and\ \bibinfo {author} {\bibfnamefont {I.}~\bibnamefont {Yavin}},\
  }\href {\doibase 10.1103/PhysRevD.89.075014} {\bibfield  {journal} {\bibinfo
  {journal} {Phys. Rev.}\ }\textbf {\bibinfo {volume} {D89}},\ \bibinfo {pages}
  {075014} (\bibinfo {year} {2014})},\ \Eprint {http://arxiv.org/abs/1401.2459}
  {arXiv:1401.2459 [hep-ph]} \BibitemShut {NoStop}%
\bibitem [{\citenamefont {Esteban}\ \emph {et~al.}(2017)\citenamefont
  {Esteban}, \citenamefont {Gonzalez-Garcia}, \citenamefont {Maltoni},
  \citenamefont {Martinez-Soler},\ and\ \citenamefont
  {Schwetz}}]{Esteban:2016qun}%
  \BibitemOpen
  \bibfield  {author} {\bibinfo {author} {\bibfnamefont {I.}~\bibnamefont
  {Esteban}}, \bibinfo {author} {\bibfnamefont {M.~C.}\ \bibnamefont
  {Gonzalez-Garcia}}, \bibinfo {author} {\bibfnamefont {M.}~\bibnamefont
  {Maltoni}}, \bibinfo {author} {\bibfnamefont {I.}~\bibnamefont
  {Martinez-Soler}}, \ and\ \bibinfo {author} {\bibfnamefont {T.}~\bibnamefont
  {Schwetz}},\ }\href {\doibase 10.1007/JHEP01(2017)087} {\bibfield  {journal}
  {\bibinfo  {journal} {JHEP}\ }\textbf {\bibinfo {volume} {01}},\ \bibinfo
  {pages} {087} (\bibinfo {year} {2017})},\ \Eprint
  {http://arxiv.org/abs/1611.01514} {arXiv:1611.01514 [hep-ph]} \BibitemShut
  {NoStop}%
\bibitem [{\citenamefont {Fukuda}\ \emph {et~al.}(1998)\citenamefont {Fukuda}
  \emph {et~al.}}]{Fukuda:1998mi}%
  \BibitemOpen
  \bibfield  {author} {\bibinfo {author} {\bibfnamefont {Y.}~\bibnamefont
  {Fukuda}} \emph {et~al.} (\bibinfo {collaboration} {Super-Kamiokande}),\
  }\href {\doibase 10.1103/PhysRevLett.81.1562} {\bibfield  {journal} {\bibinfo
   {journal} {Phys. Rev. Lett.}\ }\textbf {\bibinfo {volume} {81}},\ \bibinfo
  {pages} {1562} (\bibinfo {year} {1998})},\ \Eprint
  {http://arxiv.org/abs/hep-ex/9807003} {arXiv:hep-ex/9807003 [hep-ex]}
  \BibitemShut {NoStop}%
\bibitem [{\citenamefont {Bilenky}\ \emph {et~al.}(1980)\citenamefont
  {Bilenky}, \citenamefont {Hosek},\ and\ \citenamefont
  {Petcov}}]{Bilenky:1980cx}%
  \BibitemOpen
  \bibfield  {author} {\bibinfo {author} {\bibfnamefont {S.~M.}\ \bibnamefont
  {Bilenky}}, \bibinfo {author} {\bibfnamefont {J.}~\bibnamefont {Hosek}}, \
  and\ \bibinfo {author} {\bibfnamefont {S.~T.}\ \bibnamefont {Petcov}},\
  }\href {\doibase 10.1016/0370-2693(80)90927-2} {\bibfield  {journal}
  {\bibinfo  {journal} {Phys. Lett.}\ }\textbf {\bibinfo {volume} {94B}},\
  \bibinfo {pages} {495} (\bibinfo {year} {1980})}\BibitemShut {NoStop}%
\bibitem [{\citenamefont {Huber}\ \emph {et~al.}(2004)\citenamefont {Huber},
  \citenamefont {Lindner}, \citenamefont {Rolinec}, \citenamefont {Schwetz},\
  and\ \citenamefont {Winter}}]{Huber:2004ug}%
  \BibitemOpen
  \bibfield  {author} {\bibinfo {author} {\bibfnamefont {P.}~\bibnamefont
  {Huber}}, \bibinfo {author} {\bibfnamefont {M.}~\bibnamefont {Lindner}},
  \bibinfo {author} {\bibfnamefont {M.}~\bibnamefont {Rolinec}}, \bibinfo
  {author} {\bibfnamefont {T.}~\bibnamefont {Schwetz}}, \ and\ \bibinfo
  {author} {\bibfnamefont {W.}~\bibnamefont {Winter}},\ }\href {\doibase
  10.1103/PhysRevD.70.073014} {\bibfield  {journal} {\bibinfo  {journal} {Phys.
  Rev.}\ }\textbf {\bibinfo {volume} {D70}},\ \bibinfo {pages} {073014}
  (\bibinfo {year} {2004})},\ \Eprint {http://arxiv.org/abs/hep-ph/0403068}
  {arXiv:hep-ph/0403068 [hep-ph]} \BibitemShut {NoStop}%
\bibitem [{\citenamefont {Abe}\ \emph {et~al.}(2011)\citenamefont {Abe} \emph
  {et~al.}}]{Abe:2011ks}%
  \BibitemOpen
  \bibfield  {author} {\bibinfo {author} {\bibfnamefont {K.}~\bibnamefont
  {Abe}} \emph {et~al.} (\bibinfo {collaboration} {T2K}),\ }\href {\doibase
  10.1016/j.nima.2011.06.067} {\bibfield  {journal} {\bibinfo  {journal} {Nucl.
  Instrum. Meth.}\ }\textbf {\bibinfo {volume} {A659}},\ \bibinfo {pages} {106}
  (\bibinfo {year} {2011})},\ \Eprint {http://arxiv.org/abs/1106.1238}
  {arXiv:1106.1238 [physics.ins-det]} \BibitemShut {NoStop}%
\bibitem [{\citenamefont {Ayres}\ \emph {et~al.}(2004)\citenamefont {Ayres}
  \emph {et~al.}}]{Ayres:2004js}%
  \BibitemOpen
  \bibfield  {author} {\bibinfo {author} {\bibfnamefont {D.~S.}\ \bibnamefont
  {Ayres}} \emph {et~al.} (\bibinfo {collaboration} {NOvA}),\ }\href@noop {}
  {\bibfield  {journal} {\bibinfo  {journal} {n/a}\ } (\bibinfo {year}
  {2004})},\ \Eprint {http://arxiv.org/abs/hep-ex/0503053}
  {arXiv:hep-ex/0503053 [hep-ex]} \BibitemShut {NoStop}%
\bibitem [{\citenamefont {An}\ \emph {et~al.}(2012)\citenamefont {An} \emph
  {et~al.}}]{An:2012eh}%
  \BibitemOpen
  \bibfield  {author} {\bibinfo {author} {\bibfnamefont {F.~P.}\ \bibnamefont
  {An}} \emph {et~al.} (\bibinfo {collaboration} {Daya Bay}),\ }\href {\doibase
  10.1103/PhysRevLett.108.171803} {\bibfield  {journal} {\bibinfo  {journal}
  {Phys. Rev. Lett.}\ }\textbf {\bibinfo {volume} {108}},\ \bibinfo {pages}
  {171803} (\bibinfo {year} {2012})},\ \Eprint {http://arxiv.org/abs/1203.1669}
  {arXiv:1203.1669 [hep-ex]} \BibitemShut {NoStop}%
\bibitem [{\citenamefont {Ahn}\ \emph {et~al.}(2012)\citenamefont {Ahn} \emph
  {et~al.}}]{Ahn:2012nd}%
  \BibitemOpen
  \bibfield  {author} {\bibinfo {author} {\bibfnamefont {J.~K.}\ \bibnamefont
  {Ahn}} \emph {et~al.} (\bibinfo {collaboration} {RENO}),\ }\href {\doibase
  10.1103/PhysRevLett.108.191802} {\bibfield  {journal} {\bibinfo  {journal}
  {Phys. Rev. Lett.}\ }\textbf {\bibinfo {volume} {108}},\ \bibinfo {pages}
  {191802} (\bibinfo {year} {2012})},\ \Eprint {http://arxiv.org/abs/1204.0626}
  {arXiv:1204.0626 [hep-ex]} \BibitemShut {NoStop}%
\bibitem [{\citenamefont {Ardellier}\ \emph {et~al.}(2006)\citenamefont
  {Ardellier} \emph {et~al.}}]{Ardellier:2006mn}%
  \BibitemOpen
  \bibfield  {author} {\bibinfo {author} {\bibfnamefont {F.}~\bibnamefont
  {Ardellier}} \emph {et~al.} (\bibinfo {collaboration} {Double Chooz}),\
  }\href@noop {} {\bibfield  {journal} {\bibinfo  {journal} {n/a}\ } (\bibinfo
  {year} {2006})},\ \Eprint {http://arxiv.org/abs/hep-ex/0606025}
  {arXiv:hep-ex/0606025 [hep-ex]} \BibitemShut {NoStop}%
\bibitem [{\citenamefont {Bilenky}\ \emph {et~al.}(2001)\citenamefont
  {Bilenky}, \citenamefont {Pascoli},\ and\ \citenamefont
  {Petcov}}]{Bilenky:2001rz}%
  \BibitemOpen
  \bibfield  {author} {\bibinfo {author} {\bibfnamefont {S.~M.}\ \bibnamefont
  {Bilenky}}, \bibinfo {author} {\bibfnamefont {S.}~\bibnamefont {Pascoli}}, \
  and\ \bibinfo {author} {\bibfnamefont {S.~T.}\ \bibnamefont {Petcov}},\
  }\href {\doibase 10.1103/PhysRevD.64.053010} {\bibfield  {journal} {\bibinfo
  {journal} {Phys. Rev.}\ }\textbf {\bibinfo {volume} {D64}},\ \bibinfo {pages}
  {053010} (\bibinfo {year} {2001})},\ \Eprint
  {http://arxiv.org/abs/hep-ph/0102265} {arXiv:hep-ph/0102265 [hep-ph]}
  \BibitemShut {NoStop}%
\bibitem [{\citenamefont {Bilenky}\ \emph {et~al.}(1996)\citenamefont
  {Bilenky}, \citenamefont {Giunti}, \citenamefont {Kim},\ and\ \citenamefont
  {Petcov}}]{Bilenky:1996cb}%
  \BibitemOpen
  \bibfield  {author} {\bibinfo {author} {\bibfnamefont {S.~M.}\ \bibnamefont
  {Bilenky}}, \bibinfo {author} {\bibfnamefont {C.}~\bibnamefont {Giunti}},
  \bibinfo {author} {\bibfnamefont {C.~W.}\ \bibnamefont {Kim}}, \ and\
  \bibinfo {author} {\bibfnamefont {S.~T.}\ \bibnamefont {Petcov}},\ }\href
  {\doibase 10.1103/PhysRevD.54.4432} {\bibfield  {journal} {\bibinfo
  {journal} {Phys. Rev.}\ }\textbf {\bibinfo {volume} {D54}},\ \bibinfo {pages}
  {4432} (\bibinfo {year} {1996})},\ \Eprint
  {http://arxiv.org/abs/hep-ph/9604364} {arXiv:hep-ph/9604364 [hep-ph]}
  \BibitemShut {NoStop}%
\bibitem [{\citenamefont {Alfonso}\ \emph {et~al.}(2015)\citenamefont {Alfonso}
  \emph {et~al.}}]{Alfonso:2015wka}%
  \BibitemOpen
  \bibfield  {author} {\bibinfo {author} {\bibfnamefont {K.}~\bibnamefont
  {Alfonso}} \emph {et~al.} (\bibinfo {collaboration} {CUORE}),\ }\href
  {\doibase 10.1103/PhysRevLett.115.102502} {\bibfield  {journal} {\bibinfo
  {journal} {Phys. Rev. Lett.}\ }\textbf {\bibinfo {volume} {115}},\ \bibinfo
  {pages} {102502} (\bibinfo {year} {2015})},\ \Eprint
  {http://arxiv.org/abs/1504.02454} {arXiv:1504.02454 [nucl-ex]} \BibitemShut
  {NoStop}%
\bibitem [{\citenamefont {Klapdor-Kleingrothaus}\ \emph
  {et~al.}(2000)\citenamefont {Klapdor-Kleingrothaus} \emph
  {et~al.}}]{KlapdorKleingrothaus:2000sz}%
  \BibitemOpen
  \bibfield  {author} {\bibinfo {author} {\bibfnamefont {H.~V.}\ \bibnamefont
  {Klapdor-Kleingrothaus}} \emph {et~al.},\ }in\ \href@noop {} {\emph {\bibinfo
  {booktitle} {{Proceedings, 3rd International Heidelberg Conference on Dark
  matter in astro- and particle physics (DARK 2000): Heidelberg, Germany, July
  10-14, 2000}}}}\ (\bibinfo {year} {2000})\ pp.\ \bibinfo {pages}
  {520--533}\BibitemShut {NoStop}%
\bibitem [{\citenamefont {Agostini}\ \emph {et~al.}(2018)\citenamefont
  {Agostini} \emph {et~al.}}]{Agostini:2018tnm}%
  \BibitemOpen
  \bibfield  {author} {\bibinfo {author} {\bibfnamefont {M.}~\bibnamefont
  {Agostini}} \emph {et~al.} (\bibinfo {collaboration} {GERDA}),\ }\href
  {\doibase 10.1103/PhysRevLett.120.132503} {\bibfield  {journal} {\bibinfo
  {journal} {Phys. Rev. Lett.}\ }\textbf {\bibinfo {volume} {120}},\ \bibinfo
  {pages} {132503} (\bibinfo {year} {2018})},\ \Eprint
  {http://arxiv.org/abs/1803.11100} {arXiv:1803.11100 [nucl-ex]} \BibitemShut
  {NoStop}%
\bibitem [{\citenamefont {Bongrand}(2015)}]{Bongrand:2015yfa}%
  \BibitemOpen
  \bibfield  {author} {\bibinfo {author} {\bibfnamefont {M.}~\bibnamefont
  {Bongrand}} (\bibinfo {collaboration} {SuperNEMO}),\ }\bibfield  {booktitle}
  {\emph {\bibinfo {booktitle} {{Proceedings, 26th International Conference on
  Neutrino Physics and Astrophysics (Neutrino 2014): Boston, Massachusetts,
  United States, June 2-7, 2014}}},\ }\href {\doibase 10.1063/1.4915592}
  {\bibfield  {journal} {\bibinfo  {journal} {AIP Conf. Proc.}\ }\textbf
  {\bibinfo {volume} {1666}},\ \bibinfo {pages} {170002} (\bibinfo {year}
  {2015})}\BibitemShut {NoStop}%
\bibitem [{\citenamefont {Yang}(2017)}]{Yang:2017lnf}%
  \BibitemOpen
  \bibfield  {author} {\bibinfo {author} {\bibfnamefont {L.}~\bibnamefont
  {Yang}} (\bibinfo {collaboration} {nEXO, EXO-200}),\ }\href {\doibase
  10.1088/1742-6596/888/1/012032} {\bibfield  {journal} {\bibinfo  {journal}
  {J. Phys. Conf. Ser.}\ }\textbf {\bibinfo {volume} {888}},\ \bibinfo {pages}
  {012032} (\bibinfo {year} {2017})}\BibitemShut {NoStop}%
\bibitem [{\citenamefont {Ouellet}(2016)}]{Ouellet:2016moa}%
  \BibitemOpen
  \bibfield  {author} {\bibinfo {author} {\bibfnamefont {J.}~\bibnamefont
  {Ouellet}} (\bibinfo {collaboration} {KamLAND-Zen}),\ }\bibfield  {booktitle}
  {\emph {\bibinfo {booktitle} {{Proceedings, 38th International Conference on
  High Energy Physics (ICHEP 2016): Chicago, IL, USA, August 3-10, 2016}}},\
  }\href@noop {} {\bibfield  {journal} {\bibinfo  {journal} {PoS}\ }\textbf
  {\bibinfo {volume} {ICHEP2016}},\ \bibinfo {pages} {492} (\bibinfo {year}
  {2016})}\BibitemShut {NoStop}%
\bibitem [{\citenamefont {Pattavina}(2016)}]{Pattavina:2016kqn}%
  \BibitemOpen
  \bibfield  {author} {\bibinfo {author} {\bibfnamefont {L.}~\bibnamefont
  {Pattavina}} (\bibinfo {collaboration} {LUCIFER}),\ }\bibfield  {booktitle}
  {\emph {\bibinfo {booktitle} {{Proceedings, 14th International Conference on
  Topics in Astroparticle and Underground Physics (TAUP 2015): Torino, Italy,
  September 7-11, 2015}}},\ }\href {\doibase 10.1088/1742-6596/718/6/062048}
  {\bibfield  {journal} {\bibinfo  {journal} {J. Phys. Conf. Ser.}\ }\textbf
  {\bibinfo {volume} {718}},\ \bibinfo {pages} {062048} (\bibinfo {year}
  {2016})}\BibitemShut {NoStop}%
\bibitem [{\citenamefont {Yu}\ \emph {et~al.}(2018)\citenamefont {Yu} \emph
  {et~al.}}]{Yu:2018due}%
  \BibitemOpen
  \bibfield  {author} {\bibinfo {author} {\bibfnamefont {C.~H.}\ \bibnamefont
  {Yu}} \emph {et~al.} (\bibinfo {collaboration} {Majorana})\ }(\bibinfo {year}
  {2018})\ \Eprint {http://arxiv.org/abs/1803.11220} {arXiv:1803.11220
  [nucl-ex]} \BibitemShut {NoStop}%
\bibitem [{\citenamefont {Gomez-Cadenas}\ \emph {et~al.}(2014)\citenamefont
  {Gomez-Cadenas} \emph {et~al.}}]{Gomez-Cadenas:2013lta}%
  \BibitemOpen
  \bibfield  {author} {\bibinfo {author} {\bibfnamefont {J.~J.}\ \bibnamefont
  {Gomez-Cadenas}} \emph {et~al.} (\bibinfo {collaboration} {NEXT}),\ }\href
  {\doibase 10.1155/2014/907067} {\bibfield  {journal} {\bibinfo  {journal}
  {Adv. High Energy Phys.}\ }\textbf {\bibinfo {volume} {2014}},\ \bibinfo
  {pages} {907067} (\bibinfo {year} {2014})},\ \Eprint
  {http://arxiv.org/abs/1307.3914} {arXiv:1307.3914 [physics.ins-det]}
  \BibitemShut {NoStop}%
\bibitem [{\citenamefont {Lozza}(2016)}]{Lozza:2016rwo}%
  \BibitemOpen
  \bibfield  {author} {\bibinfo {author} {\bibfnamefont {V.}~\bibnamefont
  {Lozza}} (\bibinfo {collaboration} {SNO+}),\ }in\ \href {\doibase
  10.3204/DESY-PROC-2016-05/5} {\emph {\bibinfo {booktitle} {{Proceedings,
  Magellan Workshop: Connecting Neutrino Physics and Astronomy: Hamburg,
  Germany, March 17-18, 2016}}}}\ (\bibinfo {year} {2016})\ pp.\ \bibinfo
  {pages} {87--94}\BibitemShut {NoStop}%
\bibitem [{\citenamefont {Torre}(2014)}]{Torre:2014tma}%
  \BibitemOpen
  \bibfield  {author} {\bibinfo {author} {\bibfnamefont {S.}~\bibnamefont
  {Torre}} (\bibinfo {collaboration} {NEMO}),\ }in\ \href
  {http://inspirehep.net/record/1338135/files/Pages_from_C14-03-15--1_149.pdf}
  {\emph {\bibinfo {booktitle} {{Proceedings, 49th Rencontres de Moriond on
  Electroweak Interactions and Unified Theories: La Thuile, Italy, March 15-22,
  2014}}}}\ (\bibinfo {year} {2014})\ pp.\ \bibinfo {pages}
  {149--154}\BibitemShut {NoStop}%
\bibitem [{\citenamefont {Schechter}\ and\ \citenamefont
  {Valle}(1980)}]{Schechter:1980gr}%
  \BibitemOpen
  \bibfield  {author} {\bibinfo {author} {\bibfnamefont {J.}~\bibnamefont
  {Schechter}}\ and\ \bibinfo {author} {\bibfnamefont {J.~W.~F.}\ \bibnamefont
  {Valle}},\ }\href {\doibase 10.1103/PhysRevD.22.2227} {\bibfield  {journal}
  {\bibinfo  {journal} {Phys. Rev.}\ }\textbf {\bibinfo {volume} {D22}},\
  \bibinfo {pages} {2227} (\bibinfo {year} {1980})}\BibitemShut {NoStop}%
\bibitem [{\citenamefont {Pascoli}\ \emph
  {et~al.}(2002{\natexlab{a}})\citenamefont {Pascoli}, \citenamefont {Petcov},\
  and\ \citenamefont {Wolfenstein}}]{Pascoli:2001by}%
  \BibitemOpen
  \bibfield  {author} {\bibinfo {author} {\bibfnamefont {S.}~\bibnamefont
  {Pascoli}}, \bibinfo {author} {\bibfnamefont {S.~T.}\ \bibnamefont {Petcov}},
  \ and\ \bibinfo {author} {\bibfnamefont {L.}~\bibnamefont {Wolfenstein}},\
  }\href {\doibase 10.1016/S0370-2693(01)01403-4} {\bibfield  {journal}
  {\bibinfo  {journal} {Phys. Lett.}\ }\textbf {\bibinfo {volume} {B524}},\
  \bibinfo {pages} {319} (\bibinfo {year} {2002}{\natexlab{a}})},\ \Eprint
  {http://arxiv.org/abs/hep-ph/0110287} {arXiv:hep-ph/0110287 [hep-ph]}
  \BibitemShut {NoStop}%
\bibitem [{\citenamefont {Pascoli}\ and\ \citenamefont
  {Petcov}(2002)}]{Pascoli:2002xq}%
  \BibitemOpen
  \bibfield  {author} {\bibinfo {author} {\bibfnamefont {S.}~\bibnamefont
  {Pascoli}}\ and\ \bibinfo {author} {\bibfnamefont {S.~T.}\ \bibnamefont
  {Petcov}},\ }\href {\doibase 10.1016/S0370-2693(02)02510-8} {\bibfield
  {journal} {\bibinfo  {journal} {Phys. Lett.}\ }\textbf {\bibinfo {volume}
  {B544}},\ \bibinfo {pages} {239} (\bibinfo {year} {2002})},\ \Eprint
  {http://arxiv.org/abs/hep-ph/0205022} {arXiv:hep-ph/0205022 [hep-ph]}
  \BibitemShut {NoStop}%
\bibitem [{\citenamefont {Pascoli}\ \emph
  {et~al.}(2002{\natexlab{b}})\citenamefont {Pascoli}, \citenamefont {Petcov},\
  and\ \citenamefont {Rodejohann}}]{Pascoli:2002qm}%
  \BibitemOpen
  \bibfield  {author} {\bibinfo {author} {\bibfnamefont {S.}~\bibnamefont
  {Pascoli}}, \bibinfo {author} {\bibfnamefont {S.~T.}\ \bibnamefont {Petcov}},
  \ and\ \bibinfo {author} {\bibfnamefont {W.}~\bibnamefont {Rodejohann}},\
  }\href {\doibase 10.1016/S0370-2693(02)02852-6} {\bibfield  {journal}
  {\bibinfo  {journal} {Phys. Lett.}\ }\textbf {\bibinfo {volume} {B549}},\
  \bibinfo {pages} {177} (\bibinfo {year} {2002}{\natexlab{b}})},\ \Eprint
  {http://arxiv.org/abs/hep-ph/0209059} {arXiv:hep-ph/0209059 [hep-ph]}
  \BibitemShut {NoStop}%
\bibitem [{\citenamefont {Pascoli}\ \emph {et~al.}(2003)\citenamefont
  {Pascoli}, \citenamefont {Petcov},\ and\ \citenamefont
  {Rodejohann}}]{Pascoli:2002ae}%
  \BibitemOpen
  \bibfield  {author} {\bibinfo {author} {\bibfnamefont {S.}~\bibnamefont
  {Pascoli}}, \bibinfo {author} {\bibfnamefont {S.~T.}\ \bibnamefont {Petcov}},
  \ and\ \bibinfo {author} {\bibfnamefont {W.}~\bibnamefont {Rodejohann}},\
  }\href {\doibase 10.1016/S0370-2693(03)00275-2} {\bibfield  {journal}
  {\bibinfo  {journal} {Phys. Lett.}\ }\textbf {\bibinfo {volume} {B558}},\
  \bibinfo {pages} {141} (\bibinfo {year} {2003})},\ \Eprint
  {http://arxiv.org/abs/hep-ph/0212113} {arXiv:hep-ph/0212113 [hep-ph]}
  \BibitemShut {NoStop}%
\bibitem [{\citenamefont {Barger}\ \emph {et~al.}(2002)\citenamefont {Barger},
  \citenamefont {Glashow}, \citenamefont {Langacker},\ and\ \citenamefont
  {Marfatia}}]{Barger:2002vy}%
  \BibitemOpen
  \bibfield  {author} {\bibinfo {author} {\bibfnamefont {V.}~\bibnamefont
  {Barger}}, \bibinfo {author} {\bibfnamefont {S.~L.}\ \bibnamefont {Glashow}},
  \bibinfo {author} {\bibfnamefont {P.}~\bibnamefont {Langacker}}, \ and\
  \bibinfo {author} {\bibfnamefont {D.}~\bibnamefont {Marfatia}},\ }\href
  {\doibase 10.1016/S0370-2693(02)02177-9} {\bibfield  {journal} {\bibinfo
  {journal} {Phys. Lett.}\ }\textbf {\bibinfo {volume} {B540}},\ \bibinfo
  {pages} {247} (\bibinfo {year} {2002})},\ \Eprint
  {http://arxiv.org/abs/hep-ph/0205290} {arXiv:hep-ph/0205290 [hep-ph]}
  \BibitemShut {NoStop}%
\bibitem [{\citenamefont {Weinheimer}(2003)}]{Weinheimer:2003fj}%
  \BibitemOpen
  \bibfield  {author} {\bibinfo {author} {\bibfnamefont {C.}~\bibnamefont
  {Weinheimer}},\ }\bibfield  {booktitle} {\emph {\bibinfo {booktitle}
  {{Neutrino physics and astrophysics. Proceedings, 20th International
  Conference, Neutrino 2002, Munich, Germany, May 25-30, 2002}}},\ }\href
  {\doibase 10.1016/S0920-5632(03)01330-6} {\bibfield  {journal} {\bibinfo
  {journal} {Nucl. Phys. Proc. Suppl.}\ }\textbf {\bibinfo {volume} {118}},\
  \bibinfo {pages} {279} (\bibinfo {year} {2003})},\ \bibinfo {note}
  {[,279(2003)]}\BibitemShut {NoStop}%
\bibitem [{\citenamefont {Kraus}\ \emph {et~al.}(2005)\citenamefont {Kraus}
  \emph {et~al.}}]{Kraus:2004zw}%
  \BibitemOpen
  \bibfield  {author} {\bibinfo {author} {\bibfnamefont {C.}~\bibnamefont
  {Kraus}} \emph {et~al.},\ }\href {\doibase 10.1140/epjc/s2005-02139-7}
  {\bibfield  {journal} {\bibinfo  {journal} {Eur. Phys. J.}\ }\textbf
  {\bibinfo {volume} {C40}},\ \bibinfo {pages} {447} (\bibinfo {year}
  {2005})},\ \Eprint {http://arxiv.org/abs/hep-ex/0412056}
  {arXiv:hep-ex/0412056 [hep-ex]} \BibitemShut {NoStop}%
\bibitem [{\citenamefont {Lobashev}\ \emph {et~al.}(2001)\citenamefont
  {Lobashev} \emph {et~al.}}]{Lobashev:2001uu}%
  \BibitemOpen
  \bibfield  {author} {\bibinfo {author} {\bibfnamefont {V.~M.}\ \bibnamefont
  {Lobashev}} \emph {et~al.},\ }\bibfield  {booktitle} {\emph {\bibinfo
  {booktitle} {{Neutrino physics and astrophysics. Proceedings, 19th
  International Conference, Neutrino 2000, Sudbury, Canada, June 16-21,
  2000}}},\ }\href {\doibase 10.1016/S0920-5632(00)00952-X} {\bibfield
  {journal} {\bibinfo  {journal} {Nucl. Phys. Proc. Suppl.}\ }\textbf {\bibinfo
  {volume} {91}},\ \bibinfo {pages} {280} (\bibinfo {year} {2001})}\BibitemShut
  {NoStop}%
\bibitem [{\citenamefont {Osipowicz}\ \emph {et~al.}(2001)\citenamefont
  {Osipowicz} \emph {et~al.}}]{Osipowicz:2001sq}%
  \BibitemOpen
  \bibfield  {author} {\bibinfo {author} {\bibfnamefont {A.}~\bibnamefont
  {Osipowicz}} \emph {et~al.} (\bibinfo {collaboration} {KATRIN}),\ }\href@noop
  {} {\bibfield  {journal} {\bibinfo  {journal} {n/a}\ } (\bibinfo {year}
  {2001})},\ \Eprint {http://arxiv.org/abs/hep-ex/0109033}
  {arXiv:hep-ex/0109033 [hep-ex]} \BibitemShut {NoStop}%
\bibitem [{\citenamefont {Lopez-Pavon}\ \emph {et~al.}(2013)\citenamefont
  {Lopez-Pavon}, \citenamefont {Pascoli},\ and\ \citenamefont
  {Wong}}]{LopezPavon:2012zg}%
  \BibitemOpen
  \bibfield  {author} {\bibinfo {author} {\bibfnamefont {J.}~\bibnamefont
  {Lopez-Pavon}}, \bibinfo {author} {\bibfnamefont {S.}~\bibnamefont
  {Pascoli}}, \ and\ \bibinfo {author} {\bibfnamefont {C.-f.}\ \bibnamefont
  {Wong}},\ }\href {\doibase 10.1103/PhysRevD.87.093007} {\bibfield  {journal}
  {\bibinfo  {journal} {Phys. Rev.}\ }\textbf {\bibinfo {volume} {D87}},\
  \bibinfo {pages} {093007} (\bibinfo {year} {2013})},\ \Eprint
  {http://arxiv.org/abs/1209.5342} {arXiv:1209.5342 [hep-ph]} \BibitemShut
  {NoStop}%
\bibitem [{\citenamefont {Casas}\ and\ \citenamefont
  {Ibarra}(2001)}]{Casas:2001sr}%
  \BibitemOpen
  \bibfield  {author} {\bibinfo {author} {\bibfnamefont {J.~A.}\ \bibnamefont
  {Casas}}\ and\ \bibinfo {author} {\bibfnamefont {A.}~\bibnamefont {Ibarra}},\
  }\href {\doibase 10.1016/S0550-3213(01)00475-8} {\bibfield  {journal}
  {\bibinfo  {journal} {Nucl. Phys.}\ }\textbf {\bibinfo {volume} {B618}},\
  \bibinfo {pages} {171} (\bibinfo {year} {2001})},\ \Eprint
  {http://arxiv.org/abs/hep-ph/0103065} {arXiv:hep-ph/0103065 [hep-ph]}
  \BibitemShut {NoStop}%
\bibitem [{\citenamefont {Lopez-Pavon}\ \emph {et~al.}(2015)\citenamefont
  {Lopez-Pavon}, \citenamefont {Molinaro},\ and\ \citenamefont
  {Petcov}}]{Lopez-Pavon:2015cga}%
  \BibitemOpen
  \bibfield  {author} {\bibinfo {author} {\bibfnamefont {J.}~\bibnamefont
  {Lopez-Pavon}}, \bibinfo {author} {\bibfnamefont {E.}~\bibnamefont
  {Molinaro}}, \ and\ \bibinfo {author} {\bibfnamefont {S.~T.}\ \bibnamefont
  {Petcov}},\ }\href {\doibase 10.1007/JHEP11(2015)030} {\bibfield  {journal}
  {\bibinfo  {journal} {JHEP}\ }\textbf {\bibinfo {volume} {11}},\ \bibinfo
  {pages} {030} (\bibinfo {year} {2015})},\ \Eprint
  {http://arxiv.org/abs/1506.05296} {arXiv:1506.05296 [hep-ph]} \BibitemShut
  {NoStop}%
\bibitem [{\citenamefont {Chen}\ \emph {et~al.}(2016)\citenamefont {Chen},
  \citenamefont {Ding},\ and\ \citenamefont {King}}]{Chen:2016ptr}%
  \BibitemOpen
  \bibfield  {author} {\bibinfo {author} {\bibfnamefont {P.}~\bibnamefont
  {Chen}}, \bibinfo {author} {\bibfnamefont {G.-J.}\ \bibnamefont {Ding}}, \
  and\ \bibinfo {author} {\bibfnamefont {S.~F.}\ \bibnamefont {King}},\ }\href
  {\doibase 10.1007/JHEP03(2016)206} {\bibfield  {journal} {\bibinfo  {journal}
  {JHEP}\ }\textbf {\bibinfo {volume} {03}},\ \bibinfo {pages} {206} (\bibinfo
  {year} {2016})},\ \Eprint {http://arxiv.org/abs/1602.03873} {arXiv:1602.03873
  [hep-ph]} \BibitemShut {NoStop}%
\bibitem [{\citenamefont {Pascoli}\ \emph
  {et~al.}(2007{\natexlab{a}})\citenamefont {Pascoli}, \citenamefont {Petcov},\
  and\ \citenamefont {Riotto}}]{Pascoli:2006ie}%
  \BibitemOpen
  \bibfield  {author} {\bibinfo {author} {\bibfnamefont {S.}~\bibnamefont
  {Pascoli}}, \bibinfo {author} {\bibfnamefont {S.~T.}\ \bibnamefont {Petcov}},
  \ and\ \bibinfo {author} {\bibfnamefont {A.}~\bibnamefont {Riotto}},\ }\href
  {\doibase 10.1103/PhysRevD.75.083511} {\bibfield  {journal} {\bibinfo
  {journal} {Phys. Rev.}\ }\textbf {\bibinfo {volume} {D75}},\ \bibinfo {pages}
  {083511} (\bibinfo {year} {2007}{\natexlab{a}})},\ \Eprint
  {http://arxiv.org/abs/hep-ph/0609125} {arXiv:hep-ph/0609125 [hep-ph]}
  \BibitemShut {NoStop}%
\bibitem [{\citenamefont {Pascoli}\ \emph
  {et~al.}(2007{\natexlab{b}})\citenamefont {Pascoli}, \citenamefont {Petcov},\
  and\ \citenamefont {Riotto}}]{Pascoli:2006ci}%
  \BibitemOpen
  \bibfield  {author} {\bibinfo {author} {\bibfnamefont {S.}~\bibnamefont
  {Pascoli}}, \bibinfo {author} {\bibfnamefont {S.~T.}\ \bibnamefont {Petcov}},
  \ and\ \bibinfo {author} {\bibfnamefont {A.}~\bibnamefont {Riotto}},\ }\href
  {\doibase 10.1016/j.nuclphysb.2007.02.019} {\bibfield  {journal} {\bibinfo
  {journal} {Nucl. Phys.}\ }\textbf {\bibinfo {volume} {B774}},\ \bibinfo
  {pages} {1} (\bibinfo {year} {2007}{\natexlab{b}})},\ \Eprint
  {http://arxiv.org/abs/hep-ph/0611338} {arXiv:hep-ph/0611338 [hep-ph]}
  \BibitemShut {NoStop}%
\bibitem [{\citenamefont {Pilaftsis}\ and\ \citenamefont
  {Underwood}(2004)}]{Pilaftsis:2003gt}%
  \BibitemOpen
  \bibfield  {author} {\bibinfo {author} {\bibfnamefont {A.}~\bibnamefont
  {Pilaftsis}}\ and\ \bibinfo {author} {\bibfnamefont {T.~E.~J.}\ \bibnamefont
  {Underwood}},\ }\href {\doibase 10.1016/j.nuclphysb.2004.05.029} {\bibfield
  {journal} {\bibinfo  {journal} {Nucl. Phys.}\ }\textbf {\bibinfo {volume}
  {B692}},\ \bibinfo {pages} {303} (\bibinfo {year} {2004})},\ \Eprint
  {http://arxiv.org/abs/hep-ph/0309342} {arXiv:hep-ph/0309342 [hep-ph]}
  \BibitemShut {NoStop}%
\bibitem [{\citenamefont {Blanchet}\ and\ \citenamefont
  {Di~Bari}(2009)}]{Blanchet:2008pw}%
  \BibitemOpen
  \bibfield  {author} {\bibinfo {author} {\bibfnamefont {S.}~\bibnamefont
  {Blanchet}}\ and\ \bibinfo {author} {\bibfnamefont {P.}~\bibnamefont
  {Di~Bari}},\ }\href {\doibase 10.1016/j.nuclphysb.2008.08.026} {\bibfield
  {journal} {\bibinfo  {journal} {Nucl. Phys.}\ }\textbf {\bibinfo {volume}
  {B807}},\ \bibinfo {pages} {155} (\bibinfo {year} {2009})},\ \Eprint
  {http://arxiv.org/abs/0807.0743} {arXiv:0807.0743 [hep-ph]} \BibitemShut
  {NoStop}%
\bibitem [{\citenamefont {Barbieri}\ \emph {et~al.}(2000)\citenamefont
  {Barbieri}, \citenamefont {Creminelli}, \citenamefont {Strumia},\ and\
  \citenamefont {Tetradis}}]{Barbieri:1999ma}%
  \BibitemOpen
  \bibfield  {author} {\bibinfo {author} {\bibfnamefont {R.}~\bibnamefont
  {Barbieri}}, \bibinfo {author} {\bibfnamefont {P.}~\bibnamefont
  {Creminelli}}, \bibinfo {author} {\bibfnamefont {A.}~\bibnamefont {Strumia}},
  \ and\ \bibinfo {author} {\bibfnamefont {N.}~\bibnamefont {Tetradis}},\
  }\href {\doibase 10.1016/S0550-3213(00)00011-0} {\bibfield  {journal}
  {\bibinfo  {journal} {Nucl. Phys.}\ }\textbf {\bibinfo {volume} {B575}},\
  \bibinfo {pages} {61} (\bibinfo {year} {2000})},\ \Eprint
  {http://arxiv.org/abs/hep-ph/9911315} {arXiv:hep-ph/9911315 [hep-ph]}
  \BibitemShut {NoStop}%
\bibitem [{\citenamefont {Francois-Xavier}\ \emph {et~al.}(2006)\citenamefont
  {Francois-Xavier}, \citenamefont {Losada},\ and\ \citenamefont
  {Riotto}}]{Abada:2006fw}%
  \BibitemOpen
  \bibfield  {author} {\bibinfo {author} {\bibnamefont {Francois-Xavier}},
  \bibinfo {author} {\bibfnamefont {M.}~\bibnamefont {Losada}}, \ and\ \bibinfo
  {author} {\bibfnamefont {A.}~\bibnamefont {Riotto}},\ }\href {\doibase
  10.1088/1475-7516/2006/04/004} {\bibfield  {journal} {\bibinfo  {journal}
  {JCAP}\ }\textbf {\bibinfo {volume} {0604}},\ \bibinfo {pages} {004}
  (\bibinfo {year} {2006})},\ \Eprint {http://arxiv.org/abs/hep-ph/0601083}
  {arXiv:hep-ph/0601083 [hep-ph]} \BibitemShut {NoStop}%
\bibitem [{\citenamefont {De~Simone}\ and\ \citenamefont
  {Riotto}(2007{\natexlab{a}})}]{DeSimone:2006nrs}%
  \BibitemOpen
  \bibfield  {author} {\bibinfo {author} {\bibfnamefont {A.}~\bibnamefont
  {De~Simone}}\ and\ \bibinfo {author} {\bibfnamefont {A.}~\bibnamefont
  {Riotto}},\ }\href {\doibase 10.1088/1475-7516/2007/02/005} {\bibfield
  {journal} {\bibinfo  {journal} {JCAP}\ }\textbf {\bibinfo {volume} {0702}},\
  \bibinfo {pages} {005} (\bibinfo {year} {2007}{\natexlab{a}})},\ \Eprint
  {http://arxiv.org/abs/hep-ph/0611357} {arXiv:hep-ph/0611357 [hep-ph]}
  \BibitemShut {NoStop}%
\bibitem [{\citenamefont {Blanchet}\ \emph {et~al.}(2007)\citenamefont
  {Blanchet}, \citenamefont {Di~Bari},\ and\ \citenamefont
  {Raffelt}}]{Blanchet:2006ch}%
  \BibitemOpen
  \bibfield  {author} {\bibinfo {author} {\bibfnamefont {S.}~\bibnamefont
  {Blanchet}}, \bibinfo {author} {\bibfnamefont {P.}~\bibnamefont {Di~Bari}}, \
  and\ \bibinfo {author} {\bibfnamefont {G.~G.}\ \bibnamefont {Raffelt}},\
  }\href {\doibase 10.1088/1475-7516/2007/03/012} {\bibfield  {journal}
  {\bibinfo  {journal} {JCAP}\ }\textbf {\bibinfo {volume} {0703}},\ \bibinfo
  {pages} {012} (\bibinfo {year} {2007})},\ \Eprint
  {http://arxiv.org/abs/hep-ph/0611337} {arXiv:hep-ph/0611337 [hep-ph]}
  \BibitemShut {NoStop}%
\bibitem [{\citenamefont {Blanchet}\ \emph {et~al.}(2013)\citenamefont
  {Blanchet}, \citenamefont {Di~Bari}, \citenamefont {Jones},\ and\
  \citenamefont {Marzola}}]{Blanchet:2011xq}%
  \BibitemOpen
  \bibfield  {author} {\bibinfo {author} {\bibfnamefont {S.}~\bibnamefont
  {Blanchet}}, \bibinfo {author} {\bibfnamefont {P.}~\bibnamefont {Di~Bari}},
  \bibinfo {author} {\bibfnamefont {D.~A.}\ \bibnamefont {Jones}}, \ and\
  \bibinfo {author} {\bibfnamefont {L.}~\bibnamefont {Marzola}},\ }\href
  {\doibase 10.1088/1475-7516/2013/01/041} {\bibfield  {journal} {\bibinfo
  {journal} {JCAP}\ }\textbf {\bibinfo {volume} {1301}},\ \bibinfo {pages}
  {041} (\bibinfo {year} {2013})},\ \Eprint {http://arxiv.org/abs/1112.4528}
  {arXiv:1112.4528 [hep-ph]} \BibitemShut {NoStop}%
\bibitem [{\citenamefont {Abada}\ \emph {et~al.}(2006)\citenamefont {Abada},
  \citenamefont {Davidson}, \citenamefont {Ibarra}, \citenamefont
  {Josse-Michaux}, \citenamefont {Losada},\ and\ \citenamefont
  {Riotto}}]{Abada:2006ea}%
  \BibitemOpen
  \bibfield  {author} {\bibinfo {author} {\bibfnamefont {A.}~\bibnamefont
  {Abada}}, \bibinfo {author} {\bibfnamefont {S.}~\bibnamefont {Davidson}},
  \bibinfo {author} {\bibfnamefont {A.}~\bibnamefont {Ibarra}}, \bibinfo
  {author} {\bibfnamefont {F.~X.}\ \bibnamefont {Josse-Michaux}}, \bibinfo
  {author} {\bibfnamefont {M.}~\bibnamefont {Losada}}, \ and\ \bibinfo {author}
  {\bibfnamefont {A.}~\bibnamefont {Riotto}},\ }\href {\doibase
  10.1088/1126-6708/2006/09/010} {\bibfield  {journal} {\bibinfo  {journal}
  {JHEP}\ }\textbf {\bibinfo {volume} {09}},\ \bibinfo {pages} {010} (\bibinfo
  {year} {2006})},\ \Eprint {http://arxiv.org/abs/hep-ph/0605281}
  {arXiv:hep-ph/0605281 [hep-ph]} \BibitemShut {NoStop}%
\bibitem [{\citenamefont {Buchmuller}\ and\ \citenamefont
  {Plumacher}(2001)}]{Buchmuller:2001sr}%
  \BibitemOpen
  \bibfield  {author} {\bibinfo {author} {\bibfnamefont {W.}~\bibnamefont
  {Buchmuller}}\ and\ \bibinfo {author} {\bibfnamefont {M.}~\bibnamefont
  {Plumacher}},\ }\href {\doibase 10.1016/S0370-2693(01)00614-1} {\bibfield
  {journal} {\bibinfo  {journal} {Phys. Lett.}\ }\textbf {\bibinfo {volume}
  {B511}},\ \bibinfo {pages} {74} (\bibinfo {year} {2001})},\ \Eprint
  {http://arxiv.org/abs/hep-ph/0104189} {arXiv:hep-ph/0104189 [hep-ph]}
  \BibitemShut {NoStop}%
\bibitem [{\citenamefont {Nardi}\ \emph {et~al.}(2006)\citenamefont {Nardi},
  \citenamefont {Nir}, \citenamefont {Racker},\ and\ \citenamefont
  {Roulet}}]{Nardi:2005hs}%
  \BibitemOpen
  \bibfield  {author} {\bibinfo {author} {\bibfnamefont {E.}~\bibnamefont
  {Nardi}}, \bibinfo {author} {\bibfnamefont {Y.}~\bibnamefont {Nir}}, \bibinfo
  {author} {\bibfnamefont {J.}~\bibnamefont {Racker}}, \ and\ \bibinfo {author}
  {\bibfnamefont {E.}~\bibnamefont {Roulet}},\ }\href {\doibase
  10.1088/1126-6708/2006/01/068} {\bibfield  {journal} {\bibinfo  {journal}
  {JHEP}\ }\textbf {\bibinfo {volume} {01}},\ \bibinfo {pages} {068} (\bibinfo
  {year} {2006})},\ \Eprint {http://arxiv.org/abs/hep-ph/0512052}
  {arXiv:hep-ph/0512052 [hep-ph]} \BibitemShut {NoStop}%
\bibitem [{\citenamefont {Schwaller}(2014)}]{Schwaller:2014hna}%
  \BibitemOpen
  \bibfield  {author} {\bibinfo {author} {\bibfnamefont {P.}~\bibnamefont
  {Schwaller}},\ }in\ \href
  {http://moriond.in2p3.fr/Proceedings/2014/Moriond_EW_2014.pdf} {\emph
  {\bibinfo {booktitle} {{Proceedings, 49th Rencontres de Moriond on
  Electroweak Interactions and Unified Theories: La Thuile, Italy, March 15-22,
  2014}}}}\ (\bibinfo {year} {2014})\ pp.\ \bibinfo {pages}
  {231--236}\BibitemShut {NoStop}%
\bibitem [{\citenamefont {Kiessig}\ \emph {et~al.}(2010)\citenamefont
  {Kiessig}, \citenamefont {Plumacher},\ and\ \citenamefont
  {Thoma}}]{Kiessig:2010pr}%
  \BibitemOpen
  \bibfield  {author} {\bibinfo {author} {\bibfnamefont {C.~P.}\ \bibnamefont
  {Kiessig}}, \bibinfo {author} {\bibfnamefont {M.}~\bibnamefont {Plumacher}},
  \ and\ \bibinfo {author} {\bibfnamefont {M.~H.}\ \bibnamefont {Thoma}},\
  }\href {\doibase 10.1103/PhysRevD.82.036007} {\bibfield  {journal} {\bibinfo
  {journal} {Phys. Rev.}\ }\textbf {\bibinfo {volume} {D82}},\ \bibinfo {pages}
  {036007} (\bibinfo {year} {2010})},\ \Eprint {http://arxiv.org/abs/1003.3016}
  {arXiv:1003.3016 [hep-ph]} \BibitemShut {NoStop}%
\bibitem [{\citenamefont {Giudice}\ \emph {et~al.}(2004)\citenamefont
  {Giudice}, \citenamefont {Notari}, \citenamefont {Raidal}, \citenamefont
  {Riotto},\ and\ \citenamefont {Strumia}}]{Giudice:2003jh}%
  \BibitemOpen
  \bibfield  {author} {\bibinfo {author} {\bibfnamefont {G.~F.}\ \bibnamefont
  {Giudice}}, \bibinfo {author} {\bibfnamefont {A.}~\bibnamefont {Notari}},
  \bibinfo {author} {\bibfnamefont {M.}~\bibnamefont {Raidal}}, \bibinfo
  {author} {\bibfnamefont {A.}~\bibnamefont {Riotto}}, \ and\ \bibinfo {author}
  {\bibfnamefont {A.}~\bibnamefont {Strumia}},\ }\href {\doibase
  10.1016/j.nuclphysb.2004.02.019} {\bibfield  {journal} {\bibinfo  {journal}
  {Nucl. Phys.}\ }\textbf {\bibinfo {volume} {B685}},\ \bibinfo {pages} {89}
  (\bibinfo {year} {2004})},\ \Eprint {http://arxiv.org/abs/hep-ph/0310123}
  {arXiv:hep-ph/0310123 [hep-ph]} \BibitemShut {NoStop}%
\bibitem [{\citenamefont {De~Simone}\ and\ \citenamefont
  {Riotto}(2007{\natexlab{b}})}]{DeSimone:2007gkc}%
  \BibitemOpen
  \bibfield  {author} {\bibinfo {author} {\bibfnamefont {A.}~\bibnamefont
  {De~Simone}}\ and\ \bibinfo {author} {\bibfnamefont {A.}~\bibnamefont
  {Riotto}},\ }\href {\doibase 10.1088/1475-7516/2007/08/002} {\bibfield
  {journal} {\bibinfo  {journal} {JCAP}\ }\textbf {\bibinfo {volume} {0708}},\
  \bibinfo {pages} {002} (\bibinfo {year} {2007}{\natexlab{b}})},\ \Eprint
  {http://arxiv.org/abs/hep-ph/0703175} {arXiv:hep-ph/0703175 [hep-ph]}
  \BibitemShut {NoStop}%
\bibitem [{\citenamefont {Beneke}\ \emph {et~al.}(2011)\citenamefont {Beneke},
  \citenamefont {Garbrecht}, \citenamefont {Fidler}, \citenamefont {Herranen},\
  and\ \citenamefont {Schwaller}}]{Beneke:2010dz}%
  \BibitemOpen
  \bibfield  {author} {\bibinfo {author} {\bibfnamefont {M.}~\bibnamefont
  {Beneke}}, \bibinfo {author} {\bibfnamefont {B.}~\bibnamefont {Garbrecht}},
  \bibinfo {author} {\bibfnamefont {C.}~\bibnamefont {Fidler}}, \bibinfo
  {author} {\bibfnamefont {M.}~\bibnamefont {Herranen}}, \ and\ \bibinfo
  {author} {\bibfnamefont {P.}~\bibnamefont {Schwaller}},\ }\href {\doibase
  10.1016/j.nuclphysb.2010.10.001} {\bibfield  {journal} {\bibinfo  {journal}
  {Nucl. Phys.}\ }\textbf {\bibinfo {volume} {B843}},\ \bibinfo {pages} {177}
  (\bibinfo {year} {2011})},\ \Eprint {http://arxiv.org/abs/1007.4783}
  {arXiv:1007.4783 [hep-ph]} \BibitemShut {NoStop}%
\bibitem [{\citenamefont {Anisimov}\ \emph {et~al.}(2011)\citenamefont
  {Anisimov}, \citenamefont {Buchmüller}, \citenamefont {Drewes},\ and\
  \citenamefont {Mendizabal}}]{Anisimov:2010dk}%
  \BibitemOpen
  \bibfield  {author} {\bibinfo {author} {\bibfnamefont {A.}~\bibnamefont
  {Anisimov}}, \bibinfo {author} {\bibfnamefont {W.}~\bibnamefont
  {Buchmüller}}, \bibinfo {author} {\bibfnamefont {M.}~\bibnamefont {Drewes}},
  \ and\ \bibinfo {author} {\bibfnamefont {S.}~\bibnamefont {Mendizabal}},\
  }\href {\doibase 10.1016/j.aop.2011.02.002, 10.1016/j.aop.2013.05.00}
  {\bibfield  {journal} {\bibinfo  {journal} {Annals Phys.}\ }\textbf {\bibinfo
  {volume} {326}},\ \bibinfo {pages} {1998} (\bibinfo {year} {2011})},\
  \bibinfo {note} {[Erratum: Annals Phys.338,376(2011)]},\ \Eprint
  {http://arxiv.org/abs/1012.5821} {arXiv:1012.5821 [hep-ph]} \BibitemShut
  {NoStop}%
\bibitem [{\citenamefont {Beneke}\ \emph {et~al.}(2010)\citenamefont {Beneke},
  \citenamefont {Garbrech}, \citenamefont {Herranen},\ and\ \citenamefont
  {Schwaller}}]{Beneke:2010wd}%
  \BibitemOpen
  \bibfield  {author} {\bibinfo {author} {\bibfnamefont {M.}~\bibnamefont
  {Beneke}}, \bibinfo {author} {\bibfnamefont {B.}~\bibnamefont {Garbrech}},
  \bibinfo {author} {\bibfnamefont {M.}~\bibnamefont {Herranen}}, \ and\
  \bibinfo {author} {\bibfnamefont {P.}~\bibnamefont {Schwaller}},\ }\href
  {\doibase 10.1016/j.nuclphysb.2010.05.003} {\bibfield  {journal} {\bibinfo
  {journal} {Nucl. Phys.}\ }\textbf {\bibinfo {volume} {B838}},\ \bibinfo
  {pages} {1} (\bibinfo {year} {2010})},\ \Eprint
  {http://arxiv.org/abs/1002.1326} {arXiv:1002.1326 [hep-ph]} \BibitemShut
  {NoStop}%
\bibitem [{\citenamefont {Davidson}\ \emph {et~al.}(2008)\citenamefont
  {Davidson}, \citenamefont {Nardi},\ and\ \citenamefont
  {Nir}}]{Davidson:2008bu}%
  \BibitemOpen
  \bibfield  {author} {\bibinfo {author} {\bibfnamefont {S.}~\bibnamefont
  {Davidson}}, \bibinfo {author} {\bibfnamefont {E.}~\bibnamefont {Nardi}}, \
  and\ \bibinfo {author} {\bibfnamefont {Y.}~\bibnamefont {Nir}},\ }\href
  {\doibase 10.1016/j.physrep.2008.06.002} {\bibfield  {journal} {\bibinfo
  {journal} {Phys. Rept.}\ }\textbf {\bibinfo {volume} {466}},\ \bibinfo
  {pages} {105} (\bibinfo {year} {2008})},\ \Eprint
  {http://arxiv.org/abs/0802.2962} {arXiv:0802.2962 [hep-ph]} \BibitemShut
  {NoStop}%
\bibitem [{\citenamefont {Blanchet}\ and\ \citenamefont
  {Di~Bari}(2007)}]{Blanchet:2006be}%
  \BibitemOpen
  \bibfield  {author} {\bibinfo {author} {\bibfnamefont {S.}~\bibnamefont
  {Blanchet}}\ and\ \bibinfo {author} {\bibfnamefont {P.}~\bibnamefont
  {Di~Bari}},\ }\href {\doibase 10.1088/1475-7516/2007/03/018} {\bibfield
  {journal} {\bibinfo  {journal} {JCAP}\ }\textbf {\bibinfo {volume} {0703}},\
  \bibinfo {pages} {018} (\bibinfo {year} {2007})},\ \Eprint
  {http://arxiv.org/abs/hep-ph/0607330} {arXiv:hep-ph/0607330 [hep-ph]}
  \BibitemShut {NoStop}%
\bibitem [{\citenamefont {Frossard}\ \emph {et~al.}(2013)\citenamefont
  {Frossard}, \citenamefont {Kartavtsev},\ and\ \citenamefont
  {Mitrouskas}}]{Frossard:2013bra}%
  \BibitemOpen
  \bibfield  {author} {\bibinfo {author} {\bibfnamefont {T.}~\bibnamefont
  {Frossard}}, \bibinfo {author} {\bibfnamefont {A.}~\bibnamefont
  {Kartavtsev}}, \ and\ \bibinfo {author} {\bibfnamefont {D.}~\bibnamefont
  {Mitrouskas}},\ }\href {\doibase 10.1103/PhysRevD.87.125006} {\bibfield
  {journal} {\bibinfo  {journal} {Phys. Rev.}\ }\textbf {\bibinfo {volume}
  {D87}},\ \bibinfo {pages} {125006} (\bibinfo {year} {2013})},\ \Eprint
  {http://arxiv.org/abs/1304.1719} {arXiv:1304.1719 [hep-ph]} \BibitemShut
  {NoStop}%
\bibitem [{\citenamefont {Nardi}\ \emph {et~al.}(2007)\citenamefont {Nardi},
  \citenamefont {Racker},\ and\ \citenamefont {Roulet}}]{Nardi:2007jp}%
  \BibitemOpen
  \bibfield  {author} {\bibinfo {author} {\bibfnamefont {E.}~\bibnamefont
  {Nardi}}, \bibinfo {author} {\bibfnamefont {J.}~\bibnamefont {Racker}}, \
  and\ \bibinfo {author} {\bibfnamefont {E.}~\bibnamefont {Roulet}},\ }\href
  {\doibase 10.1088/1126-6708/2007/09/090} {\bibfield  {journal} {\bibinfo
  {journal} {JHEP}\ }\textbf {\bibinfo {volume} {09}},\ \bibinfo {pages} {090}
  (\bibinfo {year} {2007})},\ \Eprint {http://arxiv.org/abs/0707.0378}
  {arXiv:0707.0378 [hep-ph]} \BibitemShut {NoStop}%
\bibitem [{\citenamefont {Garbrecht}\ and\ \citenamefont
  {Schwaller}(2014)}]{Garbrecht:2014kda}%
  \BibitemOpen
  \bibfield  {author} {\bibinfo {author} {\bibfnamefont {B.}~\bibnamefont
  {Garbrecht}}\ and\ \bibinfo {author} {\bibfnamefont {P.}~\bibnamefont
  {Schwaller}},\ }\href {\doibase 10.1088/1475-7516/2014/10/012} {\bibfield
  {journal} {\bibinfo  {journal} {JCAP}\ }\textbf {\bibinfo {volume} {1410}},\
  \bibinfo {pages} {012} (\bibinfo {year} {2014})},\ \Eprint
  {http://arxiv.org/abs/1404.2915} {arXiv:1404.2915 [hep-ph]} \BibitemShut
  {NoStop}%
\bibitem [{\citenamefont {Weckesser}(14  )}]{odeintw}%
  \BibitemOpen
  \bibfield  {author} {\bibinfo {author} {\bibfnamefont {W.}~\bibnamefont
  {Weckesser}},\ }\href {https://github.com/WarrenWeckesser/odeintw} {\enquote
  {\bibinfo {title} {{odeintw}: Complex and matrix differential equations.}}\ }
  (\bibinfo {year} {2014--})\BibitemShut {NoStop}%
\bibitem [{ode()}]{odepack}%
  \BibitemOpen
  \href {http://www.llnl.gov/CASC/nsde/pubs/u88007.pdf} {\enquote {\bibinfo
  {title} {{A. C. Hindmarsh, ``ODEPACK, A Systematized Collection of ODE
  Solvers,'' in Scientific Computing, R. S. Stepleman et al. (eds.),
  North-Holland, Amsterdam, 1983 (vol. 1 of IMACS Transactions on Scientific
  Computation), pp. 55-64.}}}\ }\BibitemShut {NoStop}%
\bibitem [{\citenamefont {Jones}\ \emph {et~al.}(01  )\citenamefont {Jones},
  \citenamefont {Oliphant}, \citenamefont {Peterson} \emph {et~al.}}]{scipy}%
  \BibitemOpen
  \bibfield  {author} {\bibinfo {author} {\bibfnamefont {E.}~\bibnamefont
  {Jones}}, \bibinfo {author} {\bibfnamefont {T.}~\bibnamefont {Oliphant}},
  \bibinfo {author} {\bibfnamefont {P.}~\bibnamefont {Peterson}},  \emph
  {et~al.},\ }\href {http://www.scipy.org/} {\enquote {\bibinfo {title}
  {{SciPy}: Open source scientific tools for {Python}},}\ } (\bibinfo {year}
  {2001--})\BibitemShut {NoStop}%
\bibitem [{\citenamefont {Feroz}\ \emph {et~al.}(2009)\citenamefont {Feroz},
  \citenamefont {Hobson},\ and\ \citenamefont {Bridges}}]{Feroz:2008xx}%
  \BibitemOpen
  \bibfield  {author} {\bibinfo {author} {\bibfnamefont {F.}~\bibnamefont
  {Feroz}}, \bibinfo {author} {\bibfnamefont {M.~P.}\ \bibnamefont {Hobson}}, \
  and\ \bibinfo {author} {\bibfnamefont {M.}~\bibnamefont {Bridges}},\ }\href
  {\doibase 10.1111/j.1365-2966.2009.14548.x} {\bibfield  {journal} {\bibinfo
  {journal} {Mon. Not. Roy. Astron. Soc.}\ }\textbf {\bibinfo {volume} {398}},\
  \bibinfo {pages} {1601} (\bibinfo {year} {2009})},\ \Eprint
  {http://arxiv.org/abs/0809.3437} {arXiv:0809.3437 [astro-ph]} \BibitemShut
  {NoStop}%
\bibitem [{\citenamefont {Feroz}\ and\ \citenamefont
  {Hobson}(2008)}]{Feroz:2007kg}%
  \BibitemOpen
  \bibfield  {author} {\bibinfo {author} {\bibfnamefont {F.}~\bibnamefont
  {Feroz}}\ and\ \bibinfo {author} {\bibfnamefont {M.~P.}\ \bibnamefont
  {Hobson}},\ }\href {\doibase 10.1111/j.1365-2966.2007.12353.x} {\bibfield
  {journal} {\bibinfo  {journal} {Mon. Not. Roy. Astron. Soc.}\ }\textbf
  {\bibinfo {volume} {384}},\ \bibinfo {pages} {449} (\bibinfo {year}
  {2008})},\ \Eprint {http://arxiv.org/abs/0704.3704} {arXiv:0704.3704
  [astro-ph]} \BibitemShut {NoStop}%
\bibitem [{\citenamefont {{Feroz}}\ \emph {et~al.}(2013)\citenamefont
  {{Feroz}}, \citenamefont {{Hobson}}, \citenamefont {{Cameron}},\ and\
  \citenamefont {{Pettitt}}}]{2013arXiv1306.2144F}%
  \BibitemOpen
  \bibfield  {author} {\bibinfo {author} {\bibfnamefont {F.}~\bibnamefont
  {{Feroz}}}, \bibinfo {author} {\bibfnamefont {M.~P.}\ \bibnamefont
  {{Hobson}}}, \bibinfo {author} {\bibfnamefont {E.}~\bibnamefont {{Cameron}}},
  \ and\ \bibinfo {author} {\bibfnamefont {A.~N.}\ \bibnamefont {{Pettitt}}},\
  }\href@noop {} {\bibfield  {journal} {\bibinfo  {journal} {ArXiv e-prints}\ }
  (\bibinfo {year} {2013})},\ \Eprint {http://arxiv.org/abs/1306.2144}
  {arXiv:1306.2144 [astro-ph.IM]} \BibitemShut {NoStop}%
\bibitem [{\citenamefont {{Buchner, J.}}\ \emph {et~al.}(2014)\citenamefont
  {{Buchner, J.}}, \citenamefont {{Georgakakis, A.}}, \citenamefont {{Nandra,
  K.}}, \citenamefont {{Hsu, L.}}, \citenamefont {{Rangel, C.}}, \citenamefont
  {{Brightman, M.}}, \citenamefont {{Merloni, A.}}, \citenamefont {{Salvato,
  M.}}, \citenamefont {{Donley, J.}},\ and\ \citenamefont {{Kocevski,
  D.}}}]{pymultinest}%
  \BibitemOpen
  \bibfield  {author} {\bibinfo {author} {\bibnamefont {{Buchner, J.}}},
  \bibinfo {author} {\bibnamefont {{Georgakakis, A.}}}, \bibinfo {author}
  {\bibnamefont {{Nandra, K.}}}, \bibinfo {author} {\bibnamefont {{Hsu, L.}}},
  \bibinfo {author} {\bibnamefont {{Rangel, C.}}}, \bibinfo {author}
  {\bibnamefont {{Brightman, M.}}}, \bibinfo {author} {\bibnamefont {{Merloni,
  A.}}}, \bibinfo {author} {\bibnamefont {{Salvato, M.}}}, \bibinfo {author}
  {\bibnamefont {{Donley, J.}}}, \ and\ \bibinfo {author} {\bibnamefont
  {{Kocevski, D.}}},\ }\href {\doibase 10.1051/0004-6361/201322971} {\bibfield
  {journal} {\bibinfo  {journal} {A\&A}\ }\textbf {\bibinfo {volume} {564}},\
  \bibinfo {pages} {A125} (\bibinfo {year} {2014})}\BibitemShut {NoStop}%
\bibitem [{\citenamefont {Fowlie}\ and\ \citenamefont
  {Bardsley}(2016)}]{Fowlie:2016hew}%
  \BibitemOpen
  \bibfield  {author} {\bibinfo {author} {\bibfnamefont {A.}~\bibnamefont
  {Fowlie}}\ and\ \bibinfo {author} {\bibfnamefont {M.~H.}\ \bibnamefont
  {Bardsley}},\ }\href {\doibase 10.1140/epjp/i2016-16391-0} {\bibfield
  {journal} {\bibinfo  {journal} {Eur. Phys. J. Plus}\ }\textbf {\bibinfo
  {volume} {131}},\ \bibinfo {pages} {391} (\bibinfo {year} {2016})},\ \Eprint
  {http://arxiv.org/abs/1603.00555} {arXiv:1603.00555 [physics.data-an]}
  \BibitemShut {NoStop}%
\bibitem [{\citenamefont {Molinaro}\ \emph {et~al.}(2008)\citenamefont
  {Molinaro}, \citenamefont {Petcov}, \citenamefont {Shindou},\ and\
  \citenamefont {Takanishi}}]{Molinaro:2007uv}%
  \BibitemOpen
  \bibfield  {author} {\bibinfo {author} {\bibfnamefont {E.}~\bibnamefont
  {Molinaro}}, \bibinfo {author} {\bibfnamefont {S.~T.}\ \bibnamefont
  {Petcov}}, \bibinfo {author} {\bibfnamefont {T.}~\bibnamefont {Shindou}}, \
  and\ \bibinfo {author} {\bibfnamefont {Y.}~\bibnamefont {Takanishi}},\ }\href
  {\doibase 10.1016/j.nuclphysb.2007.12.033} {\bibfield  {journal} {\bibinfo
  {journal} {Nucl. Phys.}\ }\textbf {\bibinfo {volume} {B797}},\ \bibinfo
  {pages} {93} (\bibinfo {year} {2008})},\ \Eprint
  {http://arxiv.org/abs/0709.0413} {arXiv:0709.0413 [hep-ph]} \BibitemShut
  {NoStop}%
\bibitem [{\citenamefont {Antusch}\ \emph {et~al.}(2012)\citenamefont
  {Antusch}, \citenamefont {Di~Bari}, \citenamefont {Jones},\ and\
  \citenamefont {King}}]{Antusch:2011nz}%
  \BibitemOpen
  \bibfield  {author} {\bibinfo {author} {\bibfnamefont {S.}~\bibnamefont
  {Antusch}}, \bibinfo {author} {\bibfnamefont {P.}~\bibnamefont {Di~Bari}},
  \bibinfo {author} {\bibfnamefont {D.~A.}\ \bibnamefont {Jones}}, \ and\
  \bibinfo {author} {\bibfnamefont {S.~F.}\ \bibnamefont {King}},\ }\href
  {\doibase 10.1103/PhysRevD.86.023516} {\bibfield  {journal} {\bibinfo
  {journal} {Phys. Rev.}\ }\textbf {\bibinfo {volume} {D86}},\ \bibinfo {pages}
  {023516} (\bibinfo {year} {2012})},\ \Eprint {http://arxiv.org/abs/1107.6002}
  {arXiv:1107.6002 [hep-ph]} \BibitemShut {NoStop}%
\end{thebibliography}%

\end{document}